\documentclass[fleqn,usenatbib]{mnras}

\usepackage{newtxtext,newtxmath}
\usepackage[T1]{fontenc}

\DeclareRobustCommand{\VAN}[3]{#2}
\let\VANthebibliography\thebibliography
\def\thebibliography{\DeclareRobustCommand{\VAN}[3]{##3}\VANthebibliography}

\usepackage{graphicx}	% Including figure files
\usepackage{amsmath}	% Advanced maths commands
\usepackage{tabularx}
 \usepackage{rotating}
 \usepackage{pdflscape}
 \usepackage{academicons}
 \usepackage{hyperref}
 \usepackage{xcolor}
 \usepackage{tikz}
 \usepackage{longtable}
\newcolumntype{L}[1]{>{\arraybackslash}p{#1}}

\newcommand{\orcid}[1]{\href{https://orcid.org/#1}{\textcolor[HTML]{A6CE39}{\aiOrcid}}}

\definecolor{lime}{HTML}{A6CE39}
\DeclareRobustCommand{\orcidicon}{%
    \begin{tikzpicture}
    \draw[lime, fill=lime] (0,0) 
    circle [radius=0.16] 
    node[white] {{\fontfamily{qag}\selectfont \tiny ID}};
    \draw[white, fill=white] (-0.0625,0.095) 
    circle [radius=0.007];
    \end{tikzpicture}
    \hspace{-2mm}
}

\newcommand{\orcidJHom}{\href{https://orcid.org/0000-0001-9994-2142}{\orcidicon}}
\newcommand{\orcidEsposito}{\href{https://orcid.org/0000-0002-0792-3719}{\orcidicon}}
\newcommand{\orcidKalas}{\href{https://orcid.org/0000-0002-6221-5360}{\orcidicon}}
\newcommand{\orcidCrotts}{\href{https://orcid.org/0000-0003-4909-256X}{\orcidicon}}
\newcommand{\orcidGonzales}{\href{https://orcid.org/0000-0003-4636-6676}{\orcidicon}}
\newcommand{\orcidLewis}{\href{https://orcid.org/0000-0002-8984-4319}{\orcidicon}}
\newcommand{\orcidMatthews}{\href{https://orcid.org/0000-0003-3017-9577}{\orcidicon}}
\newcommand{\orcidRice}{\href{https://orcid.org/0000-0002-7670-670X}{\orcidicon}}
\newcommand{\orcidWilner}{\href{https://orcid.org/0000-0003-1526-7587}{\orcidicon}}
\newcommand{\orcidDeRosa}{\href{https://orcid.org/0000-0002-4918-0247}{\orcidicon}}
\newcommand{\orcidLopez}{\href{https://orcid.org/0000-0002-2019-4995}{\orcidicon}}
\newcommand{\orcidNielsen}{\href{https://orcid.org/0000-0001-6975-9056}{\orcidicon}}
\newcommand{\orcidOppenheimer}{\href{https://orcid.org/0000-0001-7130-7681}{\orcidicon}}
\newcommand{\orcidRen}{\href{https://orcid.org/0000-0003-1698-9696}{\orcidicon}}
\newcommand{\orcidWeinberger}{\href{https://orcid.org/0000-0001-6654-7859}{\orcidicon}}
\newcommand{\orcidPadgett}{\href{https://orcid.org/0000-0001-5334-5107}{\orcidicon}}
\newcommand{\orcidMarchis}{\href{https://orcid.org/0000-0001-7016-7277}{\orcidicon}}
\newcommand{\orcidWolff}{\href{https://orcid.org/0000-0002-9977-8255}{\orcidicon}}
\newcommand{\orcidMax}{\href{https://orcid.org/0000-0001-6205-9233}{\orcidicon}}
\newcommand{\orcidMazoyer}{\href{https://orcid.org/0000-0002-9133-3091}{\orcidicon}}

\title[A Uniform Analysis of Debris Disks with GPI II, Hom et al. 2023]{A Uniform Analysis of Debris Disks with the Gemini Planet Imager II: Constraints on Dust Density Distribution Using Empirically-Informed Scattering Phase Functions}

\author[J. Hom et al.]{
Justin Hom,$^{1,2,\orcidJHom}$\thanks{E-mail: jrhom@asu.edu}
J. Patience,$^{1}$
C.~H. Chen,$^{3}$
G. Duch{\^e}ne,$^{4,5}$
J. Mazoyer,$^{6,\orcidMazoyer}$
M.~A. Millar-Blanchaer,$^{7,\orcidMax}$
\newauthor
T.~M. Esposito,$^{4,8,\orcidEsposito}$
P. Kalas,$^{4,8,9,\orcidKalas}$
K.~A. Crotts,$^{10,\orcidCrotts}$
E.~C. Gonzales,$^{11,\orcidGonzales}$
L. Kolokolova,$^{12}$
B.~L. Lewis,$^{13,\orcidLewis}$
\newauthor
B.~C. Matthews,$^{14,\orcidMatthews}$
M. Rice,$^{15,\orcidRice}$
A.~J. Weinberger,$^{16,\orcidWeinberger}$
D.~J. Wilner,$^{17,\orcidWilner}$
S.~G. Wolff,$^{2,\orcidWolff}$
S. Bruzzone,$^{18}$
\newauthor
E. Choquet,$^{19}$
J. Debes,$^{20}$
R.~J. De Rosa,$^{21,\orcidDeRosa}$
J. Donaldson,$^{16}$
Z. Draper,$^{10}$
M.~P. Fitzgerald,$^{13}$
D.~C. Hines,$^{3}$
\newauthor
S. Hinkley,$^{22}$
A.~M. Hughes,$^{23}$
R.~A. L{\'o}pez,$^{13,\orcidLopez}$
F. Marchis,$^{8,\orcidMarchis}$
S. Metchev,$^{18}$
A. Moro-Martin,$^{3}$
E. Nesvold,$^{13,24}$
\newauthor
E.~L. Nielsen,$^{25,\orcidNielsen}$
R. Oppenheimer,$^{26,\orcidOppenheimer}$
D. Padgett,$^{27,\orcidPadgett}$
M.~D. Perrin,$^{3}$
L. Pueyo,$^{3}$
F. Rantakyr{\"o},$^{28}$
\newauthor
B.~B. Ren,$^{5,29,\orcidRen}$
G. Schneider,$^{2}$
R. Soummer,$^{3}$
I. Song,$^{30}$
and C. C. Stark$^{31}$
\\
$^{1}$School of Earth and Space Exploration, Arizona State University, Tempe, AZ 85281, USA\\
$^{2}$Steward Observatory and Department of Astronomy, The University of Arizona, 933 North Cherry Avenue, Tucson, AZ 85721, USA\\
$^{3}$Space Telescope Science Institute, Baltimore, MD 21218, USA\\
$^{4}$Astronomy Department, University of California, Berkeley, CA 94720, USA\\
$^{5}$Universit{\'e} Grenoble Alpes/CNRS, Institut de Plan{\'e}tologie et d’Astrophysique de Grenoble, F-38000 Grenoble, France\\
$^{6}$LESIA, Observatoire de Paris, Universit{\'e} PSL, CNRS, Universit{\'e} Paris Cit{\'e}, Sorbonne Universit{\'e}, 5 place Jules Janssen, F-92195 Meudon, France\\
$^{7}$Department of Physics, University of California, Santa Barbara, CA 93106, USA\\
$^{8}$SETI Institute, Carl Sagan Center, 189 Bernardo Ave., Mountain View CA 94043, USA\\
$^{9}$Institute of Astrophysics, FORTH, GR-71110 Heraklion, Greece\\
$^{10}$Physics \& Astronomy Department, University of Victoria, 3800 Finnerty Rd. Victoria, BC, V8P 5C2\\
$^{11}$Department of Physics \& Astronomy, San Francisco State University, 1600 Holloway Avenue, San Francisco, CA 94132, USA\\
$^{12}$University of Maryland, College Park, MD 20742, USA\\
$^{13}$Department of Physics and Astronomy, University of California, Los Angeles, 475 Portola Plaza, Los Angeles, CA 90025, USA\\
$^{14}$Herzberg Astronomy \& Astrophysics Research Centre, National Research Council of Canada, 5071 West Saanich Road, Victoria, BC, V9E 2E7, Canada\\
$^{15}$Department of Astronomy, Yale University, New Haven, CT 06511, USA\\
$^{16}$Earth and Planets Laboratory, Carnegie Institution for Science, 5241 Broad Branch Rd NW, Washington, DC 20015\\
$^{17}$Center for Astrophysics, Harvard \& Smithsonian, 60 Garden Street, Cambridge, MA 02138, USA\\
$^{18}$Department of Physics and Astronomy, Centre for Planetary Science and Exploration, The University of Western Ontario, London, ON N6A 3K7, Canada\\
$^{19}$Aix Marseille Univ, CNRS, CNES, LAM, Marseille, France\\
$^{20}$AURA for ESA, Space Telescope Science Institute, 3700 San Martin Dr., Baltimore, MD 21218, USA\\
$^{21}$European Southern Observatory, Alonso de C\'{o}rdova 3107, Vitacura, Santiago, Chile\\
$^{22}$University of Exeter, Astrophysics Group, Physics Building, Stocker Road, Exeter, EX4 4QL, UK\\
$^{23}$Astronomy Department and Van Vleck Observatory, Wesleyan University, 96 Foss Hill Drive, Middletown, CT 06459, USA\\
$^{24}$Mani L. Bhaumik Institute for Theoretical Physics, University of California, Los Angeles, Los Angeles, CA 90095, USA\\
$^{25}$Department of Astronomy, New Mexico State University, P.O. Box 30001, MSC 4500, Las Cruces, NM 88003, USA\\
$^{26}$Department of Astrophysics, American Museum of Natural History, 79th Street at Central Park West, New York, NY 10024\\
$^{27}$Jet Propulsion Laboratory, California Institute of Technology, 4800 Oak Grove Drive, Pasadena, CA, 91109, USA\\
$^{28}$Gemini Observatory, Casilla 603, La Serena, Chile\\
$^{29}$Universit\'{e} C\^{o}te d'Azur, Observatoire de la C\^{o}te d'Azur, CNRS, Laboratoire Lagrange, Bd de l'Observatoire, CS 34229, 06304 Nice cedex 4, France\\
$^{30}$Department of Physics and Astronomy, University of Georgia, Athens, GA 30602, USA\\
$^{31}$NASA Goddard Space Flight Center, Greenbelt, MD 20771, USA\\
}

\date{Accepted 2024 January 31. Received 2024 January 30; in original form 2023 August 30}

\pubyear{2024}

\begin{document}
\label{firstpage}
\pagerange{\pageref{firstpage}--\pageref{lastpage}}
\maketitle

\clearpage

\begin{abstract}
Spatially-resolved images of debris disks are necessary to determine disk morphological properties and the scattering phase function (SPF) which quantifies the brightness of scattered light as a function of phase angle. Current high-contrast imaging instruments have successfully resolved several dozens of debris disks around other stars, but few studies have investigated trends in the scattered-light, resolved population of debris disks in a uniform and consistent manner. We have combined Karhunen-Loeve Image Projection (KLIP) with radiative-transfer disk forward modeling in order to obtain the highest quality image reductions and constrain disk morphological properties of eight debris disks imaged by the Gemini Planet Imager at \textit{H}-band with a consistent and uniformly-applied approach. In describing the scattering properties of our models, we assume a common SPF informed from solar system dust scattering measurements and apply it to all systems. We identify a diverse range of dust density properties among the sample, including critical radius, radial width, and vertical width. We also identify radially narrow and vertically extended disks that may have resulted from substellar companion perturbations, along with a tentative positive trend in disk eccentricity with relative disk width. We also find that using a common SPF can achieve reasonable model fits for disks that are axisymmetric and asymmetric when fitting models to each side of the disk independently, suggesting that scattering behavior from debris disks may be similar to Solar System dust.
\end{abstract}


%%%%%%%%%%%%%%%%%%%%%%%%%%%%%%%%%%%%%%%%%%%%%%%%%%

%%%%%%%%%%%%%%%%% BODY OF PAPER %%%%%%%%%%%%%%%%%%

\section{Introduction}
Since the first indirect detection of a circumstellar debris disk around Vega \citep{aumann1984}, numerous studies have investigated the properties and structures of these dusty systems across a wide range of wavelength regimes and angular resolutions. Resolved disks have been observed in sizes ranging from tens to hundreds of AU in diameter (e.g. \citealt{schneider1999}). Due to the short timescale processes of Poynting-Robertson drag and radiation pressure, debris disks must continually produce dust to sustain their large and extended structures. These dust-replenishing properties may indicate the presence of ongoing planet formation and/or dynamical influence, such as the collisional grinding of planetesimals \citep{backman1993} or collisions of planets \citep{cameron1997}.

Direct imaging provides significant insight into the architecture of a debris disk system. While spectral energy distributions (SEDs) can place some constraints on the radial extent, dust mass, and composition of system dust, debris disk images can more directly constrain overall properties of disk morphology such as the radial extents of dust and planetesimal belts \citep{esposito2020}. Substellar companions can directly influence the shapes of these belts, inducing features such as gaps, warps, and clumps that can be identified from resolved imaging. In addition to informing studies of disk dynamics and evolution (e.g. \citealt{lee2016}), resolved debris disk images can also constrain the gravitational interactions between planets and disks (e.g., \citealt{liou1999,kuchner2003,quillen2006,wyatt2006}).

Direct imaging studies at the scales of 0$\farcs$01-1$\arcsec$ have been explored at optical and near-IR wavelengths with instruments such as the Space Telscope Imaging Spectrograph (STIS; e.g. \citealt{schneider2014,schneider2016}) and ground-based adaptive optics (e.g. Nasmyth Adaptive Optics System-COude Near Infrared CamerA; NACO; \citealt{lenzen2003naos}; Coronagraphic High Angular Resolution Imaging Spectrograph; CHARIS; \citealt{groff2015charis}). At near-IR wavelengths, (sub)micron-sized dust grains are expected to scatter light from the host stars they surround. These observations are technically challenging, as the disk brightness from dust scattering is typically $\sim10^6$ times fainter than the host star brightness. Current generation direct imaging instruments such as the Gemini Planet Imager (GPI; \citealt{macintosh2014}) and the Spectro-Polarimetric High-contrast Exoplanet REsearch (SPHERE; \citealt{beuzit2019}) utilize adaptive optics (AO) and coronagraphy to increase the contrast by minimizing residual atmospheric turbulence (e.g. \citealt{poyneer2014}) and blocking light from the host star (e.g. \citealt{soummer2009}). Increased sensitivity and spatial resolution allow for easier identification and characterization of subtle asymmetric features such as eccentricity, warps, and clumps that could be attributed to the presence of substellar companions or ongoing planet formation.

Since its first light in 2014, GPI has spatially resolved $\sim15$ debris disks for the first time (e.g. \citealt{esposito2018,kalas2015,hom2020,hung2015,millar-blanchaer2016,currie2015}) at inner working angles and resolutions not accessible by previous generation instruments. Combined with an additional sample of newly-resolved scattered-light debris disks from SPHERE and other instruments (e.g. \citealt{thalmann2013,wahhaj2016,engler2018,engler2020,bonnefoy2021}) and previously resolved systems (e.g. \citealt{schneider1999,schneider2005,padgett2016,liu2004,choquet2016,soummer2014,hines2007,kalas2007}), the ensemble of scattered-light imaged disks provides a rich and diverse sample to investigate overall trends in system architectures of young planetary systems and the properties of dust grains around stars of different spectral types and ages.

Despite this new large sample of scattered-light resolved debris disks, few group/population studies have been conducted. \cite{ren2023} investigated debris disk color through \textit{HST} observations, identifying a predominantly blue color suggesting higher scattering efficiency at shorter wavelengths. \cite{esposito2020} first reported on the statistics and properties of the 29 circumstellar disks imaged as a part of the Gemini Planet Imager Exoplanet Survey (GS-2015A-Q-500, PI B. Macintosh). While the study reported and characterized the geometries of a few newly-resolved disks, the majority of properties of the sample were collated from previous investigations which applied different analysis approaches of individual systems, including studies utilizing observations from other instruments. These individual investigations often utilize unique approaches to data reduction and disk characterization, making direct comparisons between observations of the same system with different instruments difficult. Data reduction, particularly stellar point spread function (PSF) subtraction, can substantially affect the apparent structural appearance of a disk system \citep{milli2012}, biasing morphological characterization efforts. Different approaches to disk modeling lead to results that are not directly comparable and sometimes biased, depending on the assumptions. For example, some studies employ ellipse fitting to assess inclination and position angle, which assumes the disk is an infinitely narrow ring (e.g. \citealt{crotts2023}). The treatment of disk scattering properties is also approached with different methods. Henyey-Greenstein (HG) functions \citep{henyey1941} have long been used to described the SPF of dust populations in the solar system \citep{hong1985}. Although they are not based on any physical scattering theories, a linear combination of a few HG functions can be used to approximate a wide range of scattering patterns using only a few parameters. Other studies that opt to model grain properties more robustly are often limited in terms of complexity, having to assume a uniform shape for all dust grains (e.g. Mie theory, \citealt{mie1908}) and a limited number of grain species. Furthermore, the parameterization of grain properties has been observed to provide unrealistic results (e.g. \citealt{duchene2020}), where constrained dust properties may be unphysical or contradictory.

In the first publication of this series \citep{crotts2023}, empirical measurements of a set of disk morphological and brightness characteristics were conducted in a uniform manner to GPI polarized intensity data. In this study, we applied a uniform data reduction and radiative-transfer modeling approach to facilitate more direct comparisons between morphological properties of a sample of debris disk systems in total intensity light, investigating a separate regime of scattered-light resolved structure from \cite{crotts2023}. This approach applied a commonly-used dust density distribution function and explores consistent parameter spaces for all debris disk targets. Rather than modeling grain properties, which can be highly biased depending on the underlying assumptions and free parameters explored, we chose to utilize two empirical scattering phase functions that are not reliant on an HG formalism or underlying assumptions of grain properties. In this work, we present our modeling results of eight debris disk targets imaged by GPI. In \S \ref{sec:sample}, we describe the samples from which our observations are derived and the additional criteria employed to construct our final sample. In \S \ref{sec:observations}, we describe the properties of the observations of our target sample. In \S \ref{sec:data reduction}, we describe the data reduction approach for all datasets in our sample. In \S \ref{sec:model methodology}, we describe our forward modeling setup. In \S \ref{sec:results}, we present the images and constrained parameters of our model analysis. In \S \ref{sec:discussion}, we describe the interpretations of our results in modeling and understanding debris disk systems. In \S \ref{sec:conclusions}, we summarize our findings and discuss future implications for the results of our analysis.

\section{Target Sample} \label{sec:sample}
Selected targets in this analysis originate from three distinct observational programs. The majority of the data originate from the Gemini Planet Imager Exoplanet Survey (GPIES) campaign (GS-2015A-Q-500, PI B. Macintosh), a 600-star direct imaging survey utilizing the Gemini Planet Imager (GPI; \citealt{macintosh2014}) in the spectral integral field unit (IFU) mode at \textit{H}-band. A subset of this campaign was dedicated to observing both previously known scattered-light and/or thermal emission-resolved circumstellar disks as well as stars with notable IR-excesses exceeding $10^{-5}$ at \textit{H}-band wavelengths. Two other programs, “Debris Characterization in Exoplanetary Systems” (PI C. Chen; GS-2016A-LP-6) and “Does the HR 4796 Debris Disk Contain Icy Grains?” (PI C. Chen; GS-2015A-Q-27) were used for the remaining observational sequences of this sample and constitute a Gemini Large and Long Program (LLP). These two programs added both \textit{J}- and \textit{K1}-band wavelengths to the sample. All of these programs utilized the capabilities of GPI to image debris disks at higher contrast ($\sim10^{-6}$) and smaller angular separations (tens of mas) than were possible with previous generation instruments.

From the extensive GPIES Disk and GPI-LLP Disk samples, eight targets were selected. These targets satisfied three main criteria: (1) data from a GPI \textit{spectral} mode dataset, (2) detection in total intensity light, and (3) average signal-to-noise ratio (SNR) must exceed $\sim$5 per pixel. As an IFU, GPI only operates in two observational modes: spectral and polarimetric. As described in \S \ref{sec:forwardmodel}, a simulated PSF core unique to each dataset is necessary for our analysis, as PSF structure and intensity can vary with observing conditions and target brightness. In GPI datasets, this PSF core is modeled from photometric measurements of satellite spots created by fiducial images of the host star superimposed on the GPI pupil apodizer in each science frame. In spectral mode datasets, these spots have well-defined shapes and flux ratios. Although GPI has resolved an extensive sample of disks in polarized intensity \citep{esposito2020,crotts2023}, they are not considered in this analysis. In polarimetric datasets, the satellite spots necessary for our PSF core modeling routine are elongated and cannot be used to accurately generate a model PSF core for disk forward modeling. Finally, although polarized intensity phase functions contain notable features among scattered-light resolved disks, no common trends have been identified to warrant a model analysis of multiple systems with a common polarized intensity phase function, and studies of cometary dust have revealed different shapes to polarized intensity phase functions depending on the composition of grains \citep{frattin2019}. Additionally, polarizability curves which describe the nature of polarization throughout a resolved scattered-light disk can vary distinctly between systems and would introduce an additional parameter space of investigation that is not relevant to studying morphological properties of disks (e.g. \citealt{hadamcik2007,hadamcik2003}). Therefore, our model methodology approach cannot be applied to polarized intensity images.

All of the targets considered for this sample have \textit{H-}band images taken as a part of the GPIES campaign. From the GPI-LLP Disk Sample, only \textit{K1-}band observations of HD 32297 and HR 4796A were considered due to the SNR threshold we applied. Summary information regarding our sample targets is shown in Table \ref{tab:targetlist}.

\begin{table*}
	\centering
	\caption{Target List summary. \textit{H} and \textit{K} magnitudes originate from 2MASS photometry \citep{cutri2003}. Distances are retrieved from Gaia DR2 \citep{gaia_dr1,gaia_dr2}. Other references: 1. \citet{nielsen2019}, 2. \citet{torres2006}, 3. \citet{schneider2005}, 4. \citet{bell2015}, 5. \citet{houk1988}, 6. \citet{torres2008}, 7. \citet{soummer2014}, 8. \citet{pecaut2016}, 9. \citet{houk1975}, 10. \citet{deZeeuw1999}, 11. \citet{kalas2015}, 12. \citet{houk1978}, 13. \citet{kasper2015}, 14. \citet{padgett2016}, 15. \citet{wahhaj2016}, 16. \citet{thalmann2013}, 17. \citet{houk1982}, 18. \citet{schneider1999}}
	\label{tab:targetlist}        
	\begin{tabular}{lccccccc}
		\hline
		Name & \textit{H} & \textit{K} & d & Age & Spectral  & Moving & First Resolved\\
              & (mag) & (mag) & (pc) & (Myr) & Type & Group & Detection\\
		\hline
		  HD 32297 & 7.6 & 7.6 & 132.79$\pm$1.06 & 15-45 (1) & A0V (2) & None & (3)\\
            HD 35841 & 7.8 & 7.8 & 103.68$\pm$0.30 & 38-48 (4) & F3V (5) & Columba (6) & (7)\\
            HD 106906 & 6.8 & 6.7 & 103.33$\pm$0.46 & 12-18 (8) & F5V (9) & LCC (10) & (11)\\
            HD 110058 & 7.5 & 7.6 & 129.98$\pm$1.33 & 12-18 (8) & A0V (12) & LCC (10) & (13)\\
            HD 111520 & 7.7 & 7.7 & 108.94$\pm$0.65 & 12-18 (8) & F5/6V (12) & LCC (10) & (14)\\
            HD 114082 & 7.2 & 7.2 & 95.65$\pm$0.45 & 12-18 (8) & F3V (9) & LCC (10) & (15)\\
            HD 146897 & 7.8 & 7.8 & 131.50$\pm$0.93 & 7-13 (8) & F3V (5) & US (10) & (16)\\
            HR 4796A & 5.8 & 5.8 & 72.78$\pm$1.75 & 7-13 (4) & F3V (17) & TWA (4) & (18)\\		
		\hline
	\end{tabular} 
\end{table*}

\section{Observations} \label{sec:observations}
Observations in \textit{H}- and \textit{K1}-band were conducted over several semesters at Gemini-South with GPI as a part of the GPIES campaign and the GPI-LLP Disk program, summarized in Table \ref{tab:observations}. All observations were conducted in GPI's spectral mode and covered high field rotation ($\Delta$PA$\sim17-80\degr$) for increased effectiveness in utilizing angular differential imaging \citep[ADI;][]{lafreniere2007} for PSF subtraction. Integration times were selected such that detector readout noise did not exceed signal from the disk while also avoiding angular smearing from rotation and saturation of speckles. \textit{K1}-band observations can tolerate longer exposure sequences ($\sim$90 s) due to PSF speckles being fainter at longer wavelengths. The GPI target prioritization scheme also emphasized observing targets near transit to further achieve high field rotation.

For all observations, the primary star of each system was centered behind a focal plane mask (FPM), allowing for higher contrast to be reached in the immediate area surrounding the star. The inner working angle (IWA) of the GPI FPM is $\sim0\farcs1$, and the field-of-view (FOV) of GPI is $2.8"\times2.8"$. GPI has a spatial sampling 
of $\sim14$ mas/pixel. For \textit{K1}-band observations, five sky exposures every hour for each target were acquired for thermal/sky background subtraction.

Although the targets were observed in GPI's spectral mode which provides spatially resolved low resolution spectra, spectral properties of targets are not investigated in this work and the full set of wavelength channels per science image was collapsed into single broadband images to increase disk SNR. Faint disk surface brightness in individual wavelength channels can be challenging to measure, and \cite{esposito2020} notes that the measured \textit{H}-band spectra of GPI-imaged debris disks tend to be relatively featureless. Further details of \textit{H}-band observations are given in \cite{esposito2020}.
\begin{table*}
	\centering
	\caption{Summary of observations for the sample. Seeing estimates are not available for HD 114082 and HD 146897 observations, as the DIMM seeing monitor at Gemini-South was inoperable at the time of observation.}
	\label{tab:observations}
	\begin{tabular}{lcccccccc}
		\hline
		Name & Filter & $t_{\rm exp}$ & $N$ & $\Delta$PA & Air Mass & Seeing & Date & Program \\
             &  & (s) &  & ($\degr$) & & (arcsec) & & \\
		\hline
		  HD 32297 &  \textit{H} & 59.65 & 38 & 16.7 & 1.26--1.29 & 0.63 & 2016 Dec 20 & GS-2015B-Q-500\\
            HD 32297 &  \textit{K1} & 88.74 & 51 & 33.1 & 1.26--1.30 & 0.72 & 2015 Nov 30 & GS-2015B-LP-6\\
            HD 35841 &  \textit{H} & 59.65 & 50 & 46.9 & 1.01--1.06 & 1.01 & 2016 Feb 28 & GS-2015B-Q-500\\
            HD 106906 &  \textit{H} & 59.65 & 42 & 25.3 & 1.11--1.12 & 0.86 & 2015 May 4 & GS-2015A-Q-500 \\
            HD 110058 &  \textit{H} & 59.65 & 38 & 29.6 & 1.06 & 0.71 & 2016 Mar 19 & GS-2015B-Q-500 \\
            HD 111520 &  \textit{H} & 59.65 & 42 & 34.8 & 1.06--1.07 & 1.01 & 2015 Jul 02 & GS-2015A-Q-500\\
            HD 114082 &  \textit{H} & 59.65 & 47 & 25.8 & 1.16 & -- & 2018 Jan 29 & GS-2017B-Q-500\\
            HD 146897 &  \textit{H} & 59.65 & 38 & 59.5 & 1.01--1.02 & -- & 2018 Aug 15 & GS-2017B-Q-500\\
            HR 4796A &  \textit{H} & 59.65 & 37 & 53.0 & 1.01--1.02 & 0.73 & 2016 Mar 18 & GS-2015B-Q-500 \\
            HR 4796A &  \textit{K1} & 88.74 & 46 & 78.5 & 1.01--1.09 & 0.55 & 2015 Apr 3 & GS-2015A-Q-27 \\
		\hline
	\end{tabular}
\end{table*}

\section{Data Reduction} \label{sec:data reduction}
The GPI Data Reduction pipeline (DRP, \citealt{perrin2014,wang2018}) was used for reducing all science data from GPI. The pipeline performed dark subtraction, correlated noise cleaning, and bad pixel correction for all raw data. For spectral mode data in particular, flexure correction for satellite spots was also performed. Before a spectral sequence was observed, an Ar lamp exposure was collected for wavelength calibration. Geometric distortion was also corrected for in all data cubes and smoothed with a Gaussian kernel ($\sigma = 1$ pixel). The DRP was responsible for assembling the integral field spectrograph (IFS) spectral data cube and determining the location of the satellite spots (necessary for performing the \texttt{DiskFM} analysis, see \S \ref{sec:model methodology}). For \textit{K1}-band observations only, thermal/sky background images were subtracted. The science images were also destriped, and some spectral slices within the IFS spectral datacube were not considered for the analysis if the SNR of the satellite spots was low due to high thermal noise at \textit{K}-band wavelengths. Finally, following the approach in \cite{wang2014}, the location of the occulted primary star in each science frame was determined by performing a least-squares fit to all visible satellite spot positions to a precision of 0.7 mas. Science frames were then shifted to align with the star center located at the center of every image. The science frames were also all rotated so that north pointed upward in the image and east pointed to the left.

Spectral data cubes were further PSF-subtracted with the \texttt{pyKLIP} \citep{wang2015} implementation of the Karhunen-Lo{\`e}ve Image Projection (KLIP) algorithm \citep{soummer2012} in combination with ADI. ADI takes advantage of the fact that instrumental PSF artifacts do not rotate with the FOV in pupil-stabilized observations. Any astrophysical object within the FOV will rotate with respect to the center of the image throughout the observational sequence. A model PSF halo pattern can then be generated from the rotating frames without including astrophysical signal (except in the case of extended structures such as disks) and subsequently subtracted from images throughout the sequence. KLIP expands upon this approach by performing principal component analysis of PSF features, achieving more robust results than ADI alone. For this analysis, 3-7 Karhunen-Lo{\`e}ve modes and $4-6$\degr of minimum rotation ($N_{\delta}$ in \citealt{lafreniere2007}) between science frames were used for PSF reconstruction. The choice in the number of KL-modes and minimum rotation angles is informed by the inclination of the system, maximizing the average disk SNR, and minimizing overlap of the disk between science frames to mitigate self-subtraction \citep{milli2012}. The analysis was performed globally across a whole GPI image and not subdivided into concentric annuli or subsections, facilitating the creation of continuous and smooth reduced images. The full details of the data reduction parameters are shown in Table \ref{tab:datared}. The Karhunen-Lo{\`e}ve basis vectors were saved and projected onto disk forward models for later analysis, as described in \S \ref{sec:model methodology}.

\begin{table}
    \centering
    \begin{tabular}{lccc}
    \hline
       Name  & Filter & KL modes & $N_{\delta}$ [$\degr$] \\ \hline
       HD 32297 & \textit{H}  & 3 & 4 \\
       HD 32297 & \textit{K1}  & 5 & 4 \\
       HD 35841 & \textit{H} & 7 & 6 \\
       HD 106906 & \textit{H} & 5 & 6 \\
       HD 110058 & \textit{H} & 5 & 6 \\
       HD 111520 & \textit{H} & 5 & 6 \\
       HD 114082 & \textit{H} & 5 & 6 \\
       HD 146897 & \textit{H} & 5 & 6 \\
       HR 4796A & \textit{H} & 7 & 6 \\
       HR 4796A & \textit{K1} & 5 & 6 \\ \hline
    \end{tabular}
    \caption{KLIP reduction parameters for all targets, chosen to maximize disk SNR.}
    \label{tab:datared}
\end{table}

\section{Model Methodology} \label{sec:model methodology}
\begin{figure*}
    \centering
    \includegraphics[width=\textwidth]{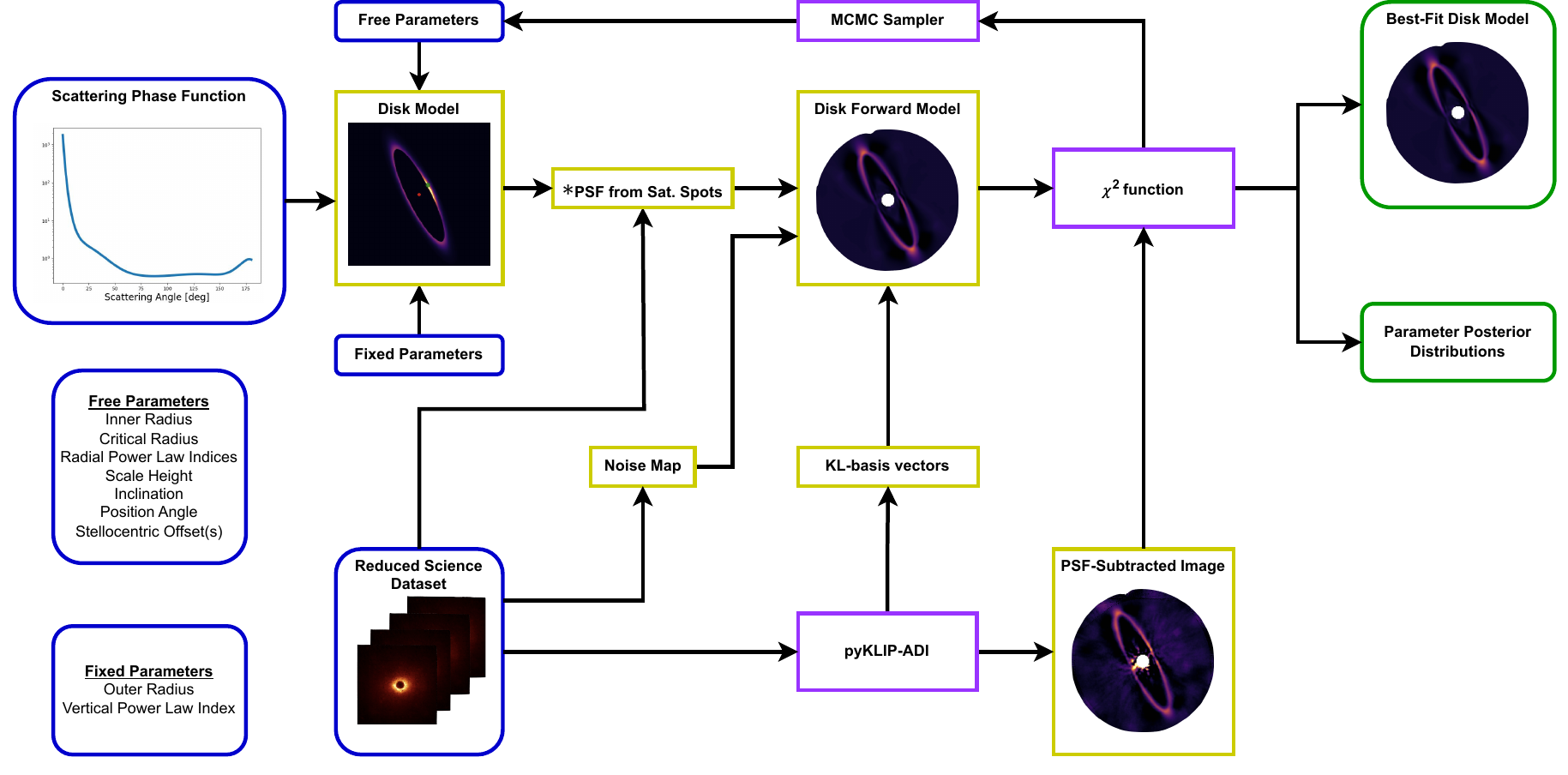}
    \caption{A diagram demonstrating the forward modeling process using \texttt{DiskFM}. A forward model is generated from a fixed scattering phase function, a set of fixed parameters, and a varying set of free parameters (blue boxes). From the reduced science images, a model PSF and representative noise map are generated. This model PSF is then convolved with all disk models, with the noise map used for likelihood calculation. The reduced science images are post-processed with \texttt{pyKLIP}-ADI, and the KL-basis vectors are saved for projection onto all forward models generated with \texttt{DiskFM}. The KL-basis vectors, model PSF, and representative noise map (gold boxes) are all used for the forward modeling of a disk given a set of inputs. A $\chi^2$ function is used to inform the MCMC sampler of forward model minimization. The outputs of \texttt{DiskFM} include the maximum likelihood best-fit disk model and posterior probability density distributions of free parameters (green boxes). Purple boxes represent calculation procedures internal to \texttt{DiskFM}.
    }
    \label{fig:flowchart}
\end{figure*}

As high contrast images of debris disks reduced with ADI often suffer from severe self-subtraction \citep{milli2012}, forward-modeling is often necessary to constrain the properties of a system. The key steps in the forward modeling methodology employed in this study are shown in Figure \ref{fig:flowchart}. To assess morphological properties, we adopt a forward modeling approach using \texttt{DiskFM} \citep{mazoyer2020} for comparison with the science data. Disk parameters are constrained using an iterative MCMC process with the affine-invariant sampler \texttt{emcee} Python package \citep{foreman-mackey2013}, generating hundreds of thousands of disk models per target. 

\subsection{Disk Model} \label{sec:diskmodel}
The disk model utilized is based on a built-in module of \texttt{DiskFM} and is originally described in \cite{millar-blanchaer2015} and \cite{millar-blanchaer2016}, with two major modifications made, related to the dust density distribution and dust scattering properties.

The updated surface density profile $\eta(r,z)$ follows a smoothly connected two power-law structure described in more detail in \cite{augereau1999} and given in Equation \ref{eq:augereau}:
\begin{equation}
    \eta(r, z) \propto R(r) Z(r, z)
    \label{eq:augereau}
\end{equation}
where $R(r)$ is given as:
\begin{equation}
    R(r) \propto \left\{ \left(\frac{r}{R_{\rm C}}\right)^{-2\alpha_{\rm in}} + \left(\frac{r}{R_{\rm C}}\right)^{-2\alpha_{\rm out}}\right\}^{-\frac{1}{2}}
    \label{eq:radialdust}
\end{equation}
where $r$ is the radial distance from the host star and $R_{\rm C}$ is the critical radius where a transition between two power law density regimes occurs. The indices of these power law density regimes are given as $\alpha_{\rm in} > 0$ and $\alpha_{\rm out}<0$ for the inner and outer regions of the disk respectively. $Z(r,z)$ is given as
\begin{equation}
    Z(r,z) \propto \exp\left(-\left(\frac{|z|}{h(r)}\right)^{\gamma_{\rm vert}}\right)
    \label{eq:verticaldust}
\end{equation}
where $z$ is the distance from the disk midplane, $\gamma_{\rm vert}$ dictates the shape of the vertical density distribution, and $h(r)$ is the height above the disk midplane as a function of $r$. The \cite{augereau1999} dust density profile was selected as it is more commonly used in dust density distribution analyses and therefore facilitates more consistent comparisons of our results to previous studies. For our analysis, we set $\gamma_{\rm vert} = 2$ for a Gaussian vertical profile and $h(r)$ is given by $a_{r}\times r$ where $a_{r}$ is the constant aspect ratio $h/r$, assuming a "bow-tie" shape for every disk. Both inner ($R_{\rm in}$) and outer ($R_{\rm out}$) cutoff radii define where the dust density $\eta(r,z)$ is treated as zero; only $R_{\rm in}$ is treated as a free parameter. All disks except for HD 110058 have evidence of extensive halo emission observed at larger FOV than GPI (e.g. \citealt{schneider2005,soummer2014}), therefore $R_{\rm out}$ is set to be the outer working angle of the GPI FOV for these systems. For HD 110058, $R_{\rm out}$ is loosely defined to exist outside of the MCMC prior range for $R_{\rm C}$. The proportionality constant is not solved for in our analysis, as we only seek to understand the overall geometry of each system and not the dust mass; a constant brightness scaling factor is used as a free parameter to achieve low residual model fits to the data.

The second significant modification is the treatment of light scattering properties of the model. In the original model framework, a choice of a one, two, or three component HG function is used for representing all scattering properties. In this study, scattering properties of models are given by an empirically-informed scattering phase function interpolated with a 3rd order spline function. 

\begin{figure}
    \centering
    \includegraphics[width=\linewidth]{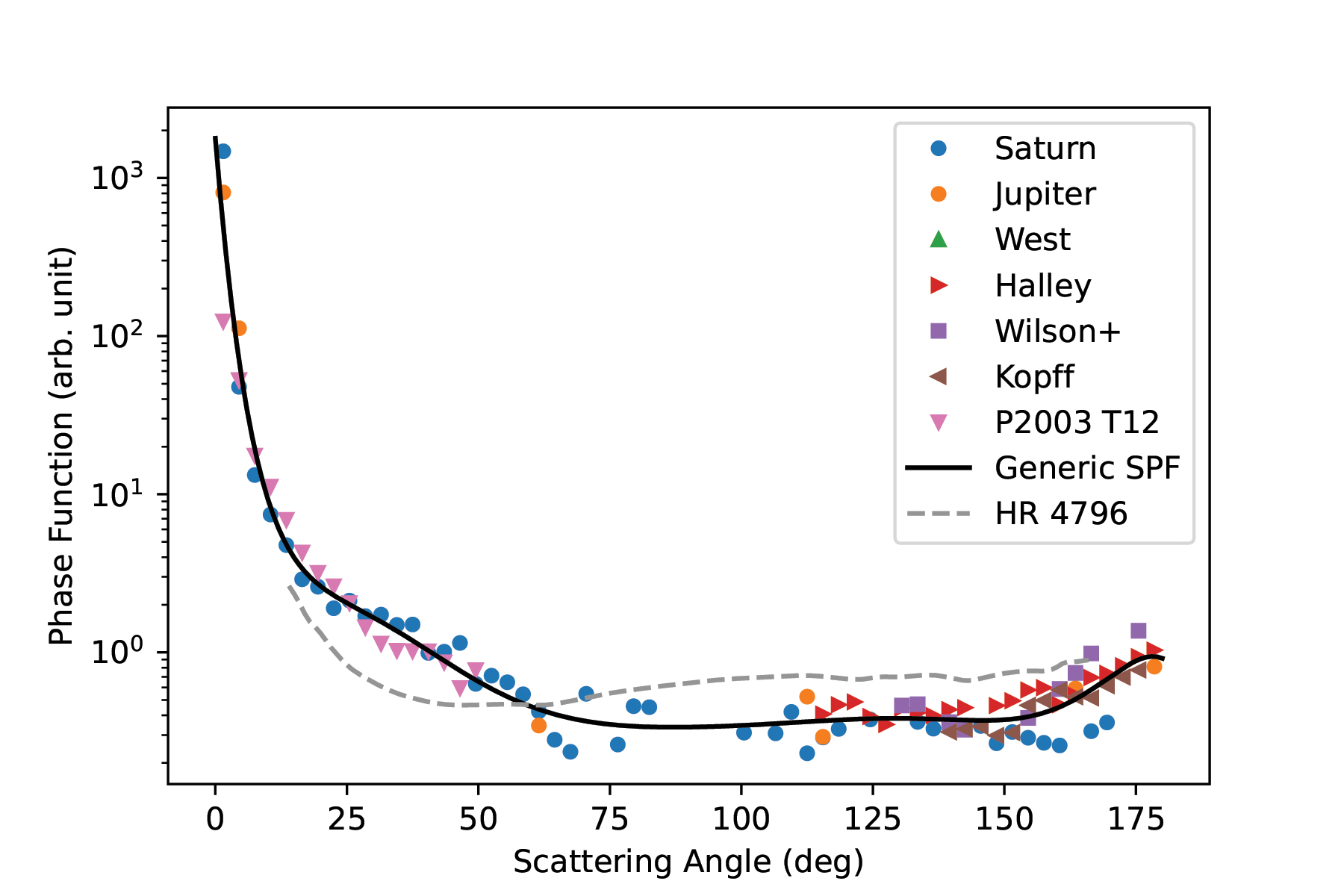}
    \caption{Our generic SPF overplotted with measured SPFs from solar system dust and the HR 4796A disk \citep{milli2017}. A 9th-degree polynomial fit is used to account for the similar shapes observed in many dust environment SPFs.}
    \label{fig:genSPF}
\end{figure}

One of the premises of this study is the observation by \citet{hughes2018} that scattering in most debris disks appears to follow a similar SPF, which itself matches qualitatively that observed in a number of solar system objects. Based on the apparent similarities of the few measured SPFs, we generated a generic scattering phase function in the following manner. First, we gathered the SPF from Saturn's D68 and G rings \citep{hedman2015}, Jupiter's ring \citep{throop2004}, and multiple comets \citep{hanner1989, schleicher1998, moreno2012, hui2013}. To avoid one particular dataset biasing the final SPF due to more intense and/or wider sampling of scattering angles, we rebinned each individual SPF to a common 3\degr\ sampling. We then renormalized each SPF to the most completely sampled SPF-that of Saturn's rings-using the median of their ratio over all overlapping bins, and averaged all resulting SPFs. Incidentally, the SPF of the Saturn D68 has been shown to have a similar shape to the measured SPF of HD 114082 from \cite{engler2023}. Finally, we fitted a 9th-degree polynomial (the lowest order that avoids overfitting) to these datapoints and the result is taken as the ``generic SPF". The generic SPF used in this study is shown in Figure \ref{fig:genSPF}, overplotted with measured SPFs from solar system dust environments and the markedly distinct SPF measured from SPHERE images of HR 4796A by \cite{milli2017}.

We note that most of the SPFs used in this process were observed in the optical \citep[the main exception being the $J$ band compilation from][]{hanner1989}. Furthermore, the various SPFs are not statistically consistent with each other, even though their overall shapes are qualitatively similar. In this context, it is intriguing that this SPF matches observed near-infrared SPFs of debris disks. 

The second empirical SPF used in this study was measured directly from SPHERE images of HR 4796A \citep{milli2017}. The HR 4796A SPF is distinct from the SPF trends seen in \cite{hughes2018} and was used as a test case for the HD 114082 and HR 4796A datasets to demonstrate the differences in best-fit model appearances and noise-scaled residuals (hereafter called residuals) when using different SPFs. Although SPFs have been measured for HD 32297 \citep{duchene2020}, HD 35841 \citep{esposito2018}, and HD 114082 \citep{engler2023}, they are not considered for this analysis as their shapes were found to be largely similar to the trends observed in \cite{hughes2018}. SPFs for the other targets in this sample have not been measured.

Physical disk models are axisymmetric and generated in a three-dimensional space and rotated according to constraints on position angle and inclination with respect to the observer to match the viewing geometry of each system. After the disk model is properly rotated, the three-dimensional disk model cube is collapsed along the line-of-sight direction, creating a two-dimensional image of the disk.

\subsection{Forward Modeling Process} \label{sec:forwardmodel}
\texttt{DiskFM} \citep{mazoyer2020} was used to perform forward modeling on all disk models. 
For every GPI integration, a grid pattern imprinted on the pupil plane mask diffracts on-axis light from the target star to create a pattern of four fainter satellite spots in the image plane. 
Each disk model, as described in the previous section, is convolved with a PSF core generated from all satellite spots in all wavelength channels of every science frame where the disk does not overlap and the satellite spot has a SNR $>3$. Finally, the KL basis vectors from the \texttt{pyKLIP}-ADI reduction of the original science dataset are projected onto the convolved model. This allows for a PSF subtraction to be applied to the forward model in an identical manner to the science data, without having to inject the model into an ``empty" observational sequence and perform KLIP-ADI again to recalculate KL basis vectors, which would have different patterns of self-subtraction artifacts than the reduced science dataset.

To generate dataset-specific noise maps for calculating forward modeling likelihood, we rotate science frames in randomly-generated orientations to create a time-collapsed datacube that medians out a PSF halo pattern while preserving radial noise properties. This approach is similar to the approach described in \cite{gerard2016}, although we choose random orientations as opposed to opposite orientations from North-up to decrease the likelihood that disk signal overlaps between individual science frames. \texttt{PyKLIP}-ADI is then utilized on the randomly-rotated science dataset to eliminate traces of the disk signal entirely and create a noise map that has been reduced with the same KLIP parameters as the non-rotated science dataset. From the randomly-rotated reduced image, the standard deviations of concentric rings 3 pixels in width are calculated. The final noise map consists of 3-pixel wide concentric rings with each ring containing the standard deviation of that same region in the "randomly-rotated" noise map. The noise map is then multiplied by a scalar of 3 for all model analyses, as described in \cite{chen2020} and \cite{mazoyer2020}. We introduce this factor as we expect our noise maps to be underestimated, due to correlated noise features that our approach cannot reproduce. This method is also used to recover accurate error bars for planet photometry (e.g. \citealt{galicher2018}). \cite{mazoyer2020} showed that in almost all cases this factor was enough to accurately recover error bars for disk parameters, although they only tested this approach for a small sample of model systems. As shown in Table \ref{tab:MCMCresultsfinal}, this factor may not be the most ideal choice for all model analyses, in some cases leading to overestimations of the noise and therefore very low $\chi^2_{\rm red}$. This noise scaling factor is likely not consistent between different datasets, but we retain the factor of 3 to be consistent to previous high contrast imaging studies and so that all disk analyses are uniform. A synopsis of the steps and components of \texttt{DiskFM} is shown in the gold and purple boxes of Figure \ref{fig:flowchart}.

\subsection{Likelihood Calculation} \label{sec:likelihood}
The forward model is compared to the data by measuring
\begin{equation}
    \chi^2 = \sum_S \frac{(Data - Forward\;Model)^2}{Uncertainty^2}
    \label{eq:chi2}
\end{equation}
where $S$ describes the region over which model likelihood was calculated.
\texttt{DiskFM} utilizes the \texttt{emcee} package \citep{foreman-mackey2013} for MCMC iterative analysis. Likelihood calculation is performed with an MCMC wrapper that maximizes $e^{-\chi^2/2}$ in a masked region unique to each disk in the sample for both the disk forward model and science data. This calculation is performed until chain convergence. For all disk target analyses, 120 walkers were used, calculated for at least 5000 iterations. More iterations were added as necessary to confirm converged behavior within the MCMC chains. After at least 300 iterations of converged behavior, all iterations up to that point were excluded as burn-in when generating the posterior distribution functions of our parameters.

Model likelihood was only calculated over specific regions where disk flux is most apparent and some background with no apparent disk signal is present. For systems where we tentatively resolve the back sides of disks--HD 35841, HD 106906, and HR 4796A--concentric rings enclosing the spine of the disk are created, with some space given between the apparent edge of the disk and the edge of the mask for the inclusion of the noise background in the likelihood calculation. If the front side of the disk passed in front of the FPM, the edge of the FPM was used for the inner boundary of the likelihood calculation mask. Regions interior to these concentric ellipses were further excluded if they contained significant amounts of noise. For close to edge-on systems HD 32297, HD 110058, and HD 111520, a flared bar shape was used as the likelihood calculation mask, with the inner radial boundary set slightly outside of the FPM. Interior to this inner boundary, noise is expected to be high due to the proximity to the host star and FPM.

The uncertainty in Equation \ref{eq:chi2} is taken from the generated noise map described in \S \ref{sec:forwardmodel}. Although spatially and spectrally-correlated noise among pixels is a concern for direct imaging IFS datasets \citep{greco2016}, the scalar factor applied to the generated noise map as described in \cite{mazoyer2020} was found to be sufficient in describing noise properties of an image.

Disk models have either 8 or 9 physical free parameters depending on disk inclination and apparent disk thickness. The 8 free parameters common to all disks are the inner cutoff radius $R_{\rm in}$, critical radius $R_{\rm C}$, surface dust density inner power law index $\alpha_{in}$, surface dust density outer power law index $\alpha_{\rm out}$, aspect ratio $a_r$, inclination $i$, position angle $PA$, stellocentric offset in the disk plane along the projected major axis $dy$ (where positive $dy$ corresponds to a stellocentric offset in au in the disk plane as defined in Table \ref{tab:MCMCparameters}). A flux normalization scaling is used as an additional free parameter but is only used for scaling values in the model images to match the data; the value of flux normalization does not have a physical interpretation; this parameter is marginalized in the corner plots for our model analyses in Appendix \ref{sec:posteriors}. For disks where we resolve the front and back sides of the disk, an additional free parameter $dx$ is parameterized (where positive $dx$ corresponds to a stellocentric offset in au in the disk plane pointing toward the observer along the minor axis), describing the stellocentric offset in the disk plane along the projected minor axis. Although we do not resolve the back side of HD 106906, we also parameterize $dx$ due to its evidence of asymmetric structure \citep{kalas2015,crotts2021}. The other targets in this sample do not contain enough spatial information to provide a meaningful constraint on $dx$.

MCMC initial parameters for $R_{\rm in}$, $R_{\rm C}$, $i$, and $PA$ are set as the constrained parameters of these disk targets as described in previous studies of scattered-light resolved imaging of disks, many of which were summarized in \cite{esposito2020}. Initial $\alpha_{\rm in}$ and $\alpha_{\rm out}$ parameters are chosen to represent median cases of power law indices, while the initial $a_r$ parameters for HD 35841, HD 114082, HD 146897, and HR 4796A are derived to be in similar orders of magnitudes of previous studies \citep{esposito2018,engler2023,goebel2018,milli2015,olofsson2022}. For HD 32297, HD 106906, HD 110058, and HD 111520 we opt to choose initial $a_r$ parameters from the apparent vertical structure evident in our reduced GPI images. MCMC prior ranges are designed to be broad for radii, power law index, aspect ratio, and flux normalization parameters and narrow for inclination, position angle, and stellocentric offsets to match gross overall morphology. The vertical density distribution power law index $\gamma_{\rm vert}$, the disk flaring index, and the outer radial cutoff $R_{\rm out}$ are fixed. A summary of initial MCMC parameters and prior ranges are shown for each disk in Table \ref{tab:MCMCparameters}. A summary of all \texttt{DiskFM} inputs (reduced science dataset, free parameters, fixed parameters, and the SPF) are shown in the blue boxes of Figure \ref{fig:flowchart}. The outputs of \texttt{DiskFM}, shown in green boxes, include the maximum likelihood best-fit model and posterior probability density distributions of free parameters.

Three disk systems -- HD 106906, HD 110058, and HD 111520 -- have significant asymmetrical structure in their scattered-light images \citep{kalas2015,kasper2015,draper2016,crotts2023}. The disk model utilized in this analysis does not have any parameterization that accounts for the strong disk asymmetries seen in these systems. Although a stellocentric offset can account for some brightness asymmetries, a reasonable offset is not sufficient to explain the significant brightness asymmetry observed in HD 111520. As seen in \S \ref{sec:uniresults}, strong residuals are present in the best model fits. As a result, we choose to conduct independent disk forward-modeling analyses on each side of these asymmetric disks (half mask models). Although we cannot fully constrain certain morphological properties because of the nature of this process, we seek to understand the value of applying the generic scattering phase function described in \S \ref{sec:diskmodel} to debris disk systems in general.

\begin{table*}
	\centering
	\caption{MCMC initial parameters and unique prior ranges for each target. For all values of $\alpha_{\rm in}$, the prior range extends from [0...10]. For $\alpha_{\rm out}$, the prior range is from [-10...0]. For all values of $a_r$, the prior range is [0.001...0.3]. For all values of $dx$ and $dy$, the prior range extends from [-5...5] au. References for initial parameter selection: 1. \citet{duchene2020}, 2. \citet{esposito2018}, 3. \citet{lagrange2016}, 4. \citet{kalas2015}, 5. \citet{kasper2015}, 6. \citet{esposito2020}, 7. \citet{draper2016}, 8. \citet{wahhaj2016}, 9. \citet{engler2017}, 10. \citet{thalmann2013}, 11. \citet{perrin2015}}
	\label{tab:MCMCparameters}
	\begin{tabular}{lcccccccccccc}
		\hline
		Name & $R_{\rm in}$ & $R_{\rm C}$ & $\alpha_{\rm in}$ & $\alpha_{\rm out}$ & $a_r$ & $i$ & PA & $dx$ & $dy$ & \\
             & [AU] & [AU] & & & & [$\degr$] & [$\degr$] & [AU] & [AU] & Pos. $dy$ \\
		\hline
		  HD 32297 & 50 [25...100] (1) & 98.4 [75...150] (1) & 5 & -5 & 0.001 & 88.4 [80...90] (1) & 47.9 [45...55] (1) & -- & 0.1 & SW\\
            HD 35841 & 60.3 [40...70] (2) & 60.3 [50...90] (2) & 5 & -5 & 0.01 & 84.9 [80...90] (2) & 165.8 [160...170] (2) & 0.1 & 0.1& SE\\
            HD 106906 & 66.6 [40...80] (3) & 72.3 [50...90] (3) & 5 & -5 & 0.001 & 84.6 [80...90] (4) & 284.2 [280...290] (4) & 0.1 & 0.1& NW\\
            HD 110058 & 39.0 [20...60] (5) & 39.0 [20...80] (5) & 5 & -5 & 0.1 & 84.0 [80...90] (6) & 155.0 [150...160]$^a$ (5) & -- & 0.1& SE\\
            HD 111520 & 71.0 [50...90] (7) & 81.0 [60...100] (7) & 5 & -5 & 0.001 & 88.0 [80...90] (7) & 165.0 [160...170] (7) & -- & 0.1 & SE\\
            HD 114082 & 28.7 [10...50] (8) & 30.7 [15...70] (8) & 5 & -5 & 0.01 & 83.3 [80...90] (8) & 105.7 [100...110] (8) & -- & 0.1& SE\\
            HD 146897 & 50.0 [25...100] (9) & 65.0 [40...115] (9) & 5 & -5 & 0.1 & 84.0 [80...90] (10) & 113.9 [110...120] (6) & -- & 0.1& SE\\
            HR 4796A & 74.4 [60...90] (11) & 78.5 [70...100] (11) & 5 & -5 & 0.01 & 76.5 [70...80] (11) & 26.1 [20...30] (11) & 0.1 & 0.1 & SW\\
		\hline
	\end{tabular}
  \vspace{2ex}
    {\raggedright $^a$This prior was expanded from [150...160] to [150...165] for models fitted to the NW side of the HD 110058 disk only, after initial analysis suggested a $PA$ greater than the prior limit.\par}
\end{table*}

\section{Results} \label{sec:results}
\subsection{Uniform Disk Model Results} \label{sec:uniresults}
\begin{figure*}
    \centering
    \includegraphics[width=\textwidth]{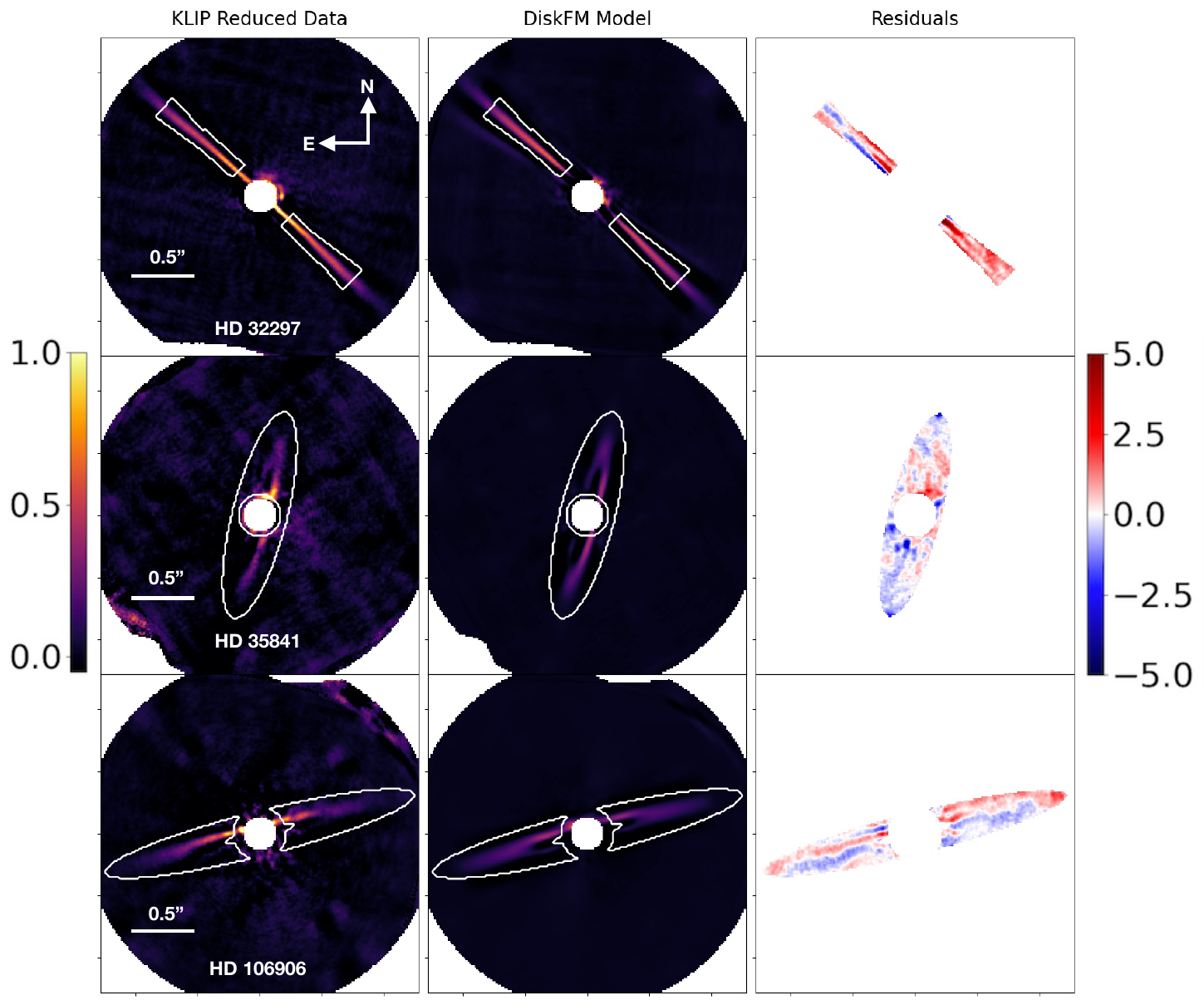}
    \caption{Modeling results from the initial uniform approach process for three targets in the sample, with the white outline representing the shape of the likelihood mask. The KLIP Reduced Data and \texttt{DiskFM} best-fit model are normalized to the maximum signal within the likelihood mask region of the KLIP Reduced Data. The noise-scaled residuals are shown in the right column. The strong residuals in the HD 32297 best-fit model are likely due to the inherent brightness of the disk, creating strong self-subtraction wings in the KLIP-ADI reduction and inducing nonlinearity in the \texttt{pyKLIP} forward modeling process. In the HD 35841 best-fit model, the strongest residuals (darkest red and blue regions in the residual map) are associated with regions closest to the FPM, where noise is expected to be high. In the HD 106906 model, the residual map appears to have structure along the spine of the disk, particularly on the SE side.
    } 
    \label{fig:uni_grid}
\end{figure*}

\begin{figure*}
    \centering
    \includegraphics[width=\textwidth]{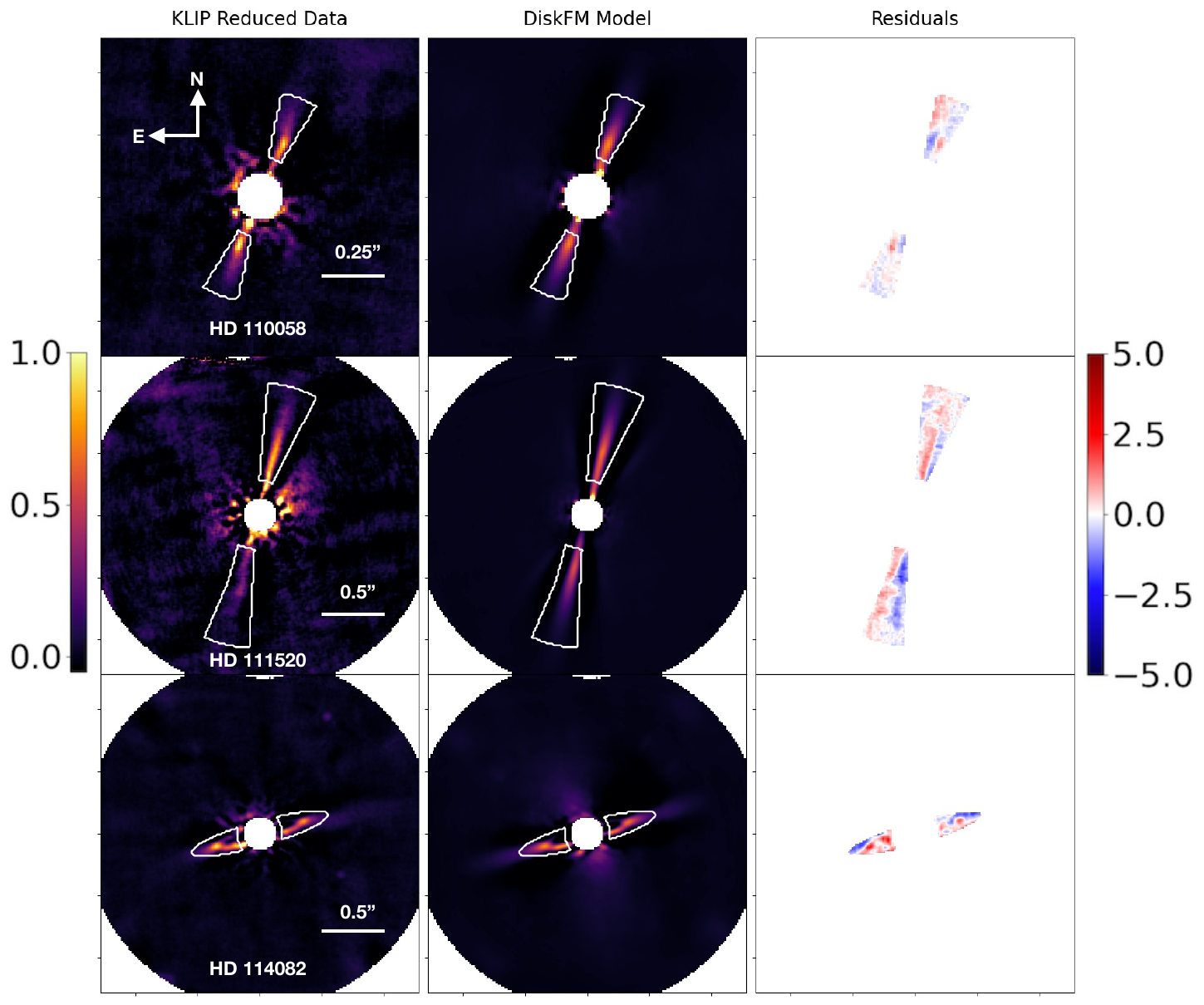}
    \caption{Same as Figure \ref{fig:uni_grid} but for three more targets in the sample. The HD 110058 residual map shows a pattern along the spine of the disk, suggesting mismatches in $PA$ for either side. The HD 111520 residual map highlights the strong brightness asymmetry first identified in \citet{draper2016}, where the best-fit model cannot reconcile the brightness between the NW and SE sides. The possible presence of a warp on the SE side also causes a $PA$ discrepancy between the data and the best-fit model. The HD 114082 best-fit model appears to have high residual features that do not overlap with the disk region and are likely associated with higher levels of noise in the immediate vicinity of the FPM.
    }
    \label{fig:uni_grid2}
\end{figure*}

\begin{figure*}
    \centering
    \includegraphics[width=\textwidth]{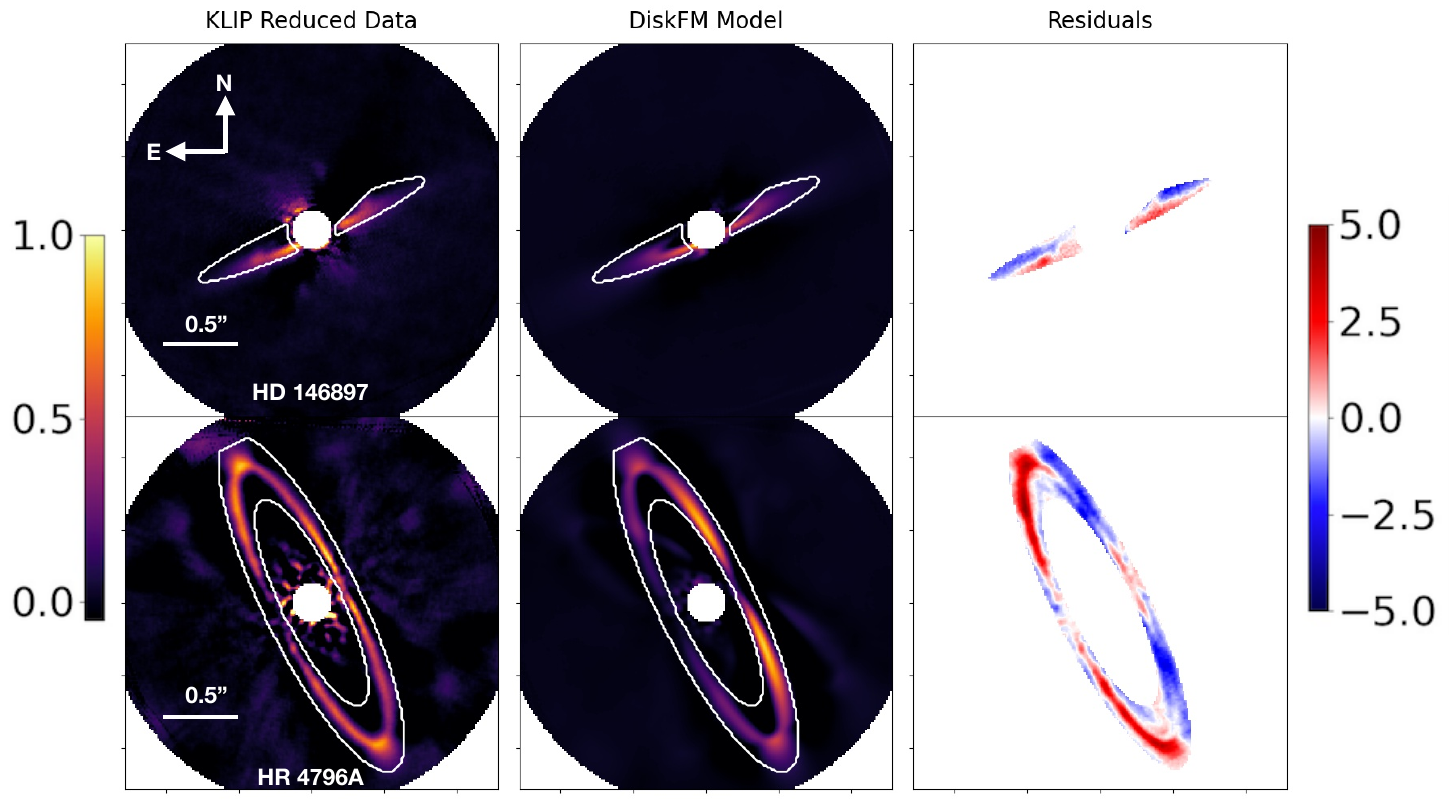}
    \caption{Same as Figure \ref{fig:uni_grid} but for two more targets in the sample. The structured and high residuals present in the HD 146897 residual map are likely due to diffuse PSF halo structure biasing the forward modeling process towards a lower inclination solution. Another possible explanation could be the choice in SPF; at a higher inclination, the model would produce too much forward scattering, and the curvature around the ansae would not be resolved. The highly structured residuals in HR 4796A are likely due to an incorrect choice of SPF, which cannot reproduce the distribution of forward and back scattering present in the data.}
    \label{fig:uni_grid3}
\end{figure*}

The best-fit models for each disk based on the uniform approach with the generic SPF, a continuous likelihood mask, and their residuals are shown in Figures \ref{fig:uni_grid}-\ref{fig:uni_grid3}. Median likelihood constrained parameters with 1$\sigma$ error bars and 3$\sigma$ upper and lower limits are reported in Table \ref{tab:MCMCresultsfinal}, except for HR 4796A\footnote{We report median likelihood parameters separately for the generic SPF analysis of HR 4796A, as the poor model fit compared to the \citet{milli2017} SPF analysis (see \S \ref{sec:SPFresults}) does not provide meaningful constraints on disk morphology: $R_{\rm in}=74.17\substack{+0.08 \\ -0.07}$ AU, $R_{\rm C}=70.14\substack{+0.22 \\ -0.10}$ AU, $\alpha_{\rm in}>1.34$, $\alpha_{\rm out}<-2.35$, $a_r=0.03\pm0.001$, $i=77\degr.01\pm0\degr.04$, $PA=26\degr.75\pm0\degr.03$, $dx=-2.13\pm0.13$ AU, $dy=0.82\pm0.06$ AU, $\chi^{2}_{\rm red}=1.90$.}. Posterior distributions for all model analyses are reported in Appendix \ref{sec:posteriors}. Figure \ref{fig:uni_grid} shows best-fit models of HD 32297, HD 35841, and HD 106906. Figure \ref{fig:uni_grid2} shows the best-fit models of HD 110058, HD 111520, and HD 114082. Figure \ref{fig:uni_grid3} shows the best-fit models of HD 146897 and HR 4796A.

Tight constraints ($1\sigma < 0.3\degr$) on $i$ were achieved for all systems except for HD 110058, for which we only find a lower limit, and HD 114082, where we find a double-peaked solution with $1\sigma > 0.3\degr$. In both of these cases, the compact nature of the system may hinder the ability of \texttt{DiskFM} to identify a clear $i$ solution. $PA$ was also tightly constrained ($1\sigma < 0.3\degr$) for all systems. In general, $i$ and $PA$ are typically the easiest parameters to constrain in a debris disk model, mostly due to clearly resolved positions.

Our constraints on the radial dust density distribution suggest two families of results: disks where the radial dust density distribution can be described by one power law ($R_{\rm C} < R_{\rm in}$ or $R_{\rm C} > R_{\rm out}$) and disks where the radial dust density distribution can be described by two power laws ($R_{\rm in} < R_{\rm C}$, $R_{\rm C}$ and $\alpha_{\rm in}$ and/or $\alpha_{\rm out}$ well-defined). 
Inner disk properties ($R_{\rm in}$ and $\alpha_{\rm in}$) were difficult to constrain for most systems, likely due to the lack of line-of-sight resolution for these regions in close-to-edge-on systems. Outer disk properties ($\alpha_{\rm out}$) were comparatively easier to constrain, as a consequence of these regions being more clearly resolved in most systems. In the case of HR 4796A, the outer radial profile prefers a steep power law index outside of our prior range, suggesting a very sharp outer edge. From our posterior distributions, we conclude that the dust density distribution of disks around HD 35841 and HD 111520 can be modeled with two power laws, while the dust density distribution of disks around HD 106906, HD 110058, HD 114082, HD 146897, and HR 4796A can be modeled with a single power law.

$1\sigma$ constraints were also found for the aspect ratios of all systems except for HD 146897 and HR 4796A, for which we achieve upper limits. The less-inclined and compact nature of HD 146897 as well as the bright PSF halo feature overlapping the front side of the disk can bias estimations of inclination and vertical extent. As HR 4796A is also relatively bright, nonlinearity in the KLIP reduction may also similarly bias estimates of some structural properties of the disk such as the vertical density distribution.

Narrow constraints ($1\sigma < 0.3$ au) on $dy$ were identified for all systems except for HD 111520 for which we find a tightly constraining lower limit. Four systems appear to have significant offsets at or beyond the 3$\sigma$ confidence level-HD 106906, HD 110058, HD 111520, and HR 4796A-consistent with previous analyses of these systems. In scenarios where $dx$ was a parameter, 1$\sigma$ constraints were found for all systems ($1\sigma < 1.1$ au) except for HD 106906, for which we find a lower limit. The lower limit suggests strong asymmetry, which has already been inferred from previous investigations of the system \citep{kalas2015,lagrange2009,crotts2021,crotts2023}. In our analyses, we identify well-fitting models that attempt to account for this asymmetry by introducing offsets along both the major ($dy$) and minor ($dx$) axes of the disk. However, residuals are still present, suggesting that stellocentric offsets alone may not be enough to account for such asymmetries.

We also calculate $\chi^{2}_{\rm red}$ for all of our maximum likelihood models. While this value is typically used as a metric for quality of the model fit, we emphasize that this quantity is highly sensitive to the shape and size of the likelihood calculation mask and should be treated as more of a guideline. The likelihood calculation regions contain pixels with no disk signal in either model or data, naturally reducing $\chi^{2}_{\rm red}$ by including additional degrees of freedom with little informational value. Additionally, this value is highly sensitive to the choice in multiplicative scaling factor applied to the noise map, with the ideal choice in scaling factor not likely to be consistent between different datasets. For some of our analyses, the noise is potentially overestimated, but reducing this factor runs the risk of underestimating error bars on our constrained parameters, a problem encountered in \cite{mazoyer2020}. We retain this scaling factor of 3 to be consistent with previous literature analyses and for uniformity in our analysis, in addition to providing conservative estimations of our constrained parameters.

Compared to other targets in this analysis, HD 35841 and HD 114082 display the smoothest and overall lowest residual maps, with the strongest residuals ($\gtrsim2\sigma$) appearing close to the FPM, likely due to deviations from Gaussian noise in this region. No strong residuals appear to overlap with disk signal.

For the asymmetric systems HD 106906, HD 110058, and HD 111520, structured, positive residuals ($\gtrsim 1\sigma$) appear in regions that overlap with disk signal. In the case of HD 106906, \texttt{DiskFM} prefers solutions favors a best-fit model that places the center of the ring SE of the FPM. For HD 111520, the best-fit model appears to prefer a solution that attempts to leverage the brightnesses of both sides, strongly preferring solutions with offsets that push against the prior boundary of 5 au NW of the star. Finally, for HD 110058, the best-fit model itself appears to have an offset in $PA$ compared to the perceived location of the disk. This offset may be the best attempt of the model to match the observed "S"-shape seen in the reduced science image. The results for all three of these systems highlight the limitation of our model setup to constrain properties of highly asymmetric systems.

For HD 32297, strong positive residuals (Res. $\gtrsim 2\sigma$) are seen in the region within 0$\farcs$2 of the inner edge of the likelihood calculation mask, centered at the midplane of the disk. This region overlaps with the highest SNR values in the reduced data, and are likely inducing nonlinearity in the KLIP-ADI reduction \citep{pueyo2016}, breaking down the linear expansion assumed in the forward modeling process. Regions where the absolute value of the residuals exceed 1 are also present far from the star and throughout most of the mask.

For HD 146897, significant positive and negative residuals are structured on the front and back sides of the disk respectively. The inability to achieve a low residual (|Res.| $\lesssim 1.5\sigma$) forward model in this case can likely be attributed to the overall noise properties present in regions near the front side, with residual PSF halo structure present that cannot be disentangled from disk signal. Because of the halo structure biasing the brightness of the front side of the disk, \texttt{DiskFM} cannot effectively identify models with a bright enough front side without making the back side too bright compared to the data using the generic SPF.

One target, HR 4796A, exhibits significant and azimuthally-structured residuals (|Res.| $\gtrsim 2\sigma$) within its likelihood calculation regions. The distribution of residuals in this case appears to be bimodal rather than normally distributed, with high positive residuals present from the ansae to the back side of the disk and high negative residuals present on the front side of the disk. This behavior is expected, given our choice in using the generic SPF. In comparing the measured SPF of HR 4796A from \cite{milli2017}, the phase function is comparatively brighter at $>75\degr$ than the generic SPF that we initially chose for the system. The generic scattering phase function is extremely forward-scattering, with less emission at the ansae of the disk and little back-scattering comparatively. Strong positive residuals (Res. $\gtrsim 2\sigma$) are shown on the back side and ansae of the HR 4796A disk.


\subsection{Half Mask Model Results} \label{sec:asymresults}
Four of the eight targets (HD 106906, HD 110058, HD 111520, and HR 4796A) have demonstrated asymmetric features in previous studies \citep{kalas2015,kasper2015,draper2016,chen2020,crotts2023}. Median likelihood parameters related to asymmetry ($dx$ and $dy$) for these four systems were also found to be greater than zero within 3$\sigma$.

\begin{figure*}
    \centering
    \includegraphics[width=\textwidth]{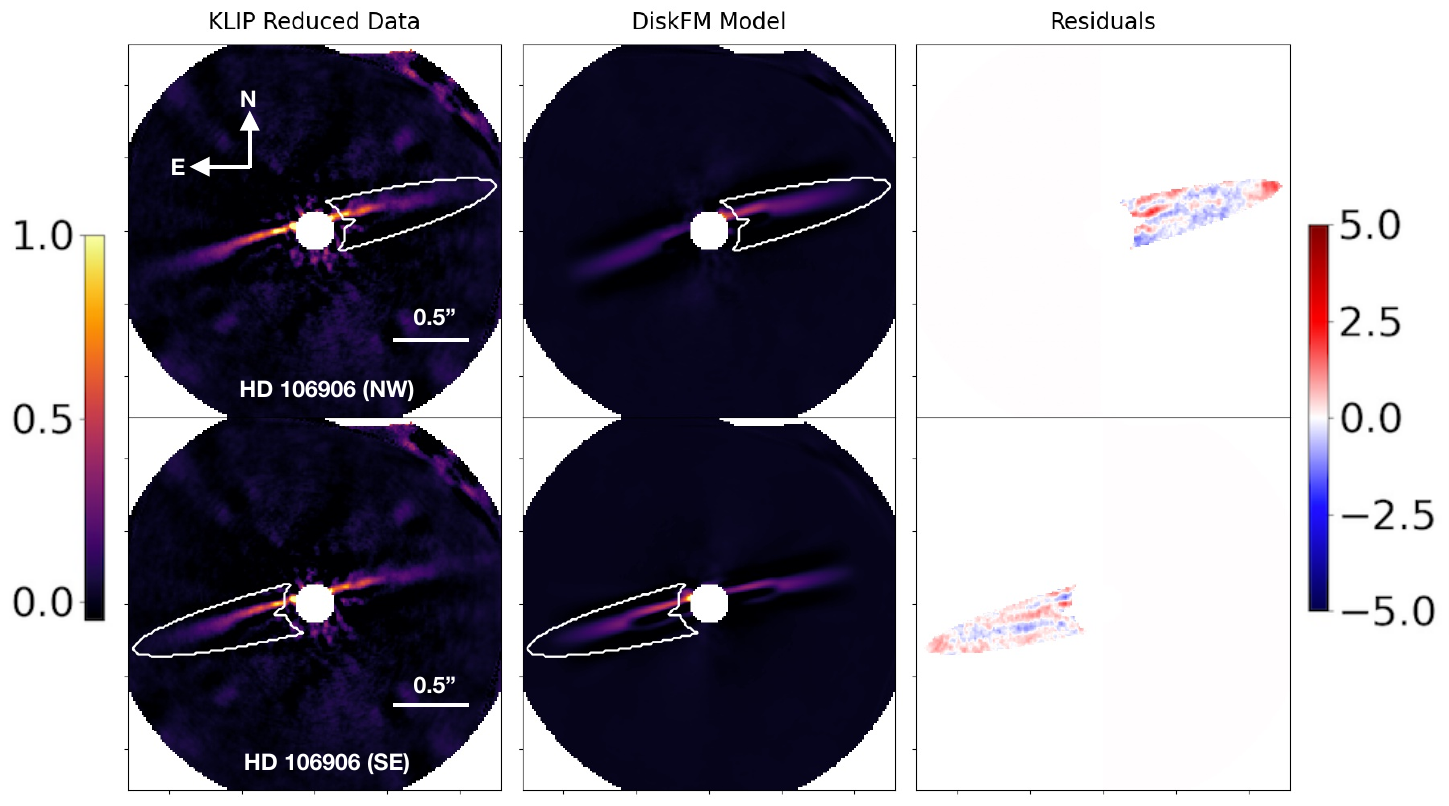}
    \caption{Best-fit modeling results for separate analyses of the NW and SE sides of HD 106906. The models are distinctly different, most notably seen in the $PA$, $i$, and vertical structure profiles of both best-fit disk models.}
    \label{fig:hd106906sep}
\end{figure*}

\begin{figure*}
    \centering
    \includegraphics[width=\textwidth]{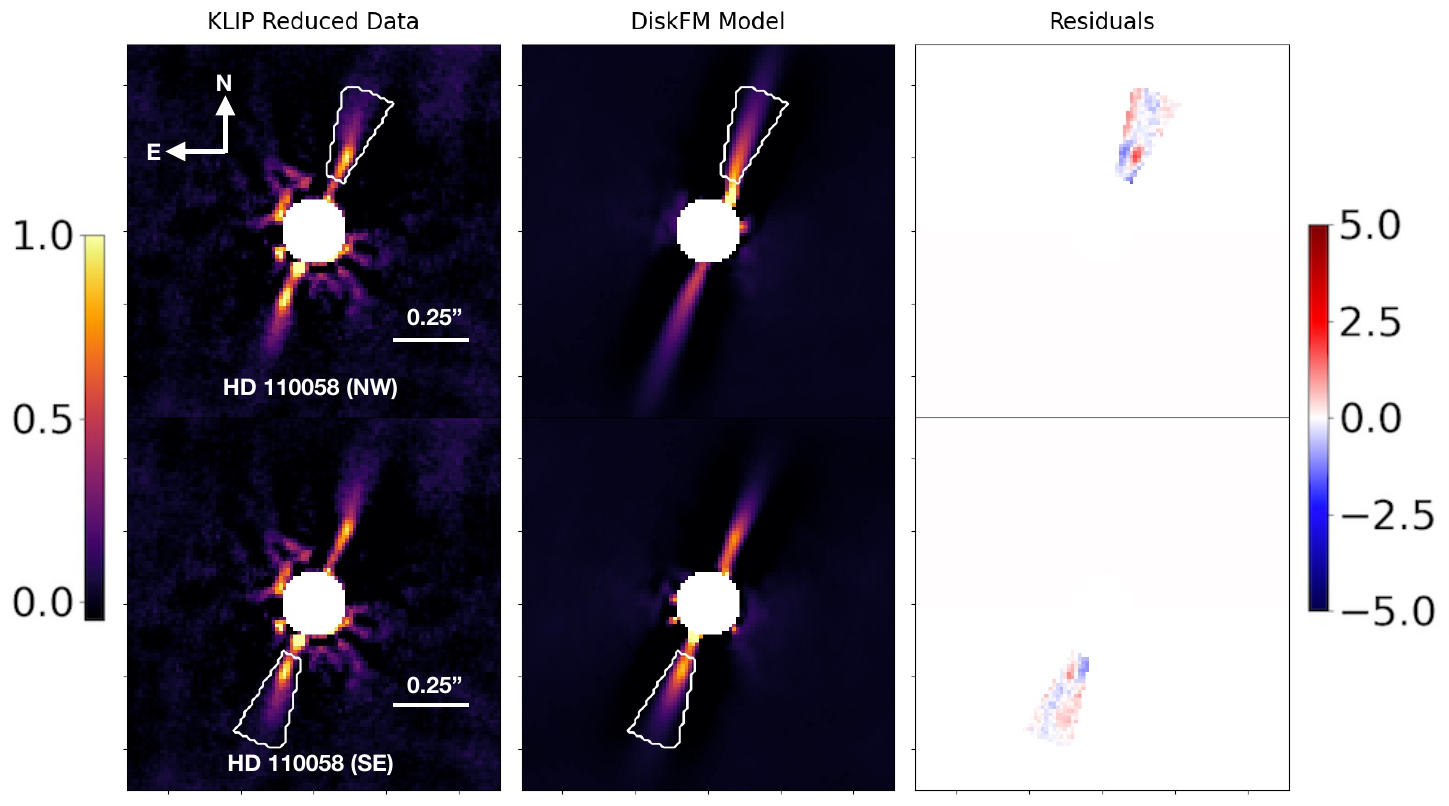}
    \caption{Best-fit modeling results for separate analyses of the NW and SE sides of HD 110058. Differences in the models are subtle but most notably seen in the $PA$ and stellocentic offsets of both best-fit disk models.}
    \label{fig:hd110058sep}
\end{figure*}

\begin{figure*}
    \centering
    \includegraphics[width=\textwidth]{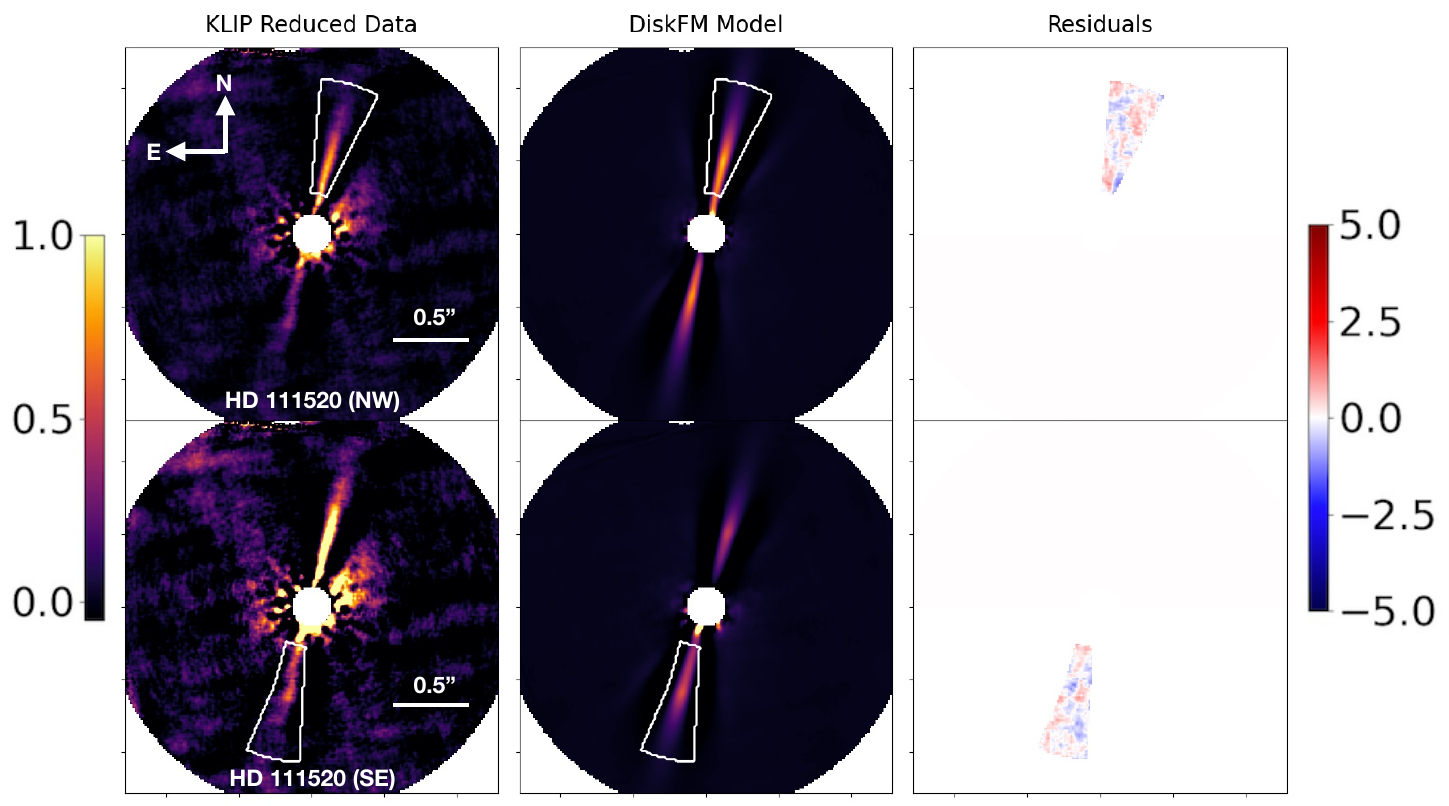}
    \caption{Best-fit modeling results for separate analyses of the NW and SE sides of HD 111520. The models are drastically different, especially in terms of overall brightness. Other notable differences include the $PA$ and overall radial extent of the models.}
    \label{fig:hd111520sep}
\end{figure*}

As test cases, we perform additional analyses on three of these asymmetric disks (HD 106906, HD 110058, and HD 111520), treating each each side of the disks separately, as seen in Figures \ref{fig:hd106906sep}-\ref{fig:hd111520sep}. These test cases assume that the disk is axisymmetric on either side of the star but are independent from each other. The large sizes of these systems allow for a sufficient region over which to calculate the likelihood, even when half of the disk is not being considered. For example, disk spine curvature is typically well-defined even when only considering half of the disk. For these analyses, the initial MCMC parameters and priors are consistent when modeling both sides of each disk, with the exception of the NW side of HD 110058, where initial investigations of the system preferred model solutions in the $PA$ outside of the initially chosen prior range. Therefore, for the NW side of HD 110058, we extend the upper bound of the prior range to $165\degr$. We do not expand the prior range for the SE side, as model solutions preferred a $PA$ well within the initially chosen prior range. The likelihood masks for these systems are split in half, allowing the likelihood calculation to ignore one side of the disk that may appear to have different morphological properties compared to the other side. We chose not to conduct a half-mask modeling analysis for HR 4796A despite its asymmetric nature, as the median likelihood value of $dy$ corresponds to an offset less than one pixel in length.

When determining best-fit models for each side separately, we are able to achieve lower residuals without systematic patterns of disk signal in the regions where likelihood was calculated. Regardless, radial and vertical dust density parameters are still found to be consistent with the initial analyses for HD 110058 and HD 111520 where the entire likelihood mask is used. These two systems do not appear to have morphological asymmetries as significant as HD 106906, with HD 110058 containing diffuse asymmetrical features around the ansae and HD 111520 containing a strong brightness asymmetry and tentative warp on the SE side \citep{draper2016}.

In modeling the two sides of HD 106906 separately, we are able to constrain positional and outer disk properties and achieve upper/lower limits for stellocentric offsets. The most significant differences are evident in the median likelihood values of $i$, $PA$, $a_r$, and $dy$. While $i$ may be somewhat degenerate with $a_r$, the distinct differences further highlight the asymmetric nature of the system, suggesting that the NW side is less inclined but vertically thicker than the SE side. Both of these models still appear to suggest large offsets, potentially suggesting that stellocentric offsets alone are still not sufficient in finding best-fit models for the disk even when analyzing both sides separately. To more robustly characterize the morphology of this system, a disk model that can create truly eccentric disks may be needed.

For HD 110058, 1$\sigma$ constraints for both sides are achieved for $\alpha_{\rm out}$. The compact nature of the system is still the most limiting factor in modeling this system, as the lack of spatial resolution makes all parameters more difficult to constrain, particularly radii and $\alpha_{\rm in}$. Otherwise, parameters for both sides appear consistent with each other within 3$\sigma$ except for $PA$, likely due to the asymmetric warps on both sides of the disk \citep{stasevic2023,lopez2023}. Appendix \ref{sec:hd110058compare} contains a more detailed discussion of the properties of the HD 110058 disk and comparisons to previous literature.

In the two-sided model analysis of the HD 111520 disk, residuals no longer contain structure and are less (|Res.| $\lesssim 1\sigma$) compared to modeling the whole disk. 1$\sigma$ constraints in $\alpha_{\rm out}$, $a_r$, and $PA$ were found for both sides. Constrained parameters for both analyses are consistent with each other except for $PA$, likely due to the tentative warp observed on the SE side.

While constraints on some density properties of modeling either side of a disk are consistent with each other, namely $\alpha_{\rm out}$ and $a_r$, the greatest differences are highlighted in the constraints of $PA$ and $i$ (and $a_r$ in the case of HD 106906). The inconsistencies further highlight the asymmetric nature of these systems, as it suggests that a single model approach cannot unify both sides of a strongly asymmetric system; additional model mechanisms should be explored to treat such systems.

\subsection{Different Scattering Phase Function Approach Results} \label{sec:SPFresults}
\begin{figure*}
    \centering
    \includegraphics[width=\textwidth]{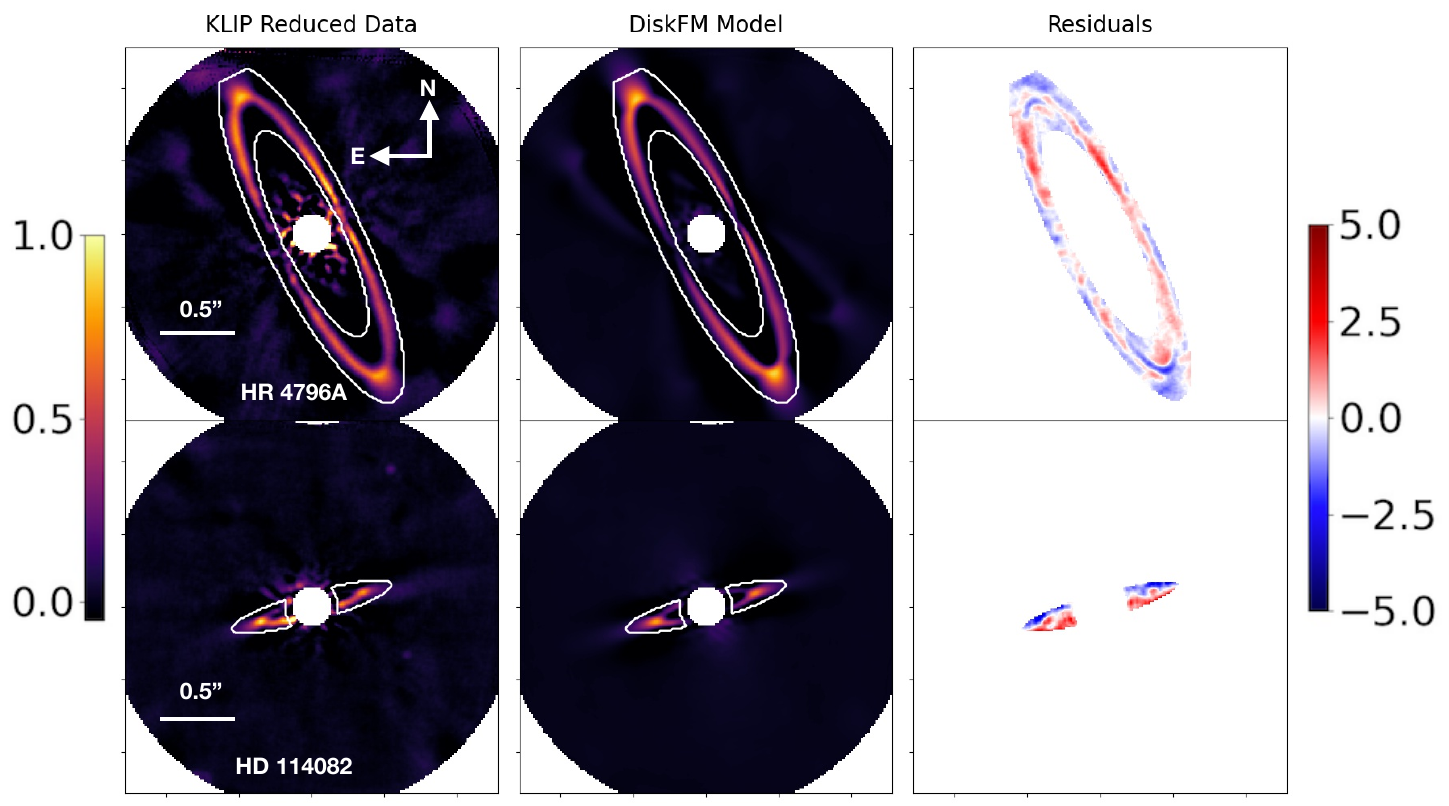}
    \caption{Best-fit modeling results for analysis of HR 4796A and HD 114082 with the measured SPF from \citet{milli2017}. While we achieve a lower-residual fit for HR 4796A, positive residuals are still present particularly along the front and back sides of the disk on the Northern sides. While a solution with a stellocentric offset is preferred in the MCMC analysis, the presence of relatively strong residuals may be explained by a dust density enhancement as observed in \citet{olofsson2019}, \citet{milli2019}, and \citet{chen2020}. To achieve a reasonable morphological model for HD 114082, the $PA$ of the disk had to be increased by 180$\degr$, as \texttt{DiskFM} prefers to align the brighter back side of the disk model with the apparent front side of the disk from the data. The forward-scattering lobe of the SPF located on the front side of the disk is also behind the FPM in this dataset, further encouraging this behavior.}
    
    \label{fig:milliplots}
\end{figure*}

Given that the previously measured SPF for HR 4796A is distinct from other systems, we performed an additional model analysis with the measured SPF from \cite{milli2017}. The initial parameters and priors are kept consistent from the previous analysis, and our best-fit model is shown in Figure \ref{fig:milliplots}.

The improvement in the best-fit model for HR 4796A is evident from the significantly lower residuals of the best-fit model, particularly at the back side and ansae of the disk (|Res.| $\lesssim 2\sigma$), in addition to the relatively lower $\chi^2_{\rm red}$ (0.41 for the \citealt{milli2017} SPF compared to 1.90 for the generic SPF). While this result supports that the generic SPF is not applicable to all dust systems, it bolsters the methodology of this approach, namely that well-constrained morphological properties of disks can still be determined with a fixed SPF. Additionally, all median likelihood parameters and limits except for the aspect ratio were consistent between both modeling analyses, suggesting that using an inaccurate SPF may still be able to provide a first-order estimate of some morphological properties for a well-resolved and bright disk system. Our findings suggest much steeper radial profiles than our prior limits, which is consistent with the sharply defined edges observed in the disk. The vertical structure profile of the disk is also difficult to constrain, likely due to either a lack of resolution along the vertical direction and/or the removal of astrophysical signal from self-subtraction.

To further assess the impact of SPF choice in constraining morphological parameters, we also conduct an analysis of the fainter HD 114082 ring with the HR 4796A SPF, even though the generic SPF analysis of this system provided a good model fit. The prior ranges, initial model parameters, and likelihood mask are kept consistent from the generic SPF analysis (except for the prior range and initial $PA$, as discussed in this section). Given the high degree of back-scattering in the HR 4796A SPF compared to our generic SPF, the faint and marginally-resolved back side of the disk, and the forward scattering peak of the disk being located behind the FPM, a reasonable model for this system would require ``flipping" the orientation of the disk such that the apparent front side of the disk aligns with the back-scattering regime of the HR 4796A SPF. To account for this, we added 180$\degr$ to both the prior range and initial $PA$, simulating placing the front side of the disk on the North side. While we achieve a reasonable maximum likelihood model (Figure \ref{fig:milliplots}), the fit is still poor compared to our initial analysis with the generic SPF and the front side of the disk on the South side, with a $\chi^2_{\rm red}$ of 1.99 compared to 0.90 with the generic SPF. Furthermore, to achieve this model, we had to place the front side of the disk on the Northern side of the image, inconsistent with previous studies of the system \citep{crotts2023,engler2023}. Additionally, we find that most constrained parameters are consistent between the generic and HR 4796A SPF analyses, aside from $i$, $\alpha_{\rm in}$, and $a_r$. 

Both of the HD 114082 and HR 4796A analyses with the \cite{milli2017} SPF suggest that choosing inaccurate scattering properties will lead to poorer model fits. We stress however that these are only two test cases, and we leave a deeper investigation of SPF choice for multiple systems to a future study. Given the significant differences between the HR 4796A SPF and our generic SPF, we cannot make a conclusion on how much the model fits are affected by more subtle changes in the fixed SPF used; this investigation is also left as a future study.

\subsection{\textit{K1}-band Modeling Results} \label{sec:K1results}
\begin{figure*}
    \centering
    \includegraphics[width=\textwidth]{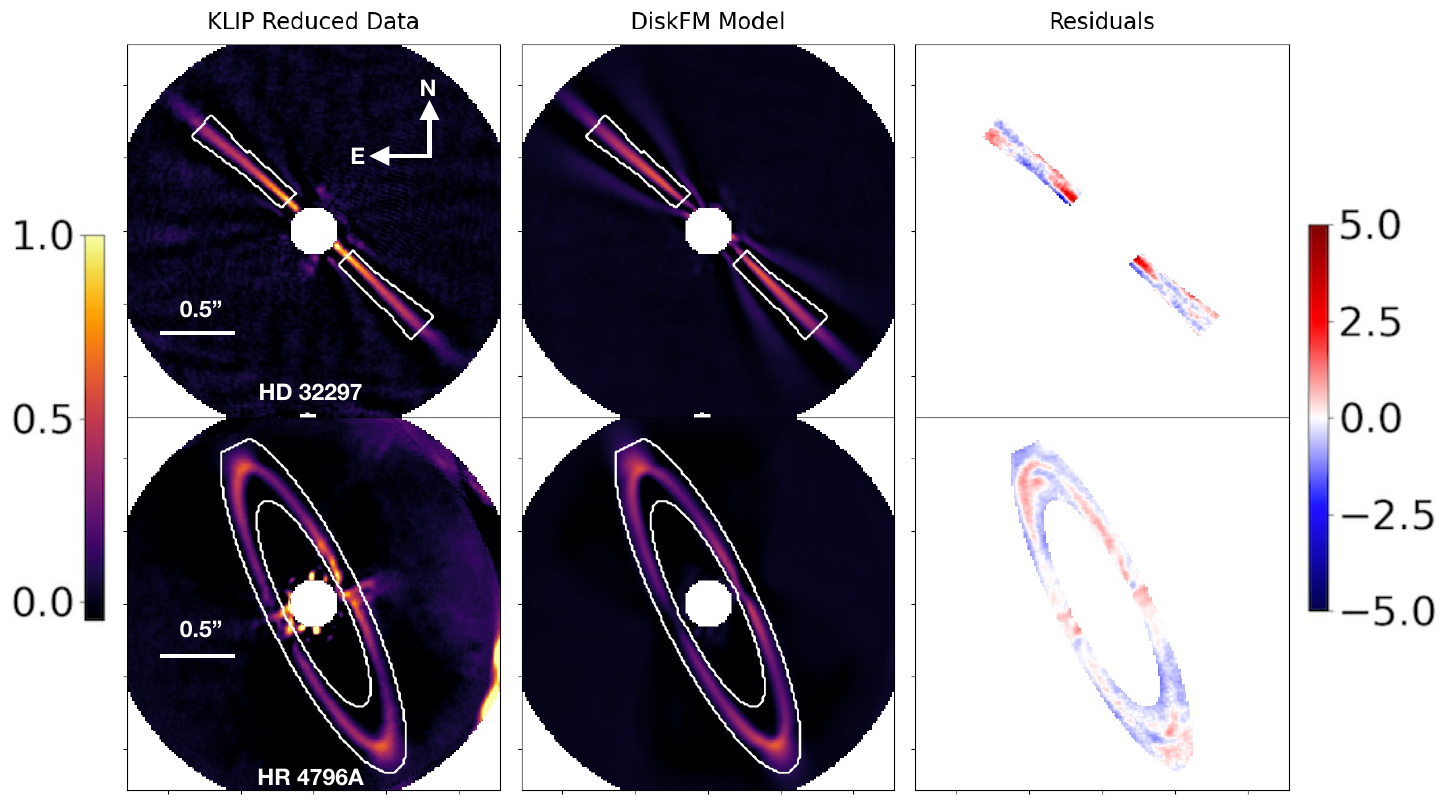}
    \caption{Best-fit modeling results for analyses of \textit{K1}-band images of HD 32297 and HR 4796A. The generic SPF was used for HD 32297 and the \citet{milli2017} SPF was used for HR 4796A. Residuals appear cleaner in both systems compared to their \textit{H}-band images, potentially due to the fainter disk brightesses at \textit{K}. Our best-fit model for HD 32297 still has some significant residuals close to the edge of the FPM, and nonlinearity in the KLIP reduction may still be evident to a lesser extent. 
    }
    \label{fig:K1grid}
\end{figure*}

For the two systems with \textit{K1}-band datasets meeting our SNR threshold, best-fit model results are shown in Figure \ref{fig:K1grid}. For HD 32297, we apply the generic scattering phase function as the proxy for grain properties. As we have already demonstrated that the generic scattering phase function does not provide a low residual model for HR 4796A, we apply the measured scattering phase function for HR 4796A from \cite{milli2017}. 

Compared to the \textit{H}-band reduced image of HD 32297 and best-fit model, the \textit{K1}-band reduced image of HD 32297 and best-fit model appear less affected by the effects of self-subtraction from PSF subtraction and nonlinearity in forward modeling, although high residuals (|Res.| $\gtrsim 2\sigma$) are still present. As the emission from the disk at \textit{K1}-band is relatively less than emission at \textit{H}-band, self-subtraction and nonlinearity in the forward modeling process is likely to be less severe. Additionally, similarities in relative errors among parameters with a lower disk SNR could suggest lower residual amplitudes. The broader PSF at \textit{K1}-band may also contribute to the smoothing of finer structural features in the disk, allowing the model to more easily match the data. Regardless, our \textit{K1}-band model fit still has significant residuals within $0\farcs1$ of the inner edge of our likelihood mask.

The HR 4796A best-fit \textit{K1}-band model achieves a low residual (|Res.| $\lesssim 1\sigma$) model fit to the \textit{K1}-band reduced science image. Some positive residuals persist on the NE side of the disk, particularly around the ansae and the front side of the disk. A brighter NE side has been observed in previous studies (e.g. \citealt{milli2017}) suggesting that HR 4796A is an eccentric system, and while our disk model can produce modest eccentricities by inducing stellocentric offsets, we are unable to reproduce the observed asymmetry, suggesting that a dust density enhancement may be present that cannot be created by our model setup. We are able to tightly constrain most of our free parameters, with the exception of $\alpha_{\rm in}$ and the aspect ratio. $\alpha_{\rm in}$ can only be constrained while $R_{\rm C}>R_{\rm in}$, and the lack of distinction between $R_{\rm in}$ and $R_{\rm C}$ suggests that the inner radial profile of the disk is sharply defined. Similar to the \textit{H}-band result, our posteriors suggest steep radial density profiles.

\begin{landscape}
\begin{table}
	\centering
	\caption{MCMC median likelihood parameters with $\pm1\sigma$ error bars and $3\sigma$ upper/lower limits.}  
	\label{tab:MCMCresultsfinal}
	\begin{tabular}{@{\extracolsep{\fill}}lcccccccccc} 
		\hline
		Name & $R_{\rm in}$ & $R_{\rm C}$ & $\alpha_{\rm in}$ & $\alpha_{\rm out}$ & $a_r$ & $i$ & $PA$ & $dx$ & $dy$ & $\chi^{2}_{\rm red}$\\
             & [AU] & [AU] & & & & [$\degr$] & [$\degr$] & [AU] & [AU] &  \\
		\hline
            \multicolumn{10}{c}{Generic Scattering Phase Function} \\ \hline
		  HD 32297 & $45.24\substack{+0.64\\-0.94}$ & $>142.54$ & $<0.26$ & $-7.40\substack{+0.43 \\ -0.41}$ & $0.01\pm0.001$ & $87.63\pm0.03$ & $47.29\pm0.01$ & -- & $-0.46\pm0.15$ & 2.20\\
            HD 32297 (K1) & $40.34\substack{+1.34\\-1.51}$ & $>142.07$ & $<0.27$ & $-5.07\pm0.29$ & $<0.01$ & $87.58\substack{+0.05 \\ -0.04}$ & $47.13\pm0.02$ & -- & $0.89\pm0.28$& 0.86\\
            HD 35841 & $<55.97$ & $62.15\substack{+1.32 \\ -1.61}$ & $2.28\substack{+1.18 \\ -0.79}$ & $<-6.09$ & $0.04\substack{+0.004 \\ -0.005}$ & $82.69\substack{+0.30 \\ -0.28}$ & $165.49\pm0.15$ & $1.94\substack{+1.11 \\ -1.09}$ & $0.04\substack{+0.49 \\ -0.48}$& 0.54\\
            HD 106906 & $<49.02$ & $<75.88$ & $3.28\substack{+2.23\\-1.66}$ & $-1.97\substack{+0.22 \\ -0.29}$ & $0.03\substack{+0.004 \\ -0.005}$ & $83.94\substack{+0.21 \\ -0.19}$ & $285.06\substack{+0.08 \\ -0.09}$ & $>4.16$ & $-2.38\substack{+0.49\\-0.52}$& 0.43\\
            HD 110058 & $<42.18$ & $<67.10$ & Unconst. & $-3.07\substack{+0.47 \\ -0.68}$ & $0.12\pm0.01$ & $>84.62$ & $158.67\substack{+0.27\\-0.26}$ & -- & $-2.04\substack{+0.59\\-0.62}$& 0.18\\
            HD 111520 & $<54.13$ & $72.12\substack{+7.48 \\ -7.12}$ & $<6.77$ & $-4.19\substack{+0.63 \\ -0.96}$ & $0.07\pm0.004$ & $>89.82$ & $166.40\pm0.09$ & -- & $<-4.64$& 0.33\\
            HD 114082 & $31.11\substack{+0.26 \\ -15.10}$ & $<33.51$ & $>0.04^a$ & $-2.99\substack{+0.12 \\ -0.13}$ & $0.06\pm0.004$ & $83.07\substack{+0.19 \\ -0.71}$ & $106.16\pm0.07$ & -- & $0.74\pm0.11$& 0.90\\
            HD 146897 & $<49.25$ & $50.10\substack{+1.03 \\ -6.87}$ & $>0.75^a$ & $-2.45\substack{+0.12 \\ -0.13}$ & $<0.04$ & $82.51\substack{+0.27 \\ -0.26}$ & $115.00\pm0.17$ & -- & $0.69\substack{+0.35 \\ -0.34}$& 1.09\\ 
            \hline
            \multicolumn{10}{c}{Half Mask Model Analysis} \\ \hline
            HD 106906 (NW) & $<43.24$ & $67.84\substack{+7.95 \\ -8.59}$ & $<4.36$ & $-2.73\substack{+0.55 \\ -0.67}$ & $0.06\pm0.005$ & $83.07\pm0.22$ & $287.28\pm0.14$ & $<-1.92$ & $>2.23$& 0.37\\
            HD 106906 (SE) & $55.21\substack{+1.48 \\ -5.04}$ & $<74.60$ & Unconst. & $-2.56\substack{+0.27 \\ -0.32}$ & $<0.01$ & $85.21\pm0.11$ & $283.32\pm0.09$ & $<-1.40$ & $<-3.75$& 0.20\\
            HD 110058 (NW) & $<40.57$ & $<68.37$ & $<9.98^a$ & $-2.35\substack{+0.59 \\ -1.00}$ & $<0.14$ & $85.05\substack{+0.64 \\ -0.77}$ & $159.97\substack{+0.52\\-0.56}$ & -- & $<4.60$& 0.29\\
            HD 110058 (SE) & $34.83\substack{+3.64 \\ -5.44}$ & $<57.59$ & Unconst. & $-3.77\substack{+0.60 \\ -0.76}$ & $0.12\substack{+0.02 \\ -0.01}$ & $>84.17$ & $157.96\substack{+0.34 \\ -0.35}$ & -- & $0.08\substack{+2.98 \\ -2.91}$& 0.14\\
            HD 111520 (NW) & $<56.64$ & $<90.35$ & $<9.89^a$ & $-3.39\substack{+0.31 \\ -0.49}$ & $0.06\substack{+0.004 \\ -0.003}$ & $>89.62$ & $167.08\pm0.08$ & -- & $-1.11\substack{+2.24 \\ -2.21}$& 0.17\\
            HD 111520 (SE) & $<74.06$ & $<92.60$ & Unconst. & $-4.33\substack{+0.80 \\ -1.02}$ & $0.07\pm0.01$ & $>88.11$ & $164.00\pm0.23$ & -- & Unconst.& 0.14\\ \hline
            \multicolumn{10}{c}{\cite{milli2017} Scattering Phase Function} \\ \hline
            HR 4796A & $73.86\substack{+0.32 \\ -0.09}$ & $73.79\substack{+0.54 \\ -2.11}$ & $>2.41$ & $<-9.81$ & $<0.004$ & $76.89\pm0.03$ & $26.75\pm0.02$ & $-2.20\pm0.09$ & $0.84\pm0.05$& 0.41\\
            HR 4796A (K1) & $74.43\pm0.10$ & $<73.05$ & $>1.96$ & $<-9.78$ & $<0.007$ & $76.95\pm0.05$ & $26.55\pm0.04$ & $-2.41\pm0.16$ & $1.79\pm0.08$& 0.21\\
		\hline
  \multicolumn{11}{l}{$^a$These upper/lower limits span most of the prior range but have low probability tails.}
	\end{tabular}
\end{table}
\end{landscape}

\section{Discussion} \label{sec:discussion}
\subsection{Limitations of the Forward-Modeling Approach} \label{sec:KLIPnonlinear}
From our analysis, not all systems yield low residual model fits. The linear approximation in Equation 4 of \cite{pueyo2016} breaks down in the presence of astrophysical sources that are brighter than or of a similar brightness to local speckles, particularly when using "aggressive" KLIP parameters. 
\cite{mazoyer2020} found that this forward modeling nonlinearity can be worse for bright and extended disks. Nonlinearity in the KLIP reduction can hinder the ability of \texttt{DiskFM} to properly treat self-subtraction and PSF convolution and is likely the cause of our poor model fits for HD 32297, the brightest disk in our sample. To mitigate this issue for future characterization of bright debris disks, a different PSF post-processing technique (e.g. NMF; \citealt{ren2018}) may be necessary. Our inability to achieve a low residual model for HD 32297 prevents us from making any definitive conclusions on its geometry and dust density distribution properties, and will be excluded from our discussion and comparisons of morphological properties between the other systems in our sample. As one of the primary goals of this study is to compare dust density properties among our sample, we present ensemble results in \S \ref{sec:vertextents} and \S \ref{sec:eccentricity}. For a detailed discussion of constrained dust density properties to previous studies on a system-by-system basis, we refer the reader to Appendix \ref{sec:pastcomparisons}.

Although estimations of $i$ and $PA$ appear mostly consistent with past studies (see Appendix \ref{sec:pastcomparisons}) despite a variety of observations from different instruments and near-IR filters, inconsistencies become apparent in estimating dust density distribution properties, evident in the midplane density profiles of previous studies and this work. The uniqueness of approaches and associated limitations in analyzing disks also prevent direct comparisons of structural properties, further complicating ensemble studies of debris disks.

Finally, the manner in which disk morphological and dust density distribution properties have been explored has been relatively limited up until recent studies within the past few years. Many studies create model grids in order to determine the best-fitting parameters and do not explore full parameter spaces. The use of MCMC and multinest sampling analysis have only recently been utilized in disk image modeling (e.g. \citealt{duchene2020,esposito2018,crotts2021,engler2023,chen2020,olofsson2022}), allowing a more thorough search of parameter spaces.

As we obtain more high-resolution images of debris disk systems, consistent methodologies (e.g. \citealt{crotts2023,olofsson2022}) are needed to ensure more direct comparisons between systems with unique properties such as host star spectral type and age. While a uniform modeling approach may limit the ability to obtain the lowest residual model fits, the general consistency we find in our analysis to previous studies suggests that loss of fit quality does not severely bias the results of dust morphological modeling and that true astrophysical trends can still be identified.

\subsection{\textit{H}- and \textit{K1}-band Model Comparisons}
Excluding the poor model fits for HD 32297, we find that the constrained parameters of HR 4796A between \textit{H}- and \textit{K1}-band are consistent to within 3$\sigma$, suggesting that the two wavelengths are probing similar scattering grain populations of dust. This suggests that our methodology is not too sensitive to the choice in dataset as long as the SPF remains largely unchanged between different wavelengths.

Interestingly, use of the measured \textit{H}-band SPF for HR 4796A can still achieve a best-fit model at \textit{K1}-band with reasonable residuals. This is supported by findings from \cite{chen2020}, where the extracted HR 4796A scattering phase functions from \textit{J}-, \textit{H}-, \textit{K1}-, and \textit{K2}-band observations appear consistent with each other within uncertainties.

\subsection{Vertical and Radial Extents} \label{sec:vertextents}
Overall, we find that vertical aspect ratios are all $<0.14$. Although most results are consistent with previous findings, we note that our determination of $a_r$ appears to suggest extremely narrow profiles for HR 4796A (vertical FWHM $\sim$ 0.1 au at $R_{\rm C}$), unlike what was found in \cite{olofsson2022} (vertical FWHM $\sim$ 4 au at $R_{\rm C}$). Additionally, the shape of the vertical density distribution may affect the aspect ratio, as \cite{olofsson2022} kept $\gamma_{\rm vert}$ as a free parameter whereas we fix ours. Our estimated aspect ratios are also notably smaller than what was found in \cite{crotts2023}, but it should be noted that the method used to measure vertical FWHM and aspect ratio in \cite{crotts2023} did not account for disk projection effects and confusion between radial and vertical extents.

Within this sample, we find no obvious correlations between aspect ratios and spectral type or approximate age. However, this may be an artifact of our small sample size that is limited to A- and F-type stars, with five of our eight systems likely sharing similar ages as members of the Scorpius-Centaurus OB association \citep{deZeeuw1999}. With a larger sample size of more diverse spectral types, \cite{crotts2023} does identify a correlation of increasing aspect ratio with increasing host star temperature in the case of radially compact disks, however. 

\cite{olofsson2022} investigated the effects of gaseous components of disks on dust density morphologies and found that intermediate levels of gas mass most strongly affect the vertical density distribution at scattered light wavelengths. From their model simulations, increases in the gas mass correlate with a lower aspect ratio or "thinner" disks as gas drag boosts the efficiency of vertical settling. Of the eight targets investigated in our sample, two disks (HD 32297; \citealt{greaves2016}, HD 110058; \citealt{hales2022}) have measured CO gas detections and one disk (HD 146897; \citealt{lieman-sifry2016}) has a tentative CO gas detection. The largest modeled aspect ratios appear to be around the star HD 110058, where $a_r$ was found to be greater than 10\% in modeling both the entire disk and the SE side independently. This may seem contradictory to the idea that a gas-bearing disk would likely be thinner, but the distance and compact nature of HD 110058 could also suggest a lack of resolution along the vertical direction and prevent the formation of a definitive conclusion on the presence of gas and disk vertical extent. Additionally, vertical stirring mechanisms (e.g. from dynamical interactions with substellar companions) may supersede the dampening of inclination from gas drag.

\begin{figure*}
    \centering
    \includegraphics[width=\linewidth]{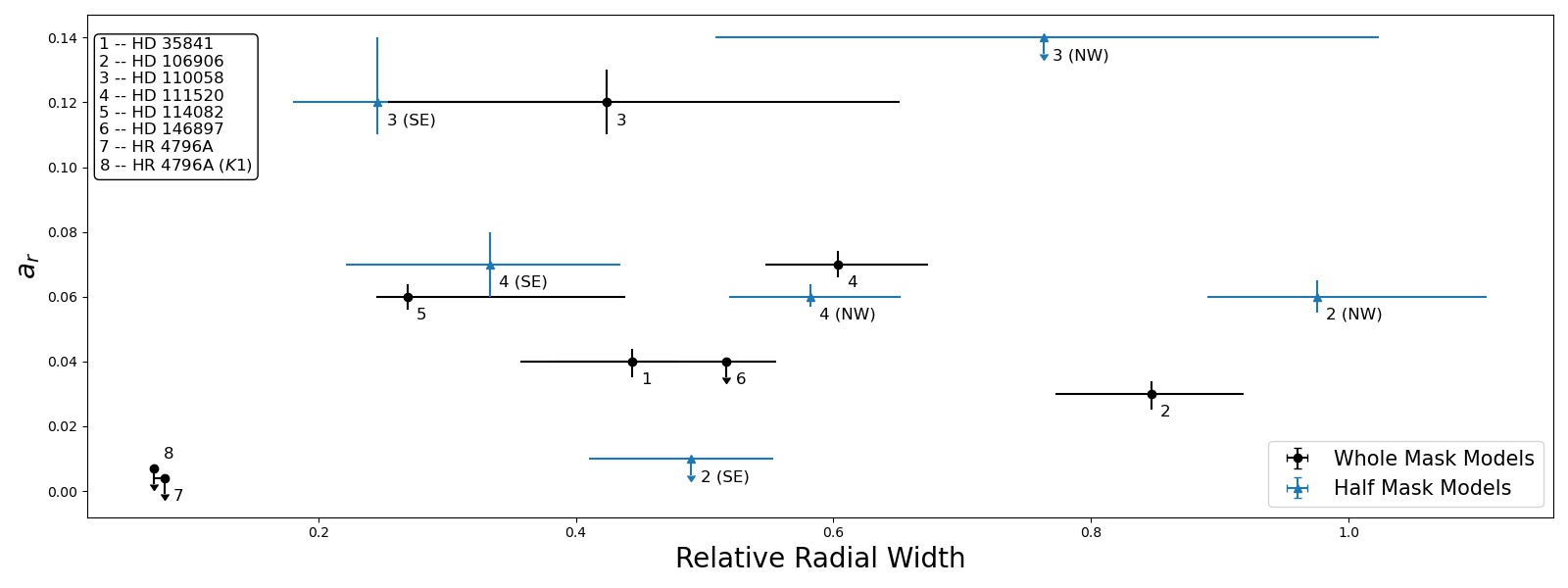}
    \caption{Aspect ratio as a function of radial disk FWHM divided by the radius of maximum dust density. No trends are observed as a function of $a_r$.}
    \label{fig:ar_radialwidths_maxdense}
\end{figure*}

To assess correlations between radial and vertical structure, we calculated the median likelihood relative radial widths from our model posterior distributions, defined as the FWHM of the radial density profile following Equation \ref{eq:radialdust}, restricted by $R_{\rm in}$ and $R_{\rm out}$ on either side and divided by the radius of maximum dust density. In Figure \ref{fig:ar_radialwidths_maxdense}, we plot $a_r$ as a function of the relative radial width. In a study of ALMA-observed debris disks, \cite{terrill2023} identified a tentative trend of increasing relative radial width and aspect ratio, with HD 110058 exhibiting the largest aspect ratio and third largest relative radial width, assuming a 90$\degr$ inclination. We do not immediately identify correlations between properties in either comparisons, although the sample size of models is relatively small. In the most narrow systems (HR 4796A and HD 114082), a few mechanisms could induce smaller radial widths, including the sculpting/shepherding of dust along the inner and outer edges of a ring \citep{boley2012}, truncation from external perturbers \citep{nesvold2017}, primordial eccentricities from instabilities \citep{kennedy2020}, and gas/dust interactions \citep{lyra2013}. As a general case, \cite{chiang2009} and \cite{rodigas2014} identified a relation between the mass of a single interior planetary sculptor and the normalized relative width of a ring, with higher mass planets linearly related to a more radially thickened exterior debris ring. 

With the exception of HD 110058, most of our constrained disk aspect ratios are consistent with "natural" scale height from radiation pressure \citep{thebault2009}, although planetary companions can still induce vertical stirring. \cite{dong2020} investigated model scenarios of massive planets influencing dust grains and found that disk aspect ratios can be enhanced most strongly ($\sim 0.05$) for multiple-planet systems where the planets lie interior to the disk. Multiple interior planets can also lead to more radially narrow structures depending on orbital and mass configurations. In a case study of HD 106906, a dynamical model analysis in \cite{nesvold2017} found that increasing the mutual inclination of an external perturbing companion increases the vertical extent of a disk as well. In the case of HD 110058, interior planetary companions could be responsible for the warps seen in scattered-light images, similar to the $\beta$ Pic system \citep{heap2000,lagrange2009}. The detection of CO gas in HD 110058 \citep{hales2022} could also contribute to the radially narrow and warped geometry observed. \cite{lopez2023} also found that both interior and exterior planetary companions can produce the geometry observed in HD 110058 through secular perturbations, while \cite{stasevic2023} concluded that only an interior companion could produce the observed warped geometry. Interestingly, the HD 106906 median likelihood whole disk and NW side radial widths are fairly broad ($\sim$50 au), despite its eccentric shape likely being caused by an external perturber \citep{nesvold2017}. Broad radial widths and heightened aspect ratios both would support dynamical models of planet scattering \citep{nesvorny2015}, although our aspect ratios for HD 106906 are not large relative to the rest of our sample. For more vertically-thin disks, self-stirring and/or secular interactions may be dominant \citep{matra2019}; this behavior may be relevant for systems such as HD 35841 and HD 146897.

\subsection{Disk Eccentricity} \label{sec:eccentricity}
Disk eccentricity in our model can only be parameterized in the form of stellocentric offsets. In general, the most significant offsets we observe are seen in the best-fit models of HD 106906, HD 110058, HD 111520, and HR 4796A, consistent with identified eccentricities and high degrees of asymmetry measured in polarimetric images in \cite{crotts2023}\footnote{\cite{crotts2023} also identifies as HD 146897 as an eccentric system, although we do not find strong evidence of this in our analysis.}. Pericenter glow \citep{wyatt1999} is postulated as a cause of the asymmetry in HR 4796A, but despite inducing a pericenter glow from an offset, a positive residual remains, suggesting that a localized density enhancement \citep{schneider2009,milli2019} may be needed to account for this effect.

In HD 106906, studies of dust dynamics suggest that a planetary perturber, either interior or exterior to the debris disk (e.g. HD 106906 b, \citealt{bailey2014}) may be the cause of the asymmetry. Additionally, catastrophic collisions from large solid bodies could also explain the needle-like morphology observed in HD 106906 (see \citealt{jones2023} and references therein). Studies of HD 110058 have been limited in scope, but the large aspect ratio found in \cite{crotts2023} and this study along with the characterized warps in \cite{stasevic2023} and \cite{lopez2023} provide tentative evidence of a vertical stirring mechanism such as a companion. Dedicated dynamical sculpting studies of HD 111520 have not been conducted, and the extremely edge-on inclination further limits our ability to assess the true eccentricity of the disk. Giant impact modeling conducted in \cite{jones2023} could also not reproduce the observed morphology of HD 111520, and suggest that an interior planet is more likely to produce the geometry observed.

\begin{figure}
    \centering
    \includegraphics[width=\linewidth]{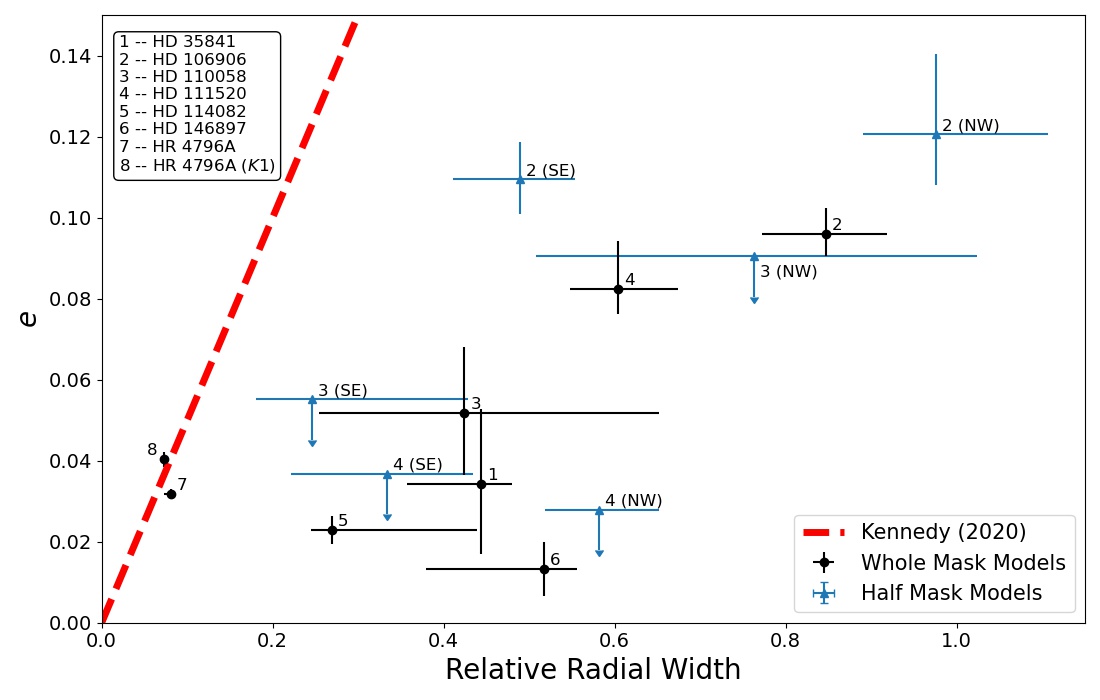}
    \caption{Eccentricities estimated using the posterior distribution of stellocentric offsets and the radii of maximum density as a function of relative radial widths. There appears to be a tentative positive trend between eccentricity and relative radial width. The slope of this trend, however, appears shallower than the slope of the dashed line in Figure 9 of \citet{kennedy2020}, shown in red. This line represents the expected radial width in a zero initial eccentricity secular perturbation scenario. The assessment from \citet{kennedy2020} suggests that our systems should be roughly circular, even though that is not the case for at least HD 106906.
    }
    \label{fig:eccVradW}
\end{figure}

In Figure \ref{fig:eccVradW}, we compare median likelihood eccentricities of all disk modeling analyses to relative radial widths. 
A tentative positive trend is seen in eccentricity versus relative radial width, and we observe that all asymmetric systems and half mask disk models appear to have higher relative radial widths in general. We also observe a large range of eccentricities over a small range of relative radial widths, similar to the large range in $a_r$ we observe. We can compare our results to findings from \cite{kennedy2020}, who observed that the highly eccentric debris disks around Fomalhaut and HD 202628 appear narrower than expected from secular perturbation models and assumptions of zero initial eccentricity. The dashed line in Figure 9 of \cite{kennedy2020} represents a predicted positive-slope relationship of eccentricity and relative radial width in a zero initial eccentricity secular perturbation condition. Although this relationship is positive along with the tentative correlation we observe, the slope of our correlation is shallower. 
Our most eccentric disk models, associated with the whole and half mask models of HD 106906, appear to disagree with the assessment from \cite{kennedy2020} that the most eccentric systems have the most radially narrow sizes, although both studies contain small samples and explore vastly different grain size populations. Finally, similar to our comparisons of radial and vertical structure, the range in eccentricities could also suggest multiple stirring mechanisms that may be responsible for eccentric shape.

\subsection{Implications for Grain Properties and Scattering Phase Functions}
Efforts in modeling the SPF for debris disk systems most commonly utilize the HG phase function \citep{henyey1941} as an analytical (but not physical) model to all scattering properties or fully parameterize dust grain properties assuming a uniform shape to dust grains such as the Mie model \citep{mie1908}. Both of these approaches, however, often fail to properly account for all observed scattering behavior and are inconsistent across multiple wavelengths and/or observing modes. In modeling analysis of HST-NICMOS observations of HD 181327, \cite{schneider2006} identified inconsistencies between a minimum grain size ($a_{\rm min}$) determined from observed mean asymmetry factor in the SPF and an $a_{\rm min}$ determined from disk color and thermal SED. A more physically-grounded basis from which to model dust grains may be in the form of complex aggregates of spherical and/or fractal shapes, identified in some studies of solar system dust (e.g. \citealt{bentley2016}). Although this treatment appears more realistic, scattered-light modeling of aggregate dust grains is a computationally daunting task \citep{tazaki2016,arnold2019}. The Distribution of Hollow Spheres (DHS; \citealt{min2016}) approach attempts to circumvent the issue of computational efficiency of aggregates, but is also not always sufficient in reproducing features seen in both SPFs and polarized intensity surface brightness profiles \citep{arriaga2020}. \cite{arnold2022} has also assessed model SPFs for Mie, DHS, and aggregate grains, demonstrating similarities between them at low scattering angles, but deviations at larger scattering angles and across multiple wavelengths. The wavelength-dependent nature observed in \cite{arnold2022} seems to contrast with many empirical SPF measurements, which show similar behavior between different wavelength regimes (e.g., HR 4796A, see \citealt{chen2020}), and multi-wavelength measurements of SPFs are likely also needed in order to better constrain the grain properties of debris disk systems.

The limited number of debris disk SPF measurements (e.g. \citealt{graham2007,duchene2020}) have been shown to follow trends in shape seen in solar system dust environments \citep{hughes2018}. This generic SPF shape has been reproduced with highly porous grains where the minimum grain size is greater than 1$\mu m$ assuming both Mie theory \citep{grynko2004} and clusters of complex aggregates \citep{bentley2016} suggesting that debris disk dust shares many similarities to cometary dust and that overall dust composition is not as important of a factor. Our generic SPF utilizes measurements from various comets and planetary rings, but other approaches in generating a generic SPF from solar system dust have also been demonstrated \citep[e.g.,][]{schleicher2011,marcus2007a,marcus2007b}. Additionally, laboratory measurements and simulated dust properties (e.g. \citealt{lolachi2023}) could also inform the general shape of a generic SPF, and future studies of debris disk SPFs could incorporate these results in their analysis.

The achievement of low residual model fits for most of the debris disk systems is a result that supports findings and conclusions from \cite{hughes2018} and previous studies that a common SPF could exist between different dust systems with only modest variations. We have demonstrated that we can achieve low residual debris disk models of the HD 35841, HD 106906, HD 110058, HD 111520, and HD 114082 systems, utilizing the same SPF informed from solar system dust. Although we did not attain a low-residual best-fit model of HD 32297, the measurement of the SPF in \cite{duchene2020} is still consistent with our generic SPF. These commonalities also support predictions that highly porous, aggregate grains may be a major contributing factor to the shape of an SPF. The robustness of a majority of our results demonstrates our methods as a uniform approach to constraining disk geometry without having to parameterize grain properties or utilizing physically limiting assumptions (e.g. ellipse fitting). 

Despite these findings, we have identified one system where we could not achieve low residual model fits using the generic SPF. HR 4796A is known to harbor a distinct SPF shape as measured in \cite{milli2017}, with relatively higher scattering at the ansae and back side of the disk than what the generic SPF allows. Modeling of the HR 4796A scattering phase function still suggests that porous aggregates are likely the types of grains inducing scattering, but at a larger minimum grain size ($>5\mu$m) than predicted for other dust systems that follow the generic SPF \citep{milli2017,chen2020}. Although our approach with the generic SPF did not achieve low residual models to HR 4796A, the analysis was still informative in supporting that not all debris disk systems have SPFs similar to the generic SPF, and that a low residual model solution cannot be achieved if the scattering properties are treated incorrectly. Additional model tests using the HR 4796A SPF for HD 114082 further supports that SPF choice is an important consideration, as the best-fit model solution preferred model solutions where the front side of the disk was on the Northern side, notably inconsistent with other studies of these systems and our own disk analysis using our generic SPF. 
This conclusion, however, is founded upon using two very different SPFs, and does not account for more subtle variations, which we leave for a future study. Although we could not achieve a low residual model fit with the generic SPF, use of the measured SPF from \cite{milli2017} achieved significantly lower residual model fits to HR 4796A, further supporting the utility of this approach in constraining morphological properties with fixed scattering properties. To further assess the sensitivity of our model fits to the choice in SPF, we demonstrated that using the HR 4796A SPF for a system fit well with the generic SPF led to poorer model fits and inconsistent morphological properties.

\section{Summary} \label{sec:conclusions}
With advances in direct imaging instrumentation, more scattered-light debris disks have been resolved at small separations than ever before. With such a large sample of imaged debris disks, consistent and uniform characterization approaches should be implemented in order to make more direct comparisons between systems. We have performed a uniform forward-modeling analysis of eight bright debris disks imaged with the Gemini Planet Imager in total intensity light. Scattering properties were determined by empirically-informed scattering phase functions and not from parameterizing Henyey-Greenstein functions or specific grain properties. From our results, we were able to identify two families of debris disks: one where the midplane density profile can be described by one power law (disks with extremely sharp inner edges) and the other where the midplane density profile can be defined by two power laws (smoother declines of dust interior and exterior to peak density of the ring). Even though we find consistent results among many prior studies of these systems, inconsistencies shed light on the sensitivity of results to approaches of data reduction and modeling.

In applying the same empirically-informed scattering phase functions used in \textit{H}-band for \textit{K1}-band images of the same system, we still achieve consistent modeling results. The range of aspect ratios we find in our modeling analyses are mostly consistent with a "natural" disk scale height associated with radiation pressure, with the exception of HR 4796A and HD 110058. The thin aspect ratio of HR 4796A is inconsistent with prior results and may be related to issues in PSF subtraction and forward modeling. The broader aspect ratio of HD 110058 along with its narrow radial width and structural warps may be indicative of complex stirring mechanisms including perturbing planetary companions. We also identify asymmetric disk systems (HD 106906, HD 110058, HD 111520, HR 4796A) but cannot fully characterize their asymmetry due to the physically-limiting assumptions in our models. The range of eccentricities compared to relative radial width further highlights the diversity of systems and potential stirring mechanisms that cause divisions in relative width and aspect ratio.

We have demonstrated that rigorous morphological modeling can be conducted with a uniform and consistent set of assumptions about grain properties. 
While it is unlikely that most disks have the same dust populations, the scale-invariant nature of highly porous aggregates may cause SPFs of different dust environments to appear similar, regardless of inherent grain properties such as grain size distribution or composition. The SPFs from aggregate grain models also appear consistent with the generic SPF we determined from Solar System dust measurements. The achievment of well-fitting models in most cases suggest links between porous aggregates, Solar System cometary dust, and extrasolar debris disks. In systems where low residual model fits were not readily found (e.g. HR 4796A), it is likely that the scattering phase function applied was not representative of the system, or another factor (e.g., KLIP nonlinearity for HD 32297) may introduce biases into the modeling approach.

To further improve the robustness of this model, more observations at higher angular resolution are needed. Additionally, other PSF subtraction algorithms should be explored to avoid complications of KLIP nonlinearity and self-subtraction in bright disks. Future observations with next-generation ELTs (e.g. TMT, GMT, E-ELT) may prove critical in resolving finer structures that may be present within these systems at greater SNR. Additionally, further studies in dust grain dynamics may improve the robustness of our treatment of the dust density distribution. Radiative-transfer models in turn will also require more complexity accounting for asymmetric features such as eccentricity and warps. Finally, more empirical measurements of debris disk scattering phase functions are needed in order to support and identify common trends and properties of grains in a variety of different environments.

\section*{Acknowledgements}
The authors acknowledge Research Computing at Arizona State University for providing the computational HPC resources from the Sol supercomputer \citep{jennewein2023} that have contributed to the research results reported within this paper. We would like to acknowledge Jason Yalim from the Research Computing Core Facilities at Arizona State University for their assistance in optimizing HPC usage. The authors would also like to acknowledge the ASU Research Computing team as a whole for their assistance in setting up and assisting in all other matters related to high power computing, along with Dr. Michael Line, Dr. Aishwarya Iyer, and Peter Smith for their perspectives and conversations related to MCMC sampling and forward modeling practices. The authors also acknowledge Julien Milli for the measured HR 4796A SPF utilized in this analysis. Finally, the authors would like to thank and acknowledge the anonymous referee for their feedback and helpful suggestions.

This work is based on observations obtained at the Gemini Observatory, which is operated by the Association of Universities for Research in Astronomy, Inc., under a cooperative agreement with the NSF on behalf of the Gemini partnership: the National Science Foundation (United States), the National Research Council (Canada), CONICYT (Chile), Ministerio de Ciencia, Tecnolog\'ia e Innovaci\'on Productiva (Argentina), and Minist\'erio da Ci\^encia, Tecnologia e Inova\c c\~ao (Brazil). This work has made use of data from the European Space Agency (ESA) mission {\it Gaia} (\url{https://www.cosmos.esa.int/gaia}), processed by the {\it Gaia} Data Processing and Analysis Consortium (DPAC, \url{https://www.cosmos.esa.int/web/gaia/dpac/consortium}). Funding for the DPAC has been provided by national institutions, in particular the institutions participating in the {\it Gaia} Multilateral Agreement. This research has made use of the SIMBAD and VizieR databases, operated at CDS, Strasbourg, France.

J.H.'s work was supported by the Space Telescope Science Institute's Director's Discretionary Fund and the Arizona State University Graduate Completion Fellowship. A.M.H.'s work is supported by the National Science Foundation under Grant No. AST-2307920. E.L.N. and F.M. are funded through the NASA 80NSSC17K0535.

%%%%%%%%%%%%%%%%%%%%%%%%%%%%%%%%%%%%%%%%%%%%%%%%%%
\section*{Data Availability}
The raw data files for all observational sequences are available through the Gemini archive (\url{https://archive.gemini.edu/searchform}). The reduced data files, modeling data files, and other materials underlying this article will be shared on reasonable request to the corresponding author. The generic SPF used for this model analysis and the updated Python scripts for \texttt{DiskFM} used in this analysis will be available through \cite{hom_diskfm_2024} and Github\footnote{https://github.com/jrhom1/DiskFMforGPIES}. Inquiries regarding the scattering phase function from \cite{milli2017} should be directed to Julien Milli.
 
%%%%%%%%%%%%%%%%%%%% REFERENCES %%%%%%%%%%%%%%%%%%

% The best way to enter references is to use BibTeX:

\bibliographystyle{mnras}
\bibliography{gpiLLP} % if your bibtex file is called example.bib

%%%%%%%%%%%%%%%%%%%%%%%%%%%%%%%%%%%%%%%%%%%%%%%%%%

%%%%%%%%%%%%%%%%% APPENDICES %%%%%%%%%%%%%%%%%%%%%

\appendix

\section{Posterior Distributions of Model Runs} \label{sec:posteriors}
Here we present the posterior distribution functions of our model run of HD 32297 ($H$-band), generated with \texttt{corner.py} \citep{foreman-mackey2017}. All dashed lines in the diagonal histograms show the 16th, 50th, and 84th percentiles. Off-diagonal plots display joint probability distributions with contour levels at the same percentiles. All other posterior distribution functions are available online as supplemental material.
\begin{figure*}
    \centering
    \includegraphics[width=\linewidth]{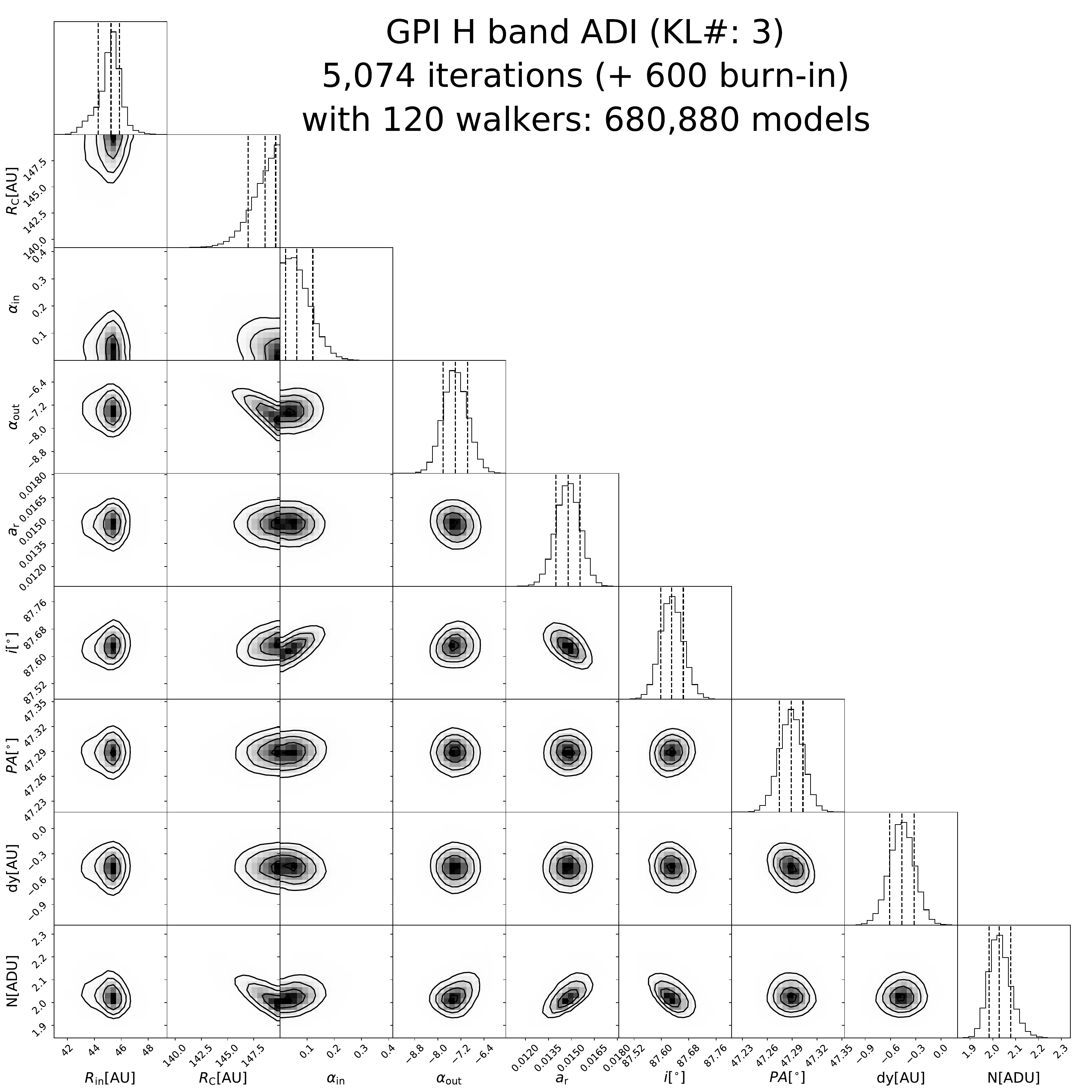}
    \caption{Posterior distribution for HD 32297 \textit{H}-band models.}
    \label{fig:hd32297post}
\end{figure*}

\section{Comparisons to Previous Scattered-Light Resolved Measurements and Model Estimations} \label{sec:pastcomparisons}
Here we compare our constrained dust density distribution parameters with properties found in previous studies. As the best model fits in both $H$- and $K1$-bands do not model the data sufficiently, we do not discuss constrained parameters for HD 32297. Additionally, constrained parameters for the HR 4796A disk analysis using our generic SPF and the HD 114082 disk analysis using the \cite{milli2017} SPF are not discussed as they provided poor best model fits. For many comparisons, caution must be given in comparing some dust density distribution properties such as radii from different approaches, as they are not always directly comparable. To best assess the comparisons to previous studies, we opt to compare midplane dust density profiles when available.

\subsection{HD 35841}
\begin{figure}
    \centering
    \includegraphics[width=\linewidth]{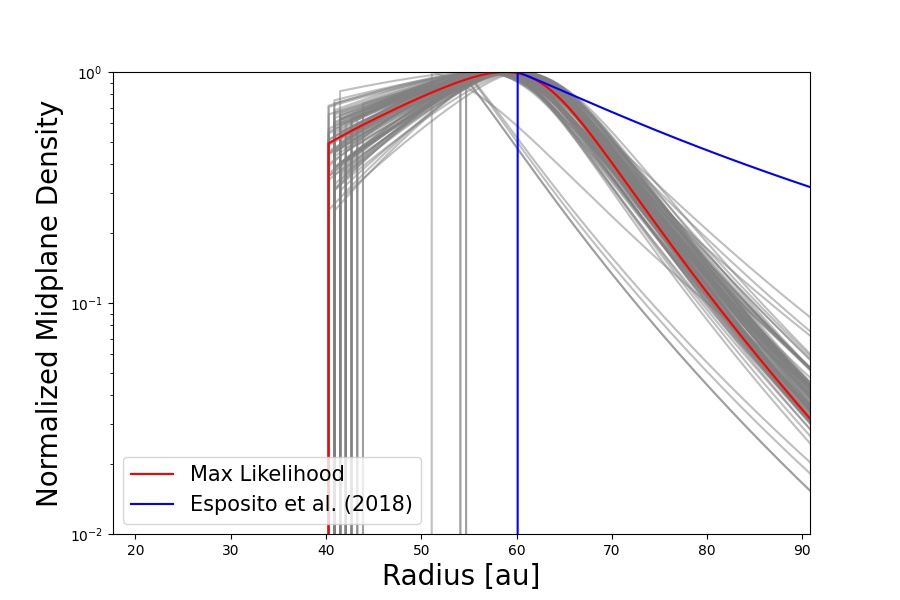}
    \caption{Normalized midplane dust density profiles from 100 randomly-selected models from our posterior distribution (gray) of HD 35841 models, the maximum likelihood of our models (red), and the maximum likelihood model from \citet{esposito2018} (blue). The range along the x-axis is restricted to within our likelihood calculation mask along the disk major axis for consistent comparison. Overall, our analysis prefers a narrow ring, two-power law solution for midplane density, where the analysis from \citet{esposito2018} prefers a single power-law solution and a relatively broader outer profile. Notably, our models prefer solutions that place dust interior to the inner edge of the profile from \citet{esposito2018}.
    }
    \label{fig:35841RAD_DENS}
\end{figure}

Overall, our best-fit model matches the data fairly well, with residual structure appearing mostly from background noise sources. Although we only find an upper limit to $R_{\rm in}$ and $\alpha_{\rm out}$, the remaining parameters are well-constrained. From our analysis, the system appears to be a symmetric ring, well-explained with a two-power law radial dust density distribution. \cite{esposito2018} was the first study to resolve and analyze this system, characterizing the disk from GPI \textit{H}-band total and polarized intensity data. Table \ref{tab:35841compare} presents comparisons of parameters determined from our analysis to \cite{esposito2018}. Overall, we find good agreement in estimations of $a_r$ within 1$\sigma$, and similar estimations for the radius of the ring. Figure \ref{fig:35841RAD_DENS} compares the radial dust midplane density profiles of 100 randomly-selected models from our posterior distribution to the median likelihood model in \cite{esposito2018}. Our analysis prefers a narrow ring with a dust density profile modeled with two power laws, while the analysis from \cite{esposito2018} appears to prefer a somewhat broader density profile modeled with a single power law. The discrepancies between our results and \cite{esposito2018} may be due to the differences in modeling approaches applied. \cite{esposito2018} fit jointly to polarized and total intensity data, where the constraints placed on grain properties from the polarized intensity data can induce distinct brightness distributions between the two images. This in turn influences the morphological parameters of the model fitting that are not present in fitting total intensity data alone. 

\begin{table}
    \centering
    \caption{Comparison of our results to the previous sub-arcsecond resolution study of HD 35841 from \citet{esposito2018}.
    }
    \begin{tabular}{c|c|c}
    \hline
         & This work & \cite{esposito2018} \\ \hline
        Instrument & GPI & GPI\\
        Filter & \textit{H} & \textit{H}\\
        Mode & Tot. & Tot \& Pol\\
        Sampler & MCMC & MCMC \\
        Scattering & Generic & Mie \\
        $R_{\rm in}$ & $<55.97$ & $59.8\substack{+1.1 \\ -2.1}$\\
        $R_{\rm C}$ & $62.15\substack{+1.32 \\ -1.61}$ & $<57$\\
        $\alpha_{\rm in}$ & $2.28\substack{+1.18 \\ -0.79}$ & $>-1.6$ \\
        $\alpha_{\rm out}$ & $<-6.09$ & $-3.0\substack{+0.2 \\ -0.2}$ \\
        $a_r$ & $0.04\substack{+0.005 \\ -0.004}$ & $0.045\substack{+0.023 \\ -0.005}$\\
        $i$ & $82.69\substack{+0.30 \\ -0.28}$ & $84.9\pm0.2$\\
        $PA$ & $165.49\pm0.15$ & $165.8\substack{+0.1\\-0.2}$\\
        $dx$ & $1.94\substack{+1.11 \\ -1.09}$ & N/A\\
        $dy$ & $0.05\substack{+0.49 \\ -0.48}$ & N/A \\
        \hline
    \end{tabular}   
    \label{tab:35841compare}
\end{table}

\subsection{HD 106906}
\begin{figure}
    \centering
    \includegraphics[width=\linewidth]{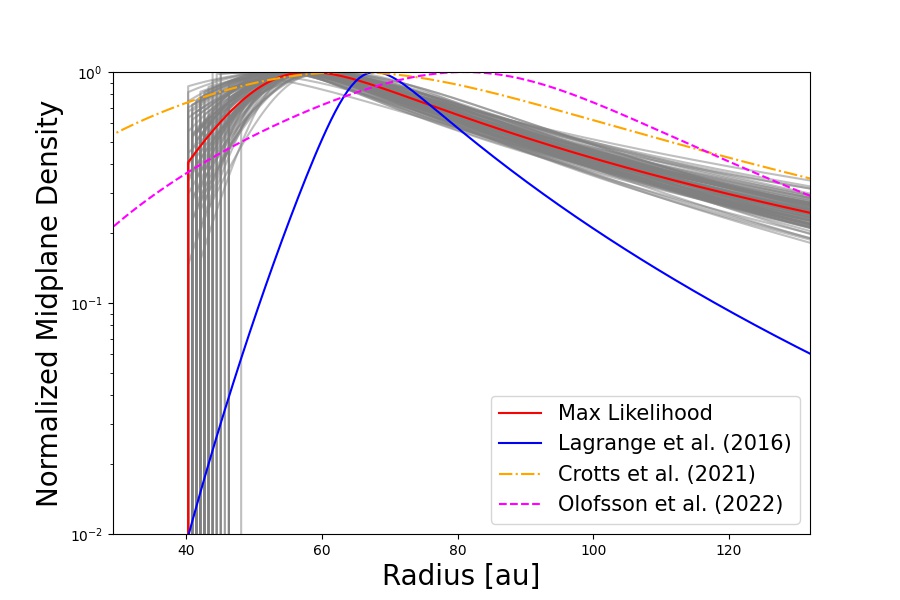}
    \caption{Same as Figure \ref{fig:35841RAD_DENS} but for 100 randomly-selected models from our posterior distribution (gray) of whole disk HD 106906 models, the maximum likelihood of our models (red), and the best fit models from \citet{lagrange2016} (blue), \citet{crotts2021} (orange), and \citet{olofsson2022} (magenta). Overall, our analysis prefers a broad-ring, single power-law, flatter inner profile, and shallow outer slope profile for midplane density, similar to \citet{crotts2021}.
    }
    \label{fig:106906RAD_DENS}
\end{figure}

Table \ref{tab:hd106906compare} presents a comparison of our constrained parameters for all modeling analyses to previous studies. In our modeling analysis of HD 106906, we find shallow radial density profiles in all modeling scenarios, except for the SE side where $\alpha_{\rm in}$ is unconstrained. Figure \ref{fig:106906RAD_DENS} compares the midplane density profiles of 100 randomly-selected models from our posterior distributions to midplane density profiles from previous studies. Our whole disk model profile is most consistent with modeling analysis from \cite{crotts2021}, highlighting a broad ring with shallow inner and outer slopes. Overall, the analysis from \cite{lagrange2016} suggests a narrow midplane density unlike what was determined in \cite{crotts2021}, \cite{olofsson2022}, and this work. This could be due to \cite{lagrange2016} fixing the inner power law index to a value of 10, forcing a narrower radial profile. Offsets are found to be significant enough to warrant our asymmetric disk modeling approach, and even larger offsets were also found in \cite{crotts2021}. In our analysis of the separate sides of HD 106906, morphological properties appear distinct from each other, with the SE model appearing to favor a larger radial extent and a much narrower aspect ratio, consistent with a needle-like asymmetry. This appears in contrast to observations of the outer halo of HD 106906 seen in \textit{HST} images of the system \citep{kalas2015}, although this may be unsurprising if a perturbing companion is influencing outer halo dust grains.

Our best-fit model for HD 106906 does not reproduce the brightness asymmetry observed in images despite containing large stellocentric offsets, likely due to the limited complexity of our model. The strong brightness asymmetry suggests that the disk could have high eccentricity and/or localized density enhancements/dearth. The best-fit models identified from analyzing the separate sides of the disk independently appear to fit the data more cleanly, but are inconsistent in $i$ and $PA$, potentially suggesting the presence of a warp and further emphasizing that the complex geometry of the system is not easily determined from a relatively simple disk model.

\begin{table*}
    \centering
    \caption{Comparison of our results to previous sub-arcsecond resolution studies of HD 106906. References: 1. \citet{lagrange2016}, 2. \citet{crotts2021}, 3. \citet{olofsson2022}}
    \begin{tabular}{c|c|c|c|c|c|c} 
    \hline
	&	This work	&	This work (NW)	&	This work (SE)	&	[1]	&	[2]	&	[3]	\\ \hline
Instrument	&	GPI	&	GPI	&	GPI	&	SPHERE	&	GPI	&	SPHERE	\\
Filter	&	\textit{H}	&	\textit{H}	&	\textit{H}	&	\textit{IRDIS-H2}	&	\textit{H}	&	\textit{IRDIS-H}	\\
Mode	&	Tot. I.	&	Tot. I.	&	Tot. I.	&	Tot. I.	&	Pol. I.	&	Pol. I.	\\
Sampler	&	MCMC	&	MCMC	&	MCMC	&	Grid	&	MCMC	&	Multinest	\\
Scattering	&	Generic	&	Generic	&	Generic	&	1-HG	&	Mie	&	see [4]	\\
$R_{\rm in}$	&	$<49.02$	&	$<43.24$	&	$55.21\substack{+1.48 \\ -5.04}$	&	N/A	&	N/A	&	N/A	\\
$R_{\rm C}$	&	$<75.88$	&	$67.84\substack{+7.95 \\ -8.59}$	&	$<74.60$	&	$66.0\pm1.8$	&	$72.21\substack{+3.10\\-3.29}$	&	$87.02\pm3.07$	\\
$\alpha_{\rm in}$	&	$3.28\substack{+2.23\\-1.66}$	&	$<4.36$	&	Unconst.	&	10 (fixed)	&	$1.03\substack{+0.29\\-0.24}$	&	$1.7\pm0.2$	\\
$\alpha_{\rm out}$	&	$-1.97\substack{+0.22 \\ -0.29}$	&	$-2.73\substack{+0.55 \\ -0.67}$	&	$-2.56\substack{+0.27 \\ -0.32}$	&	$-4.5\pm0.3$	&	$-2.26\substack{+0.10\\-0.08}$	&	$-3.7\pm0.2$	\\
$a_r$	&	$0.03\substack{+0.004 \\ -0.005}$	&	$0.06\pm0.005$	&	$<0.01$	&	0.008 (fixed)	&	$0.04\substack{+0.0004\\-0.0034}$	&	$0.047\pm0.006$	\\
$i$	&	$83.94\substack{+0.21 \\ -0.19}$	&	$83.07\pm0.22$	&	$85.21\pm0.11$	&	$85.4\pm0.1$	&	$85.57\substack{+0\farcs14\\-0\farcs15}^{a}$	&	$84.3\pm0.4$	\\
$PA$	&	$285.06\substack{+0.08 \\ -0.09}$	&	$287.28\pm0.14$	&	$283.32\pm0.09$	&	$284.4\pm0.3$	&	$284.31\substack{+0\farcs16\\-0\farcs30}^{a}$	&	$285.3\pm0.6$	\\
$dx$	&	$>4.16$ &	$<-1.92$	&	$<-1.40$ 	&	N/A	&	$-2.99\substack{+0.28\\-0.29}$	&	N/A	\\
$dy$	&	$-2.38\substack{+0.49\\-0.52}$	&	$>2.23$	&	$<-3.75$	&	N/A	&	$16.48\substack{+2.04\\-1.76}$	&	N/A	\\ \hline
    \end{tabular}  
    
    \vspace{2ex}
    {\raggedright $^a$These values were not determined as part of the dust density distribution model and were found via other techniques.\par}    
    \label{tab:hd106906compare}
\end{table*}

\subsection{HD 110058} \label{sec:hd110058compare}
HD 110058's resolved debris disk was first presented in \cite{kasper2015}. The most interesting features present in both SPHERE and GPI total intensity images of the system are symmetric warps present at about 0\farcs3 from the star oriented in opposite directions, creating an "S"-like shape. Similar to measurements in \cite{crotts2023}, our modeling analyses suggest vertically-broad profiles, which could be indicative of vertical stirring mechanisms such as perturbing planetary companions (see \S \ref{sec:vertextents}). Our estimates of modeling the whole disk achieve a similar result for $R_{\rm in}$ as $R_{\rm C}$ in \cite{kasper2015} and the measurement of $R_d$ from \cite{crotts2023}. In modeling the two sides of the disk separately, differences in $PA$ of 2$\degr$ between both analyses provide further evidence of the warps seen in images. The midplane density profiles from our models tend to prefer a narrow ring with a somewhat shallow outer slope. Regardless, our modeling results are tentative at best due to the loose constraints of the posterior distributions in all modeling scenarios of this system. The compact size of the disk and the observed ``S"-shape are expected to be the most significant limiting factors in our interpretation of the results, as our model has less information to calculate a likelihood function over and contains no formalism for producing warps. \cite{stasevic2023} provides quantitative constraints on the strength of the warps and suggest an inner perturbing planetary companion could be responsible, similar to the $\beta$ Pictoris system. \cite{lopez2023} determined that the strengths of the warps on each side of the disk are distinct, and that constraints on disk measurements are also difficult to achieve due to the compact nature of the system.

\subsection{HD 111520}
Our best-fit symmetric model solution for HD 111520 suggests a significant stellocentric offset, though it does not fully reproduce the strong brightness asymmetry present in the GPI images \citep{draper2016}. Thus, a truly eccentric geometry is likely needed to properly describe the disk. In modeling the two sides of the disk separately, an offset along the major axis could not be constrained in the SE model. The NW model constrained a nonzero offset, but with a wide posterior that overlaps with zero. \cite{crotts2022} did not identify a stellocentric offset from fitting the location of the disk spine, but a surface density asymmetry could produce the strong brightness asymmetry present. No previous studies have attempted to model radial dust density distribution properties, but we do find disk aspect ratios that are similar to other systems in our analysis, and radii roughly consistent with measurements from \cite{crotts2023}. Midplane density profiles from our posterior distributions support a narrow ring structure with shallow outer slopes. In their analysis of the GPI \textit{H}-spec total intensity and \textit{H}-pol polarized intensity images, \cite{draper2016} identified a nearly edge-on geometry from modeling the vertical offset profile but did not incorporate disk models. \cite{crotts2022} expanded this analysis by also investigating the GPI \textit{J}- and \textit{K1}-band total and polarized intensity images, applying an MCMC analysis with a simple inclined ring model. We find that our $i$ estimates for both our whole disk model and asymmetric model approaches are consistent with the estimations from \cite{crotts2022} to 3$\sigma$, while the $PA$ appears only inconsistent in analyzing the SE side of the disk. A warp does appear present along the SE side, also seen in polarized intensity images of the system in \cite{crotts2022}, but the reduced SNR on the SE side makes the existence of such a warp tentative.

\subsection{HD 114082}
\begin{figure}
    \centering
    \includegraphics[width=\linewidth]{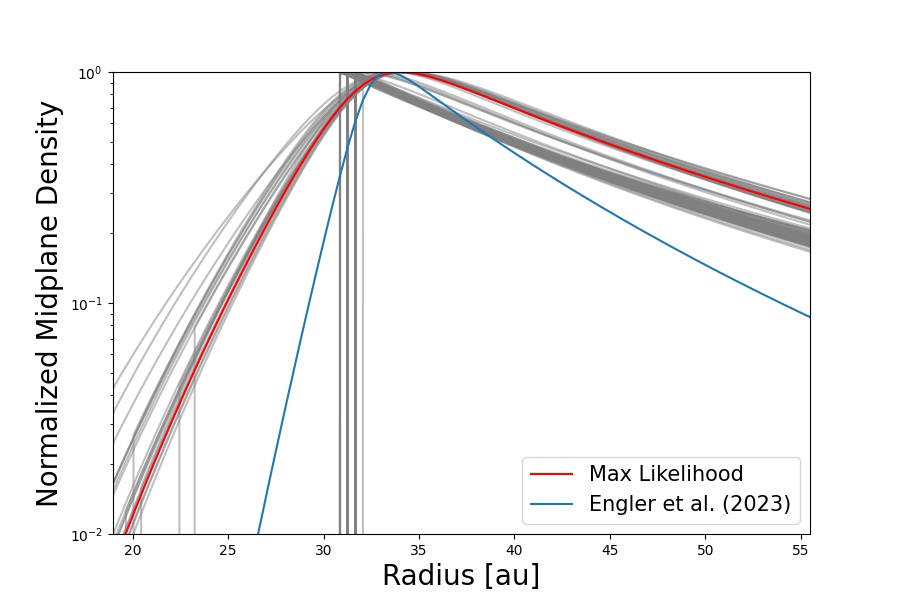}
    \caption{Same as Figure \ref{fig:35841RAD_DENS} but for 100 randomly-selected models from our posterior distribution (gray) of HD 114082 models, the maximum likelihood of our models (red), and the best fit model from \citet{engler2023} (blue). Both the analysis from \citet{engler2023} and our own work suggest steep inner truncations at $\sim$30 au, although \citet{engler2023} appears to prefer a shallower outer slope for midplane dust density.
    }
    \label{fig:114082RAD_DENS}
\end{figure}

We achieve well-constrained parameters and relatively low residual model fits for HD 114082. The median likelihood parameters suggest a compact, narrow-ring structure as observed in previous studies (e.g. \citealt{wahhaj2016,engler2023}), although the posterior distribution functions show two families of distinct profiles, one with a smaller peak radius and truncated slope (single power law-preferred) and one with a slightly higher peak radius and shallower inner slope (two power law-preferred). Figure \ref{fig:114082RAD_DENS} reveals that our model analyses preferred narrow-ring structures similar to the analysis of \cite{engler2023}, albeit with a shallower outer profile. HD 114082 is also one of a handful of debris disks with a measured SPF; \cite{engler2023} determined the measured SPF of HD 114082 to be similar in shape to other solar system dust environments and other debris disks, including zodiacal light and the Saturn D68 ring which our generic SPF is partially derived from. The aspect ratio appears much broader than what was found in \cite{engler2023}, with a vertical FWHM approximately twice as large. This discrepancy is likely related to the quality of data and differing model approaches. In the SPHERE IFS and IRDIS images of the system, the front and back sides of the ring are resolved, while GPI was only able to resolve the front side of the disk. The compact nature and inclination of the disk may also lead to degeneracies between radial and vertical density properties.

\subsection{HD 146897}
\begin{figure}
    \centering
    \includegraphics[width=\linewidth]{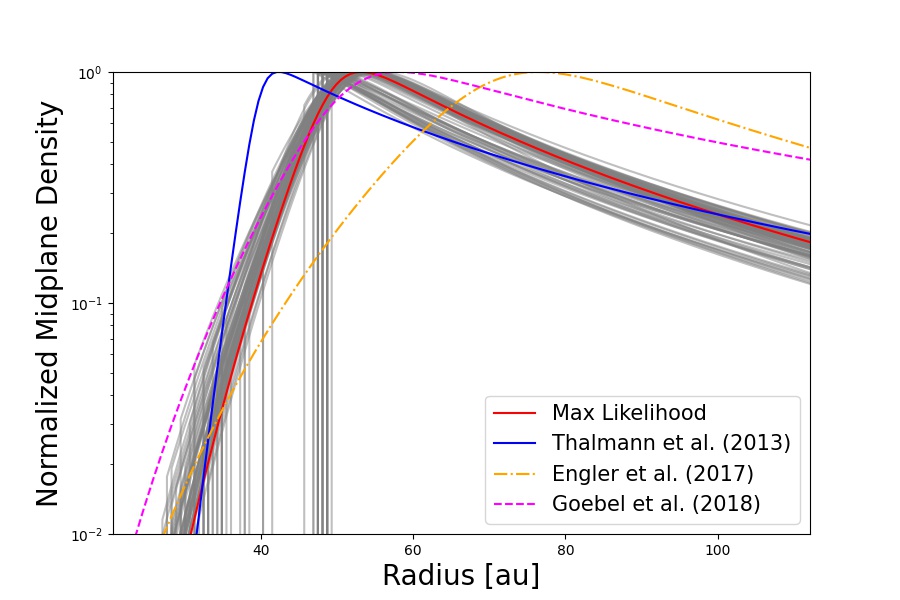}
    \caption{Same as Figure \ref{fig:35841RAD_DENS} but for 100 randomly-selected models from our posterior distribution (gray) of HD 146897 models, the maximum likelihood of our models (red), and the best fit models from \citet{thalmann2013} (blue), \citet{engler2017} (orange), and \citet{goebel2018} (magenta). Our midplane density profiles appear most similar to \citet{goebel2018} albeit with a steeper outer profile, while the slope of the outer density profile is consistent with \citet{engler2017}.
    }
    \label{fig:146897RAD_DENS}
\end{figure}

Table \ref{tab:146897compare} presents comparisons of our median likelihood model parameters to previous studies of the system. In comparing the midplane density profiles from our analysis to previous studies \citep{thalmann2013,engler2017,goebel2018} as in Figure \ref{fig:146897RAD_DENS}, we identify a broad range of profiles, suggesting that this system presents large uncertainties in general. Despite this broad range, there does appear to be a steep inner slope around 50 au seen in \cite{goebel2018} and this work. \cite{thalmann2013} also has a steep inner slope, although their inner radial density properties were fixed. The analysis from \cite{engler2017} contrasts from these other profiles, with a broader midplane density profile in general peaking at a larger radius. The compact nature of this system, overall noise levels near the FPM, and inability to resolve the back side of the disk despite the moderate inclination of the system can contribute to confusion among radial and vertical density properties. One explanation for the high residuals could be bright PSF halo features overlapping with the SW side of the disk biasing the likelihood calculation of the MCMC, resulting in a disk model that is unable to match the brightness of the front side of the disk without overestimating the brightness of the back side of the disk. Additionally, HD 146897 could have a distinct SPF similar to HR 4796A, inducing systematic residual structure between the front and back sides of the disk, although the compact nature of the system prevents a definitive conclusion on this behavior.

\begin{table*}
    \centering
    \caption{Comparison of our results to previous sub-arcsecond resolution studies of HD 146897. References: 1. \citet{thalmann2013}, 2. \citet{engler2017}, 3. \citet{goebel2018}}
    \begin{tabular}{c|c|c|c|c} \hline
	&	This work	&	[1]	&	[2]	&	[3]	\\ \hline
Instrument	&	GPI	&	HiCIAO	&	SPHERE	&	CHARIS	\\
Filter	&	\textit{H}	&	\textit{H}	&	\textit{ZIMPOL-I}	&$1.13-2.39\mu m$	\\
Mode	&	Tot. I.	&	Tot. I.	&	Pol. I.	&	Tot. I.	\\
Sampler	&	MCMC	&	Grid	&	Grid	&	Grid	\\
Scattering	&	Generic	&	1-HG	&	1-HG	&	1-HG	\\
$R_{\rm in}$	&	$<49.25$	&	N/A	&	N/A	&	N/A	\\
$R_{\rm C}$	&	$50.10\substack{+1.03 \\ -6.87}$	&	40 (fixed)	&	$73\pm8$	&	53	\\
$\alpha_{\rm in}$	&	$>0.75^a$	&	20 (fixed)	&	$5.0\pm1.4$	&	6	\\
$\alpha_{\rm out}$	&	$-2.45\substack{+0.12 \\ -0.13}$	&	-1.7	&	$-2.5\pm0.7$	&	-1.5	\\
$a_r$	&	$<0.04$	&	N/A	&	$0.03\pm0.01$	&	0.06	\\
$i$	&	$82.51\substack{+0.27 \\ -0.26}$	&	84	&	$84.6\pm0.85$	&	84.6 (fixed)	\\
$PA$	&	$115.00\pm0.17$	&	114	&	$114.5\pm0.6^{a}$	&	$114.59\pm0.4^{a}$	\\
%$dx$	&	$0.17\substack{+4.62 \\ -0.07}$	&	N/A	&	N/A	&	N/A	\\
$dy$	&	$0.69\substack{+0.35 \\ -0.34}$	&	3	&	N/A	&	N/A	\\ \hline
    \end{tabular}
    
    \vspace{2ex}
    {\raggedright $^a$These values were not determined as part of the dust density distribution model and were found via other techniques.\par}
    \label{tab:146897compare}
\end{table*}

\subsection{HR 4796A}
\begin{figure}
    \centering
    \includegraphics[width=\linewidth]{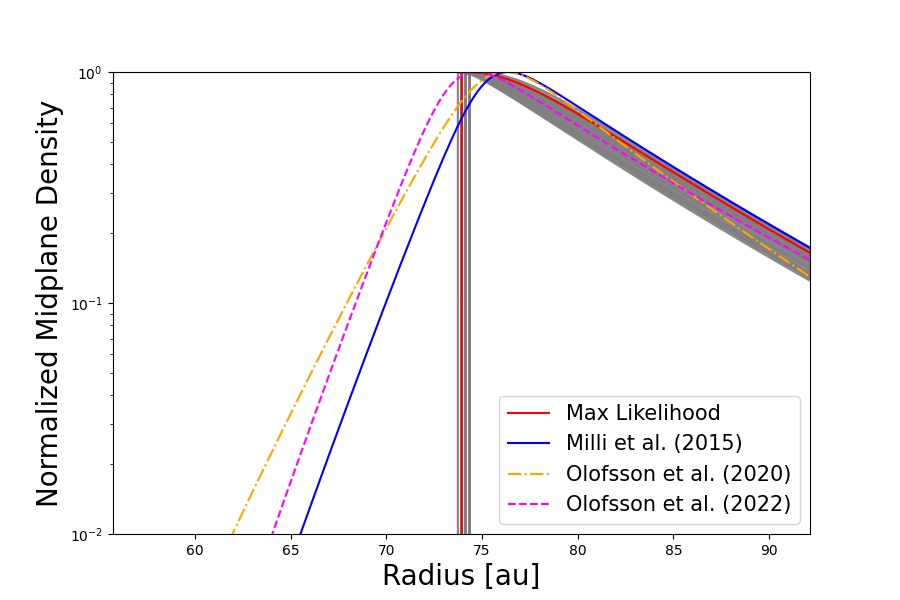}
    \caption{Same as Figure \ref{fig:35841RAD_DENS} but for 100 randomly-selected models from our posterior distribution (gray) of HR 4796A $H$-band models, the maximum likelihood of our models (red), and the best fit models from \citet{milli2015} (blue), \citet{olofsson2020} (orange), and \citet{olofsson2022} (magenta). Overall, our midplane density profiles appear consistent to all three studies, suggesting that the system is a narrow ring, although our models prefer a steeper inner cutoff.
    }
    \label{fig:4796RAD_DENS}
\end{figure}

Table \ref{tab:4796compare} presents comparisons of our median likelihood model parameters to previous studies of HR 4796A. Our best-fit model using the SPF from \cite{milli2017} fits the data reasonably well at both \textit{H}- and \textit{K1}-band, and we find consistent results among our estimations of radii and offsets to prior studies within 3$\sigma$ (e.g. \citealt{milli2015,olofsson2020,olofsson2022,crotts2023}). The posterior distributions of the radial power law indices suggest steep profiles, which is unsurprising given the sharply defined edges of the disk. Midplane density profiles from our analysis appear consistent to narrow ring width determined in previous studies, as seen in Figure \ref{fig:4796RAD_DENS}. The most significant inconsistency is seen in our determination of the aspect ratio, with our posterior distributions suggesting an unexpectedly thin ring compared to analyses by \cite{milli2017} and \cite{olofsson2022}. Unusually thin vertical structure was also identified in \cite{chen2020}, which analyzed the same datasets we have used. This could suggest systematic differences between the GPI and SPHERE analyses, such as the treatment of the instrumental PSF.

Another consideration is the apparent brightness asymmetry between the NE and SW sides of the front-side of the disk. By introducing a stellocentric offset to the ring, pericenter glow \citep{wyatt1999} may be induced. Our best-fit model appears unable to compensate for this brightness asymmetry, even after considering stellocentric offsets along both the major and minor axes of the disk. This result may support findings in \cite{olofsson2019}, \cite{milli2019}, and \cite{chen2020} that a dust density enhancement at pericenter may be responsible for the asymmetry.

Finally, the consistency of results and similarities in residual maps between \textit{H}- and \textit{K1}-band using the same measured SPF suggests that there is no difference in the SPF as a function of wavelength, despite evidence from other dust systems that more forward scattering occurs at redder wavelengths \citep{schroder2014}.

\begin{table*}
    \centering
    \caption{Comparison of our results to previous sub-arcsecond resolution studies of HR 4796A. References: 1. \citet{milli2015}, 2. \citet{olofsson2020}, 3. \citet{olofsson2022}}
    \begin{tabular}{c|c|c|c|c|c} \hline
	&	This work	&	This work	&	[1]	&	[2]	&	[3]	\\	\hline
Instrument	&	GPI	&	GPI	&	NACO	&	SPHERE	&	SPHERE	\\	
Filter	&	\textit{H}	&	\textit{K1}	&	\textit{Ks}	&	\textit{ZIMPOL-I}	&	\textit{ZIMPOL-I}	\\	
Mode	&	Tot. I.	&	Tot. I.	&	Pol. I.	&	Pol. I.	&	Pol. I.	\\	
Sampler	&	MCMC	&	MCMC	&	Grid	&	Multinest	&	Multinest	\\	
Scattering	&	Milli 2017	&	Milli 2017	&	1-HG	&	see [2]	&	see [2]	\\	
$R_{\rm in}$	&	$73.86\substack{+0.32 \\ -0.09}$	&	$74.43\pm0.10$	&	N/A	&	N/A	&	N/A	\\	
$R_{\rm C}$	&	$73.79\substack{+0.54 \\ -2.11}$	&	$<73.07$	&	$75.3\substack{+2.0 \\ -2.2}$	&	$75.44\pm0.07$	&	$73.6\pm1.42$	\\	
$\alpha_{\rm in}$	&	$>2.41$	&	$>1.98$	&	35 (fixed)	&	25 (fixed)	&	$35.0\pm0.1$	\\	
$\alpha_{\rm out}$	&	$<-9.81$	&	$<-9.78$	&	-10 (fixed)	&	$-11.78\pm0.20$	&	$-9.5\pm0.3$	\\	
$a_r$	&	$<0.004$	&	$<0.007$	&	0.013 (fixed)	&	$0.035\pm0.001$	&	$0.041\pm0.002$	\\	
$i$	&	$76.89\pm0.03$	&	$76.95\pm0.05$	&	$75.5\substack{+1.3 \\ -1.7}$	&	$77.60\pm0.06$	&	$77.7\pm0.2$	\\	
$PA$	&	$26.75\pm0.02$	&	$26.55\pm0.04$	&	$26.7\pm1.6$	&	27.9 (fixed)	&	$28.4\pm0.3$	\\	
$dx$	&	$-2.20\pm0.09$	&	$-2.41\pm0.16$ 	&	$-5.8\pm8.3$	&	N/A	&	N/A	\\	
$dy$	&	$0.84\pm0.05$	&	$1.79\pm0.08$	&	$0.8\substack{+3.5\\-3.0}$	&	N/A	&	N/A	\\	\hline
    \end{tabular}
    \label{tab:4796compare}
\end{table*}

\begin{figure*}
    \centering
    \includegraphics[width=\linewidth]{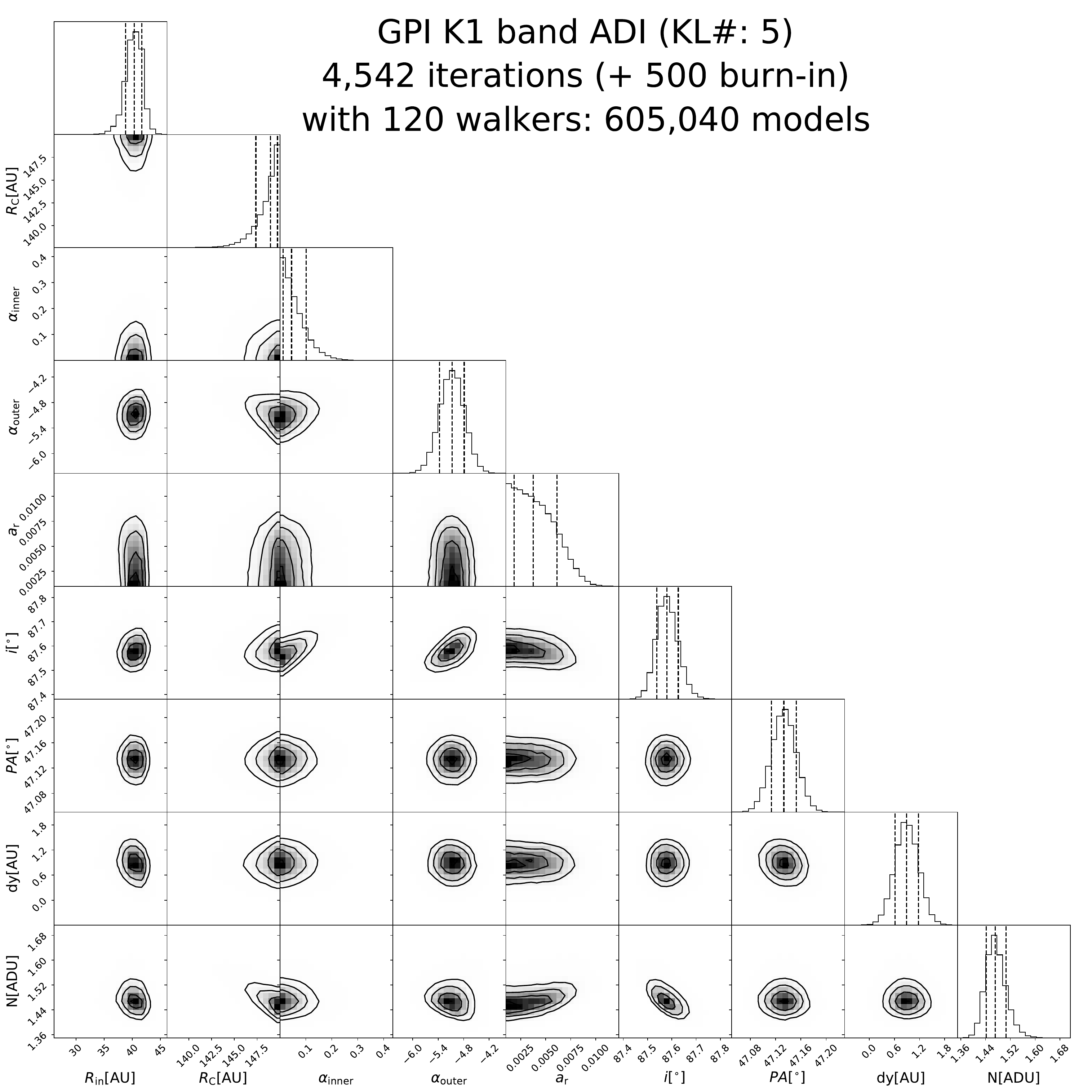}
    \caption{Posterior distribution for HD 32297 \textit{K1}-band models.}
    \label{fig:hd32297K1post}
\end{figure*}
\begin{figure*}
    \centering
    \includegraphics[width=\linewidth]{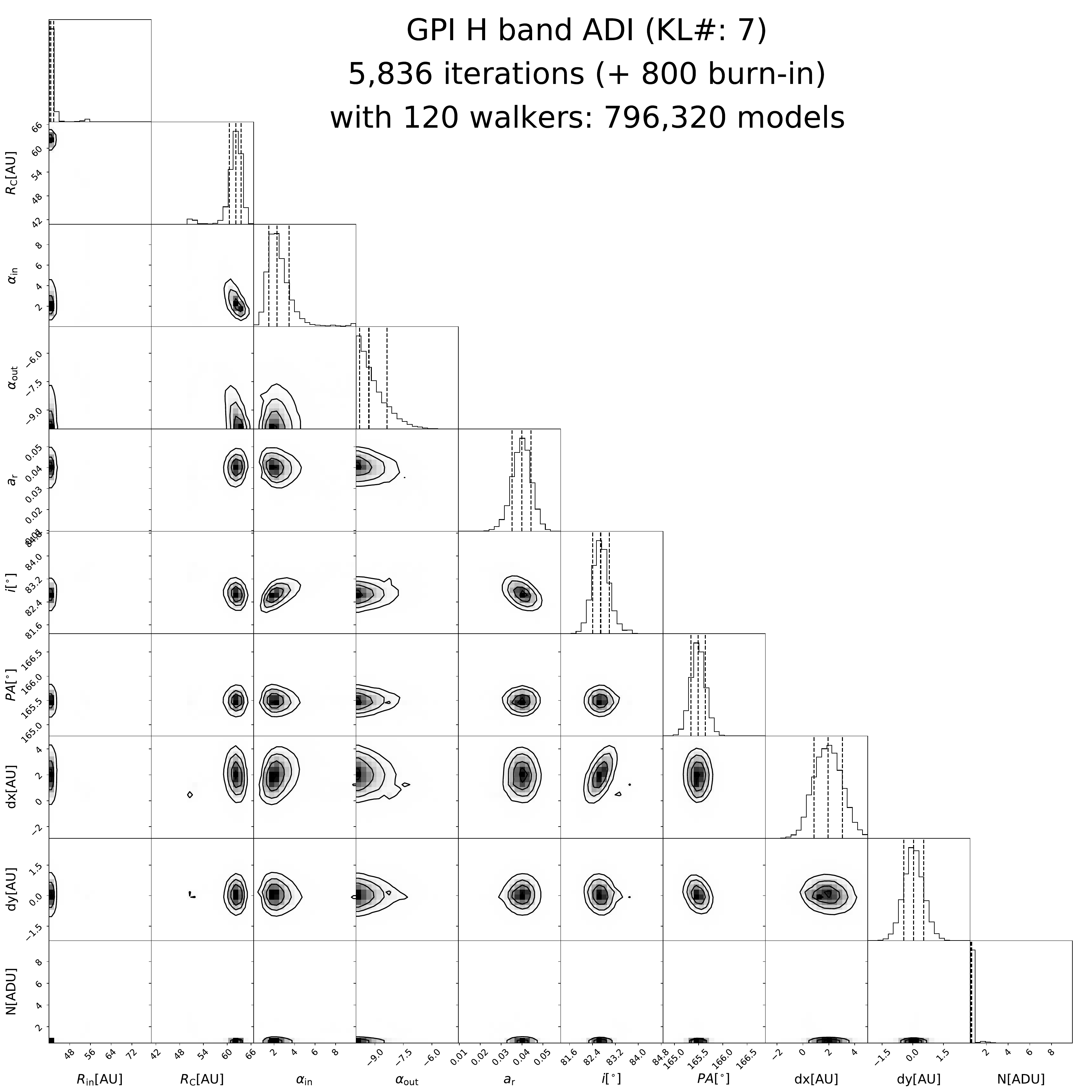}
    \caption{Posterior distribution for HD 35841 \textit{H}-band models.}
    \label{fig:hd35841post}
\end{figure*}
\begin{figure*}
    \centering
    \includegraphics[width=\linewidth]{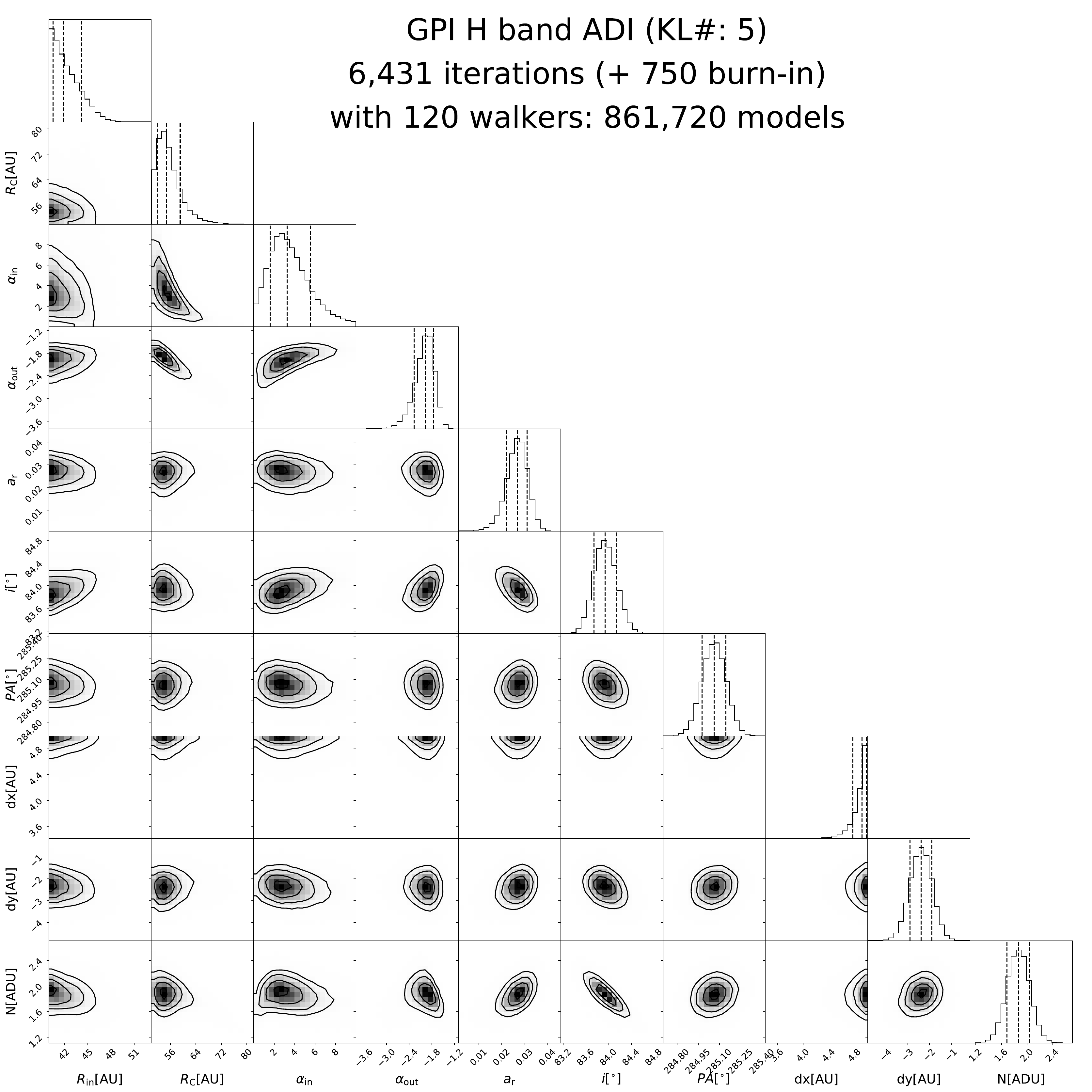}
    \caption{Posterior distribution for HD 106906 whole disk \textit{H}-band models.}
    \label{fig:hd106906post}
\end{figure*}
\begin{figure*}
    \centering
    \includegraphics[width=\linewidth]{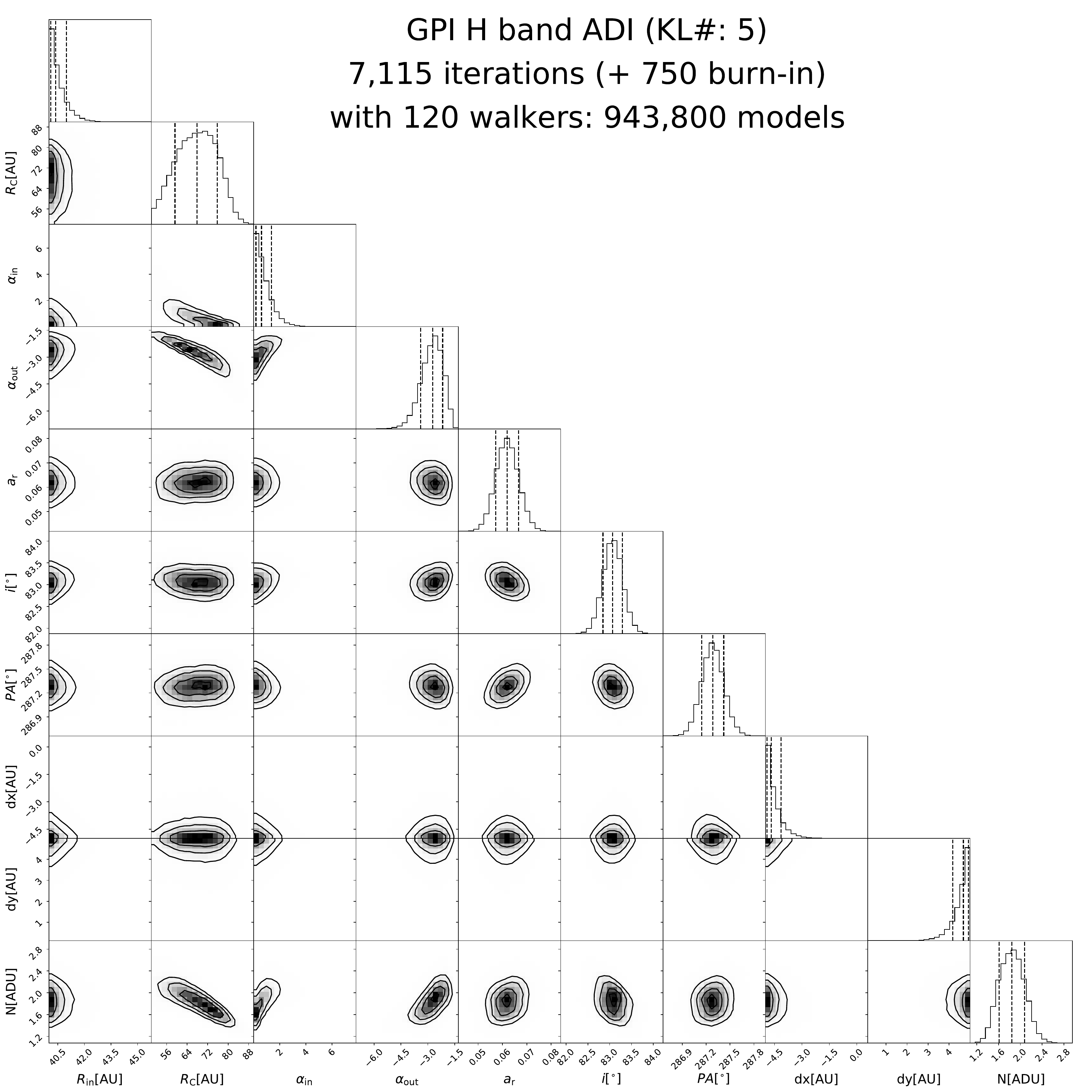}
    \caption{Posterior distribution for HD 106906 NW \textit{H}-band models.}
    \label{fig:hd106906NWpost}
\end{figure*}
\begin{figure*}
    \centering
    \includegraphics[width=\linewidth]{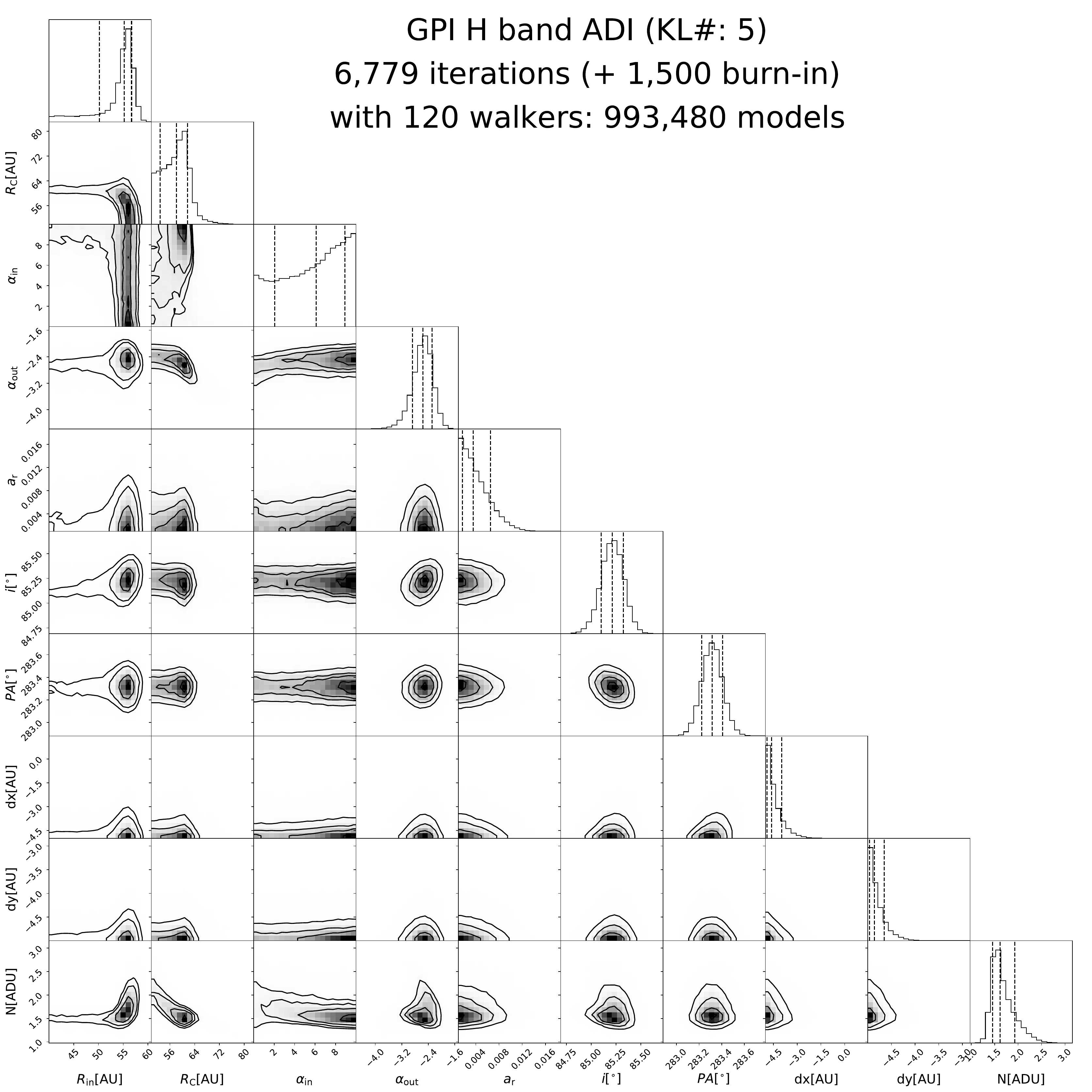}
    \caption{Posterior distribution for HD 106906 SE \textit{H}-band models.}
    \label{fig:hd106906SEpost}
\end{figure*}
\begin{figure*}
    \centering
    \includegraphics[width=\linewidth]{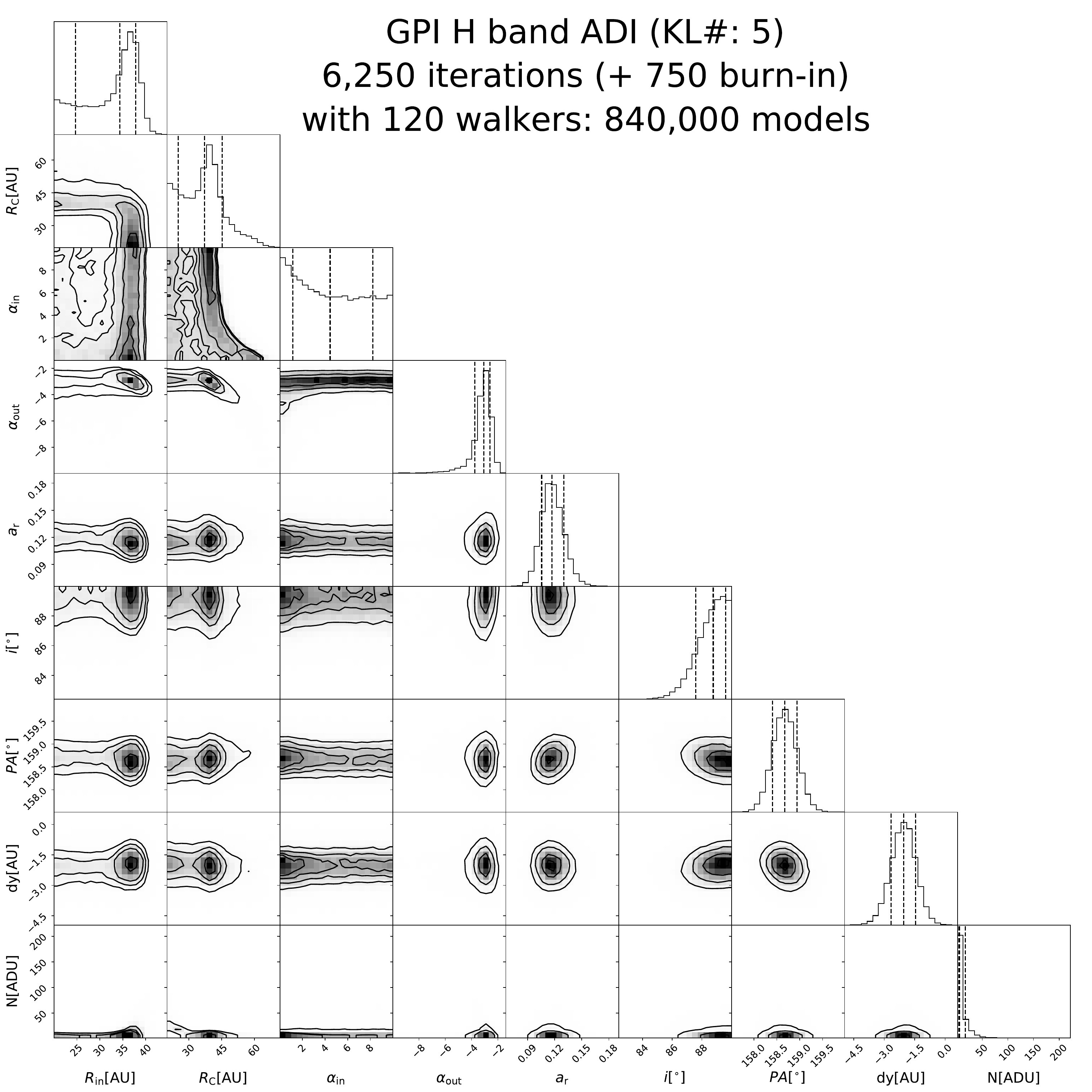}
    \caption{Posterior distribution for HD 110058 whole disk \textit{H}-band models.}
    \label{fig:hd110058post}
\end{figure*}
\begin{figure*}
    \centering
    \includegraphics[width=\linewidth]{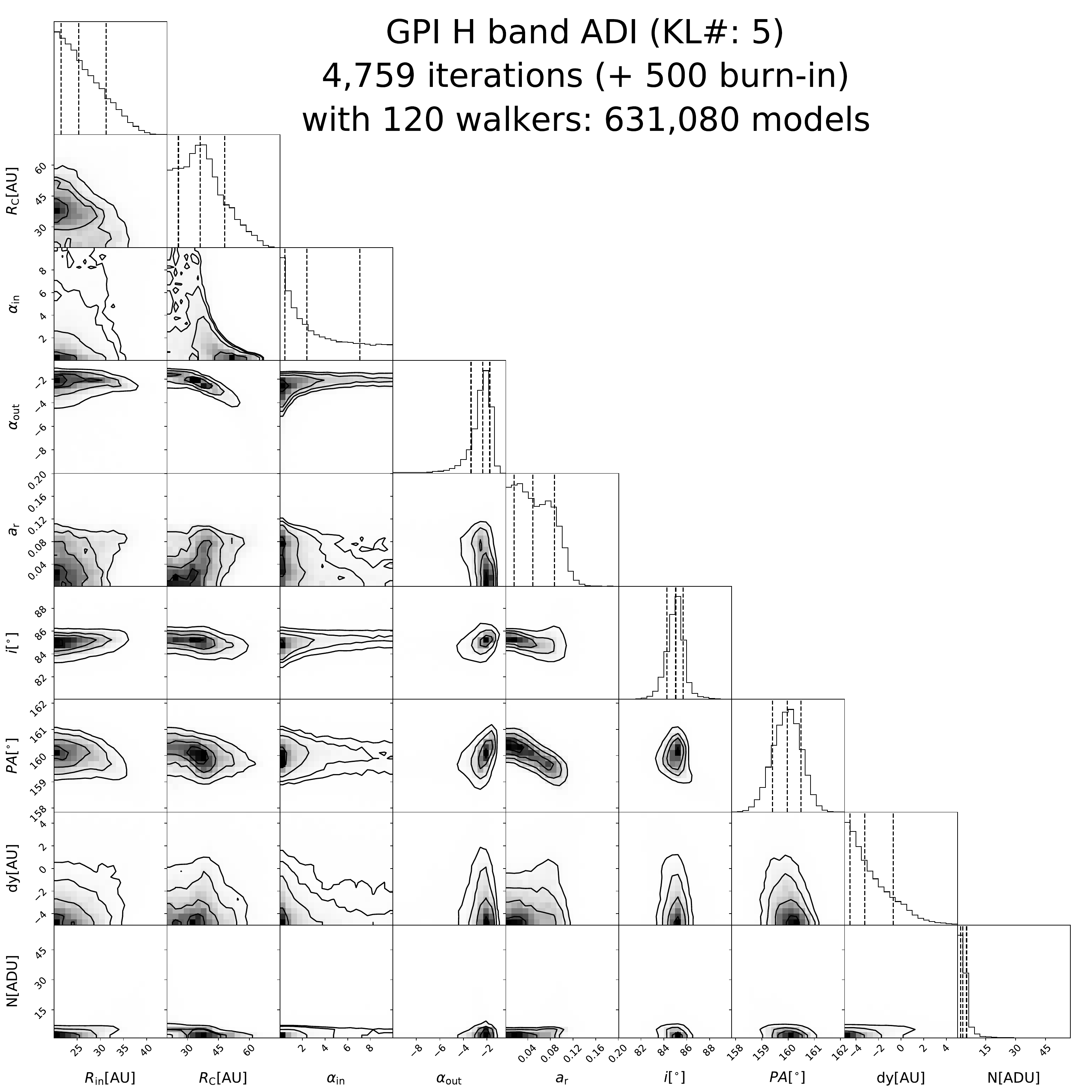}
    \caption{Posterior distribution for HD 110058 NW \textit{H}-band models.}
    \label{fig:hd110058NWpost}
\end{figure*}
\begin{figure*}
    \centering
    \includegraphics[width=\linewidth]{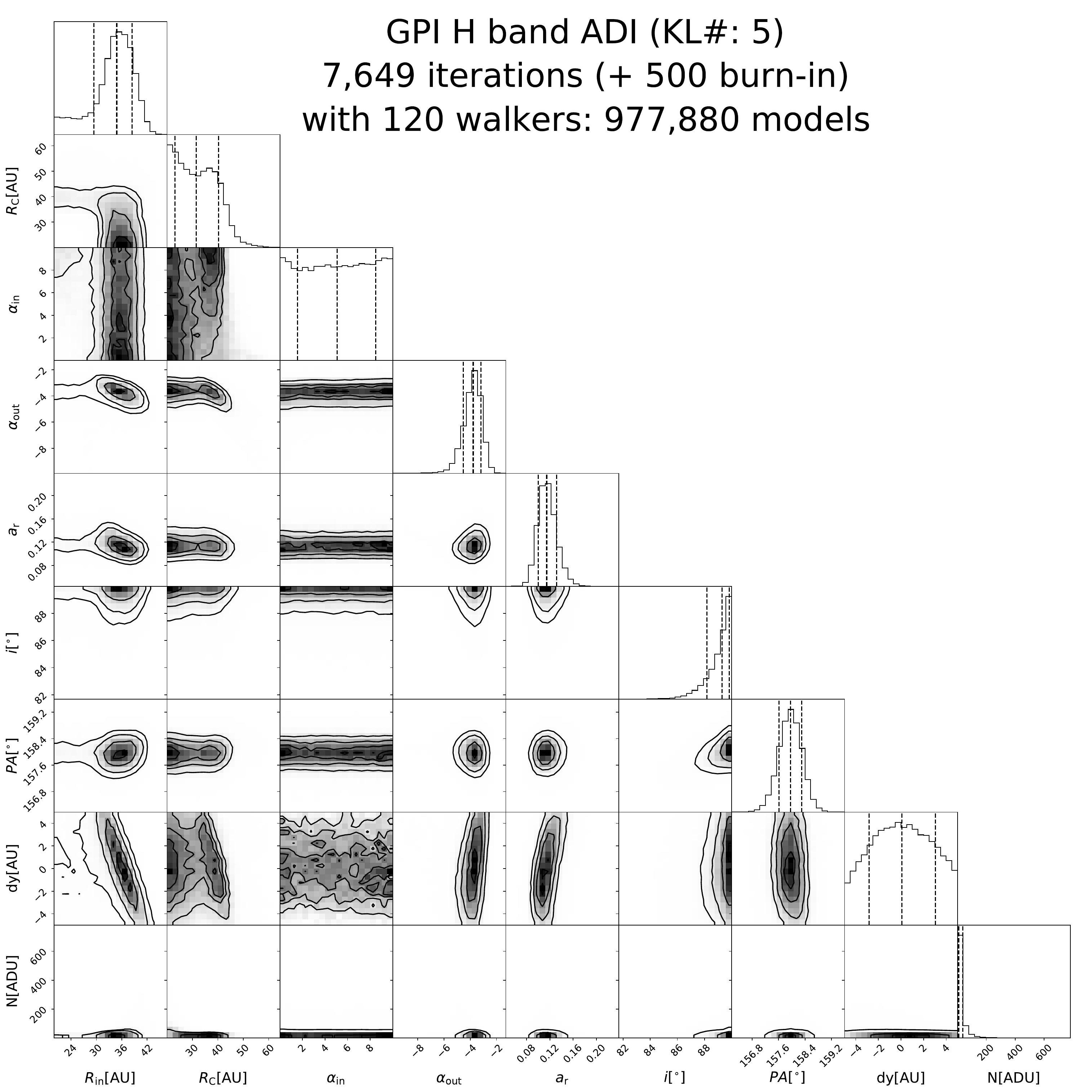}
    \caption{Posterior distribution for HD 110058 SE \textit{H}-band models.}
    \label{fig:hd110058SEpost}
\end{figure*}
\begin{figure*}
    \centering
    \includegraphics[width=\linewidth]{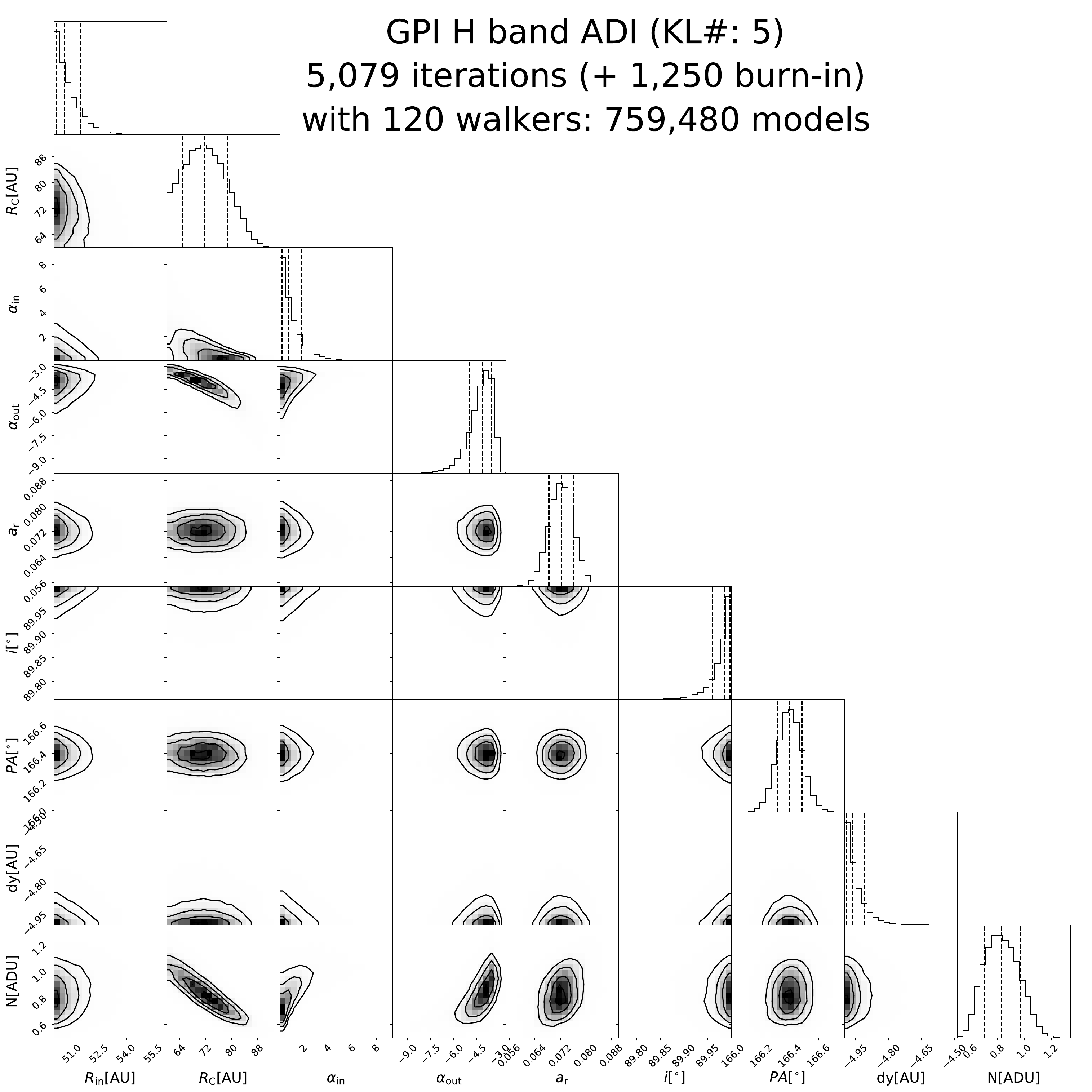}
    \caption{Posterior distribution for HD 111520 whole disk \textit{H}-band models.}
    \label{fig:hd111520post}
\end{figure*}
\begin{figure*}
    \centering
    \includegraphics[width=\linewidth]{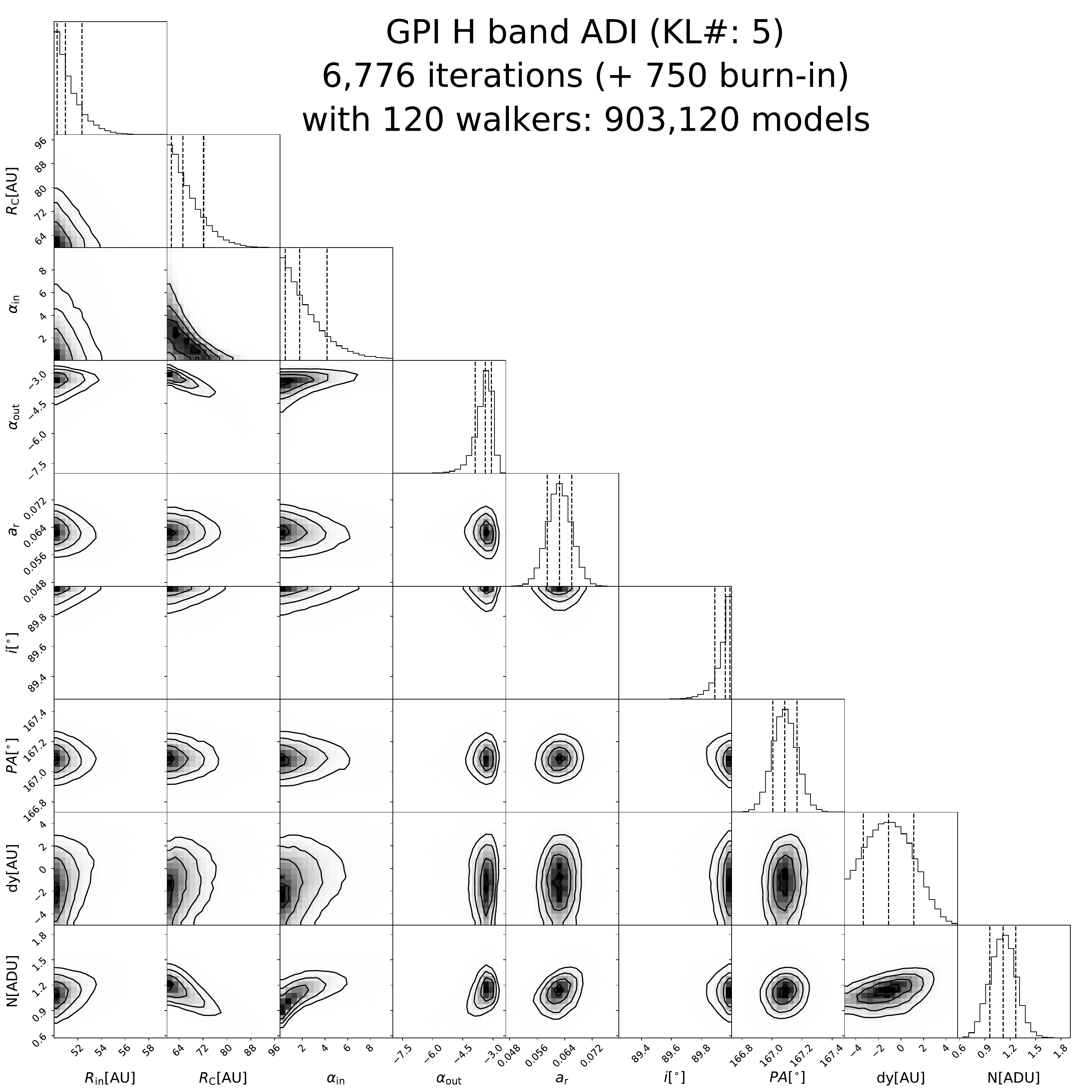}
    \caption{Posterior distribution for HD 111520 NW \textit{H}-band models.}
    \label{fig:hd111520NWpost}
\end{figure*}
\begin{figure*}
    \centering
    \includegraphics[width=\linewidth]{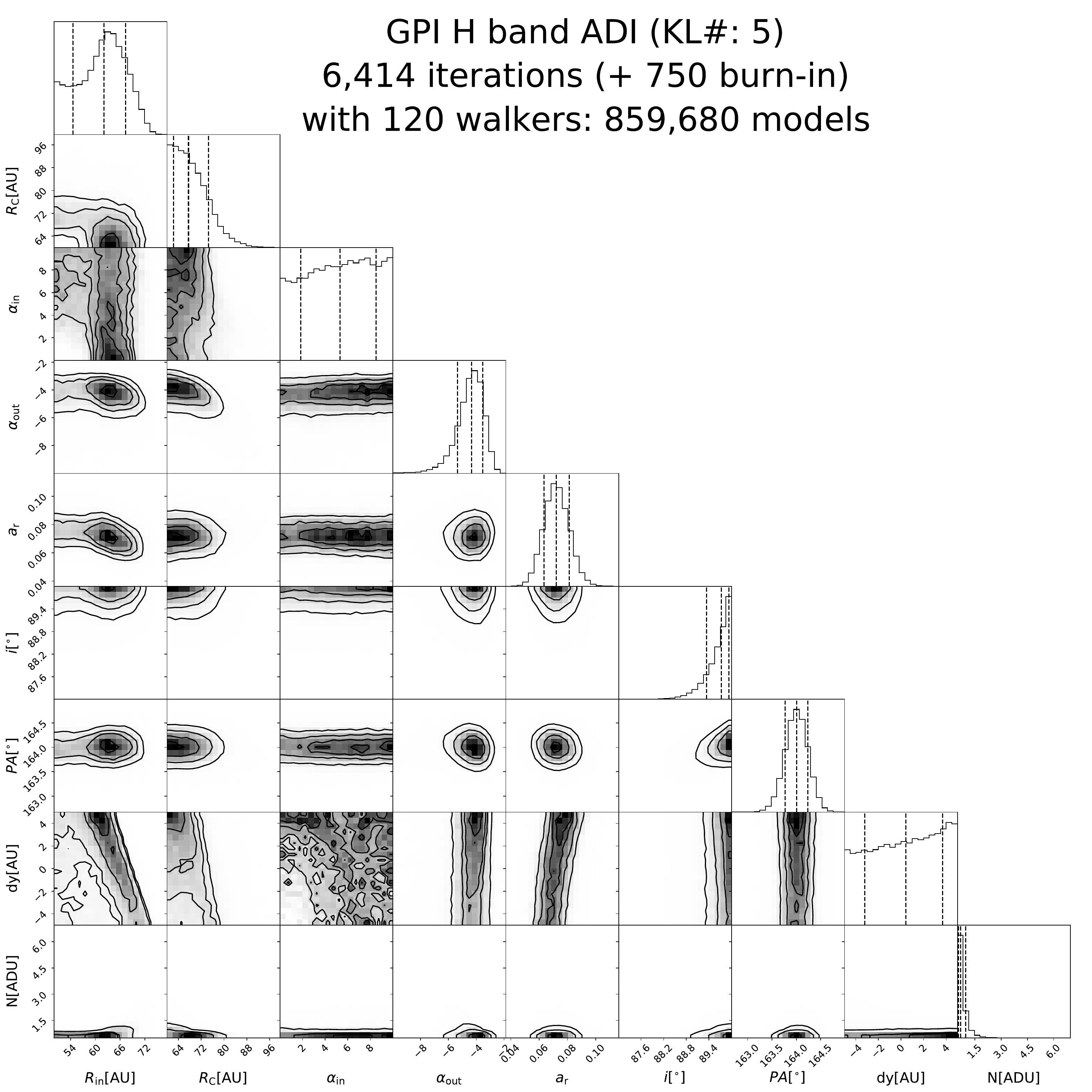}
    \caption{Posterior distribution for HD 111520 SE \textit{H}-band models.}
    \label{fig:hd111520SEpost}
\end{figure*}
\begin{figure*}
    \centering
    \includegraphics[width=\linewidth]{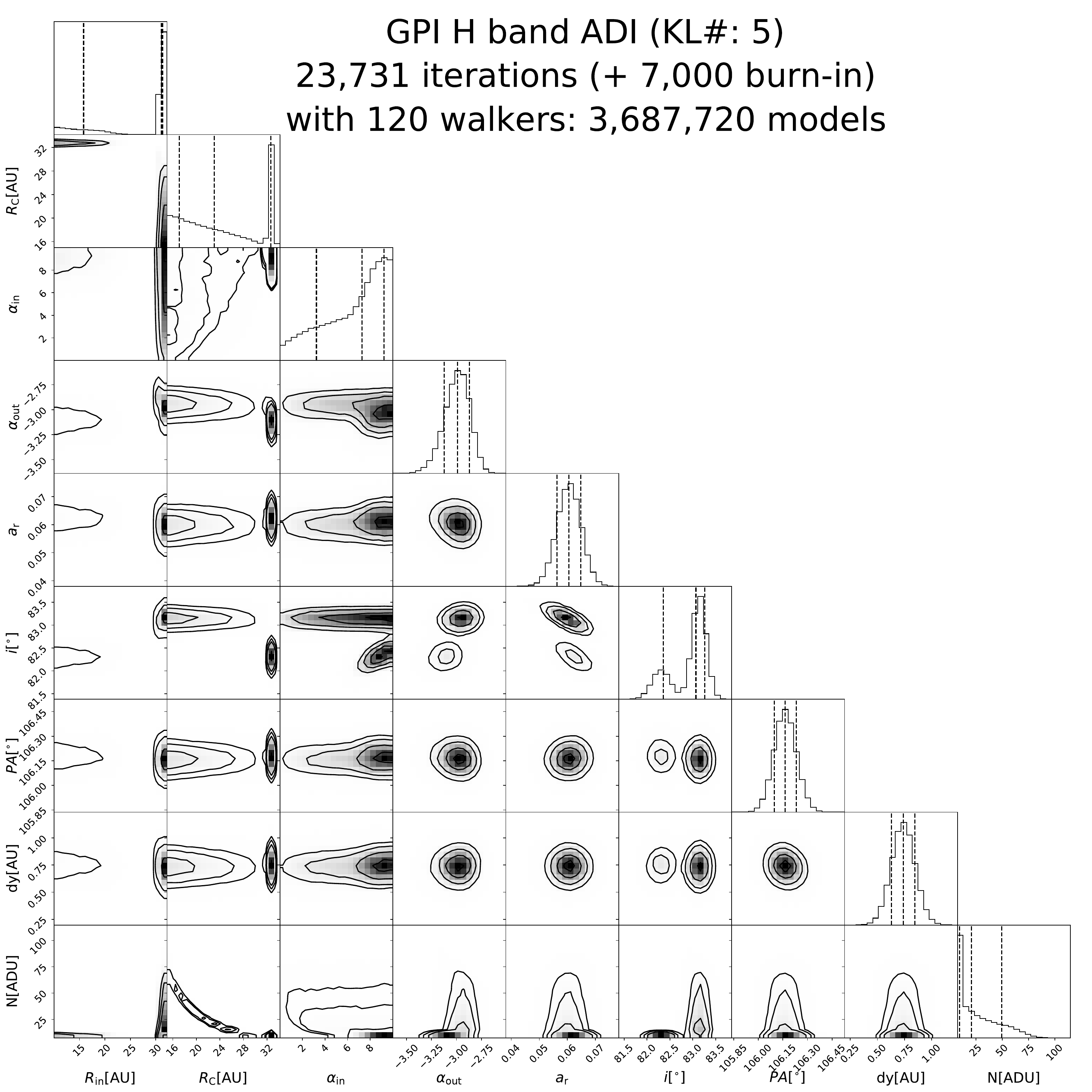}
    \caption{Posterior distribution for HD 114082 \textit{H}-band models.}
    \label{fig:hd114082post}
\end{figure*}
\begin{figure*}
    \centering
    \includegraphics[width=\linewidth]{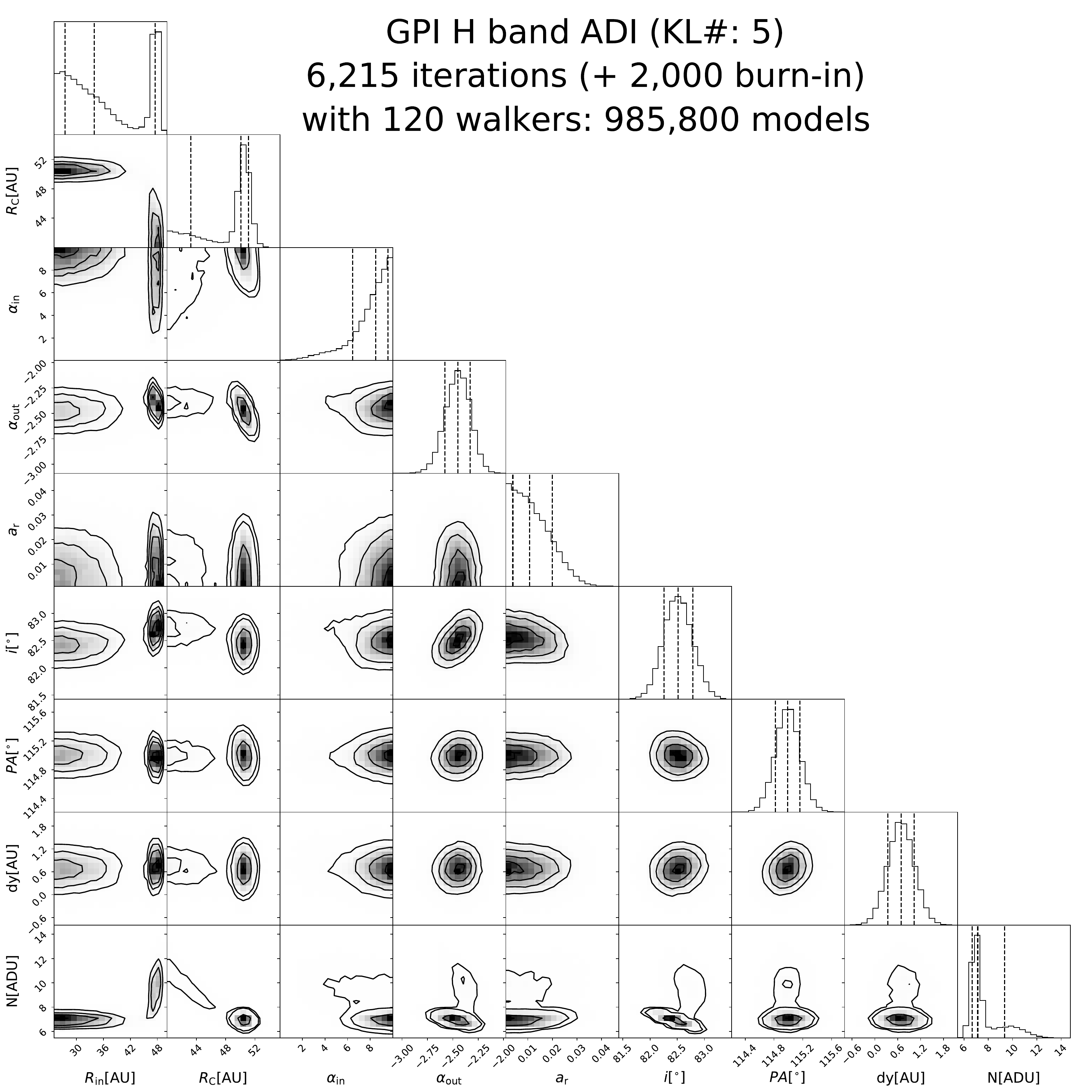}
    \caption{Posterior distribution for HD 146897 \textit{H}-band models.}
    \label{fig:hd146897post}
\end{figure*}
%\begin{figure*}
%    \centering
%    \includegraphics[width=\linewidth]{figs/hr4796_120w_1000i_newgenspf_backend_file_mcmc_pdfs.pdf}
%    \caption{\red{Posterior distribution for the generic SPF HR 4796A \textit{H}-band models.}}
%    \label{fig:hr4796AHpostgen}
%\end{figure*}
\begin{figure*}
    \centering
    \includegraphics[width=\linewidth]{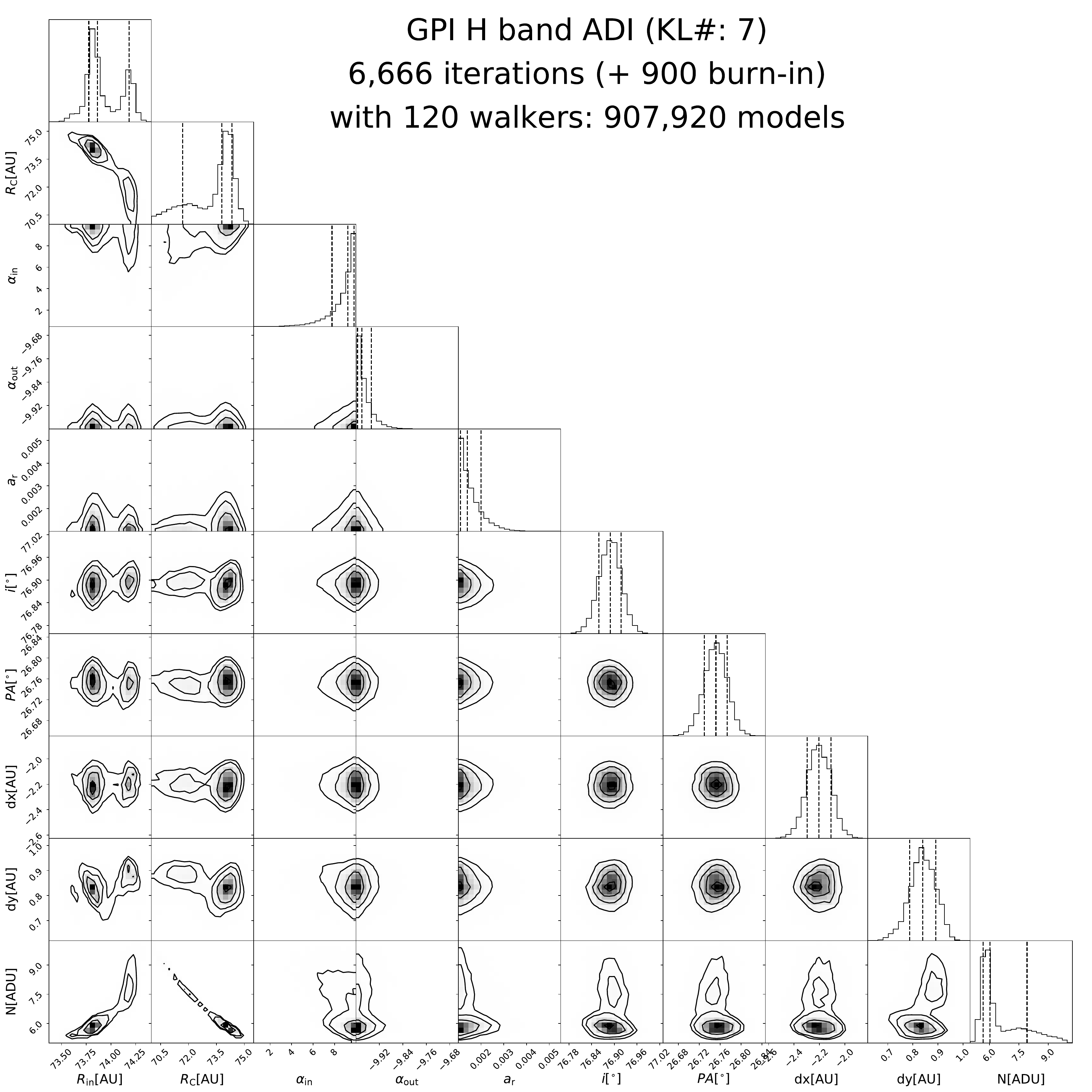}
    \caption{Posterior distribution for the \citet{milli2017} SPF HR 4796A \textit{H}-band models.}
    \label{fig:hr4796AHpost}
\end{figure*}
\begin{figure*}
    \centering
    \includegraphics[width=\linewidth]{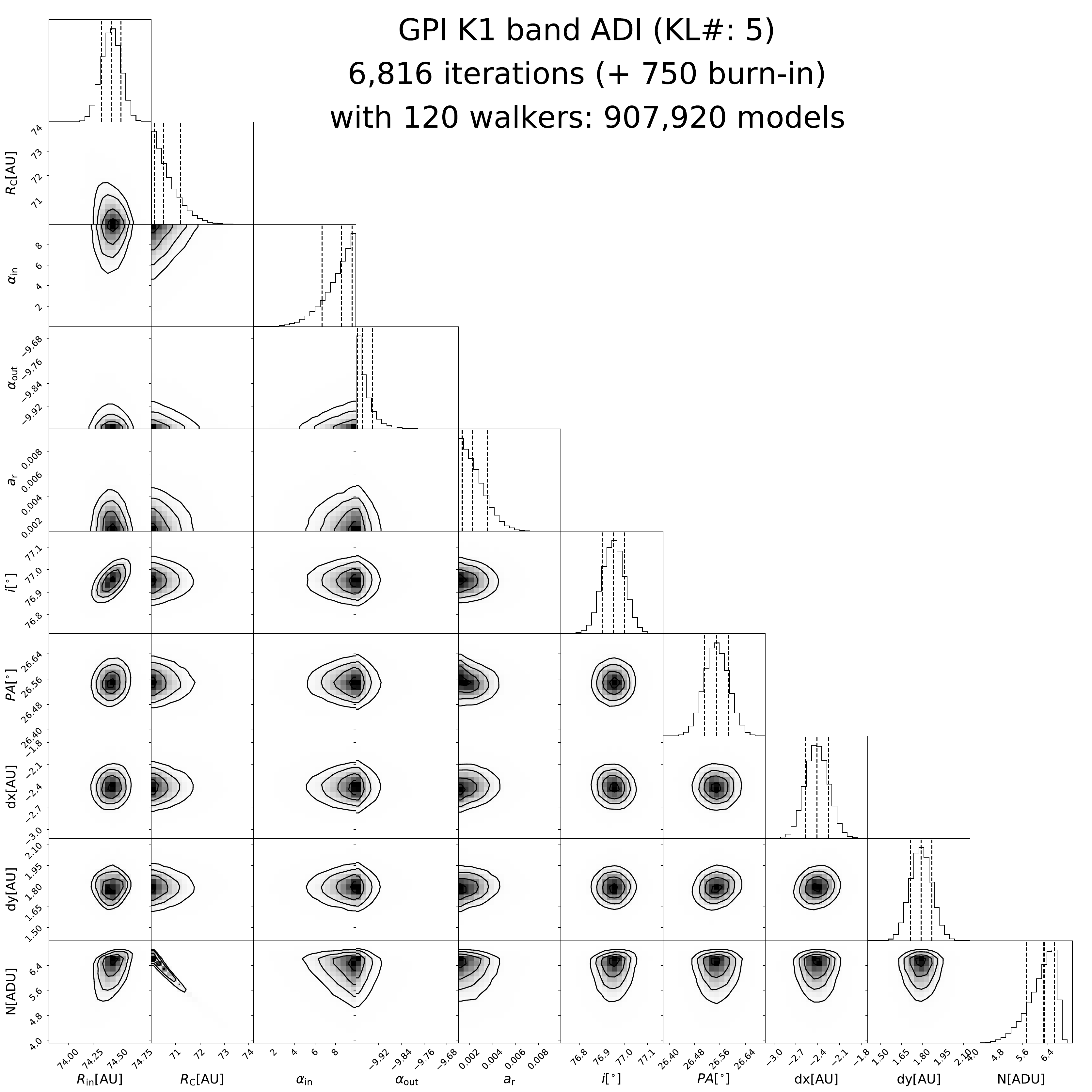}
    \caption{Posterior distribution for the \citet{milli2017} SPF HR 4796A \textit{K1}-band models.}
    \label{fig:hr4796AK1post}
\end{figure*}

%If you want to present additional material which would interrupt the flow of the main paper,
%it can be placed in an Appendix which appears after the list of references.

%%%%%%%%%%%%%%%%%%%%%%%%%%%%%%%%%%%%%%%%%%%%%%%%%%

% Don't change these lines
\bsp	% typesetting comment
\label{lastpage}
\end{document}